\documentclass[11pt]{article}
\usepackage[margin=1in]{geometry}

\usepackage[usenames,dvipsnames]{xcolor}
\definecolor{Gred}{RGB}{219, 50, 54}
\definecolor{ToCgreen}{RGB}{0, 128, 0}

\usepackage[T1]{fontenc}

\usepackage{babel}
\usepackage[spacing=true,kerning=true,babel=true,tracking=true]{microtype}

\DeclareMathAlphabet{\pazocal}{OMS}{zplm}{m}{n} 
\renewcommand{\mathcal}[1]{\pazocal{#1}}

\usepackage{makecell}
\usepackage{dsfont}

\usepackage{multirow}
\usepackage{amsmath,amsthm}
\usepackage{booktabs}   
\usepackage{bm}
\usepackage{bbm}
\usepackage{textgreek}
\usepackage{mathtools}
\usepackage{enumitem}
\usepackage[numbers,comma,sort&compress]{natbib}
\usepackage{authblk}
\usepackage{graphicx}
\usepackage{ragged2e}
\usepackage[font=small]{caption}
\usepackage[labelformat=simple]{subcaption}

\usepackage{float}
\usepackage[linesnumbered,ruled,vlined]{algorithm2e}
\SetKwInput{KwInput}{Input}
\SetKwInput{KwOutput}{Output}
\SetKwInOut{Promise}{Promise}
\SetKwInput{Goal}{Goal}
\SetKwProg{Fn}{function}{}{}
\SetKwFor{RepTimes}{repeat}{times}{}
\SetKwFunction{Ver}{Verify}
\SetKwFunction{Prep}{PrepareState}
\usepackage{algorithmic}

\usepackage{physics}
\usepackage{footnote}
\usepackage{xcolor}
\usepackage{amsfonts}
\usepackage{mathrsfs}
\usepackage{bbm}
\usepackage{braket}
\usepackage[colorlinks]{hyperref}
\usepackage[capitalize,nameinlink]{cleveref}
\hypersetup{
	colorlinks = true,
	linkcolor = red!60!black,
	citecolor = blue!60!black,
	urlcolor = blue!60!black}
\usepackage{placeins}
\usepackage{comment}
\usepackage{textgreek}

\usepackage{tikz,tkz-euclide,tikz-3dplot}
\usetikzlibrary{matrix,fit,backgrounds,3d,arrows, decorations.pathreplacing,shapes.geometric,3d, calc}

\newtheorem{theorem}{Theorem}
\newtheorem{lemma}{Lemma}

\newtheorem{remark}{Remark}
\newtheorem{definition}{Definition}

\newtheorem{problem}{Problem}

\numberwithin{proposition}{section}
\numberwithin{theorem}{section}
\numberwithin{lemma}{section}
\numberwithin{corollary}{section}
\numberwithin{remark}{section}
\numberwithin{definition}{section}
\numberwithin{equation}{section}

\begin{document}

\title{{New aspects of quantum topological data analysis:} \\ \vspace{0.2em}
\large Betti number estimation, and testing and tracking of homology and cohomology classes}

\author{
Nhat A. Nghiem\thanks{Contact author: \href{mailto:nhatanh.nghiemvu@stonybrook.edu}{\texttt{nhatanh.nghiemvu@stonybrook.edu}}, C. N. Yang Institute for Theoretical Physics, State University of New York at Stony Brook, Stony Brook, NY 11794-3840, USA}
}


\maketitle
\thispagestyle{empty}

\begin{abstract}
We introduce several new quantum algorithms for estimating homological invariants, specifically Betti numbers and persistent Betti numbers, of a simplicial complex given via a structured classical input. At the core of our algorithm lies the ability to efficiently construct the block-encoding of Laplacians (and persistent Laplacians) based on the classical description of the given complex. From such block-encodings, Betti numbers (and persistent Betti numbers) can be estimated by incorporating existing techniques, for example, stochastic rank estimation. The complexity of our method is polylogarithmic in the number of simplices in both simplex-sparse and simplex-dense regimes, thus offering an advantage over existing works. 

Moreover, prior quantum algorithms based on spectral methods incur significant overhead due to their reliance on estimating the kernel of combinatorial Laplacians, particularly when the Betti number is small. We introduce a new approach for estimating Betti numbers based on homology tracking and homology property testing, which enables exponential quantum speedups over both classical and prior quantum approaches under sparsity and structure assumptions.

We further initiate the study of homology triviality and equivalence testing as natural property testing problems in topological data analysis, and provide efficient quantum algorithms with time complexity nearly linear in the number of simplices when the rank of the boundary operator is large. In addition, we develop a cohomological approach based on block-encoded projections onto cocycle spaces, enabling rank-independent testing of homology equivalence. This yields the first quantum algorithms for constructing and manipulating $r$-cocycles in time polylogarithmic in the size of the complex. Together, these results establish a new direction in quantum topological data analysis and demonstrate that computing topological invariants can serve as a fertile ground for provable quantum advantage.
\end{abstract}

\newpage
\tableofcontents
\thispagestyle{empty}

\newpage
\setcounter{page}{1}
\section{Introduction}\label{sec: introduction}
The study of topological data analysis (TDA) aims to extract global geometric and topological structure from data. Given a collection of data points equipped with a metric, one typically forms a combinatorial object---for example, a graph or a simplicial complex---that records proximity relations among the points. The resulting topological invariants, such as Betti numbers, capture higher-dimensional features of the data, including connected components, loops, and voids. Since graphs encode only pairwise interactions, they are generally insufficient for detecting such higher-order structure. Simplicial complexes and their associated homology groups provide a natural formalism for this purpose.

Quantum algorithms for TDA have attracted considerable attention over the past several years. Beginning with the work of Lloyd, Garnerone, and Zanardi~\cite{lloyd2016quantum}, a number of papers have proposed quantum procedures for estimating Betti numbers and related topological quantities~\cite{hayakawa2022quantum, mcardle2022streamlined, berry2024analyzing, scali2024quantum, gyurik2024quantum, hayakawa2024quantum, schmidhuber2025quantum, incudini2024testing, crichigno2024clique, king2024gapped, scali2024topology, rayudu2024fermionic}. These works combine tools from algebraic topology, Hamiltonian simulation, and quantum linear algebra, and have helped clarify the algorithmic potential of quantum computation in topology.

Recent complexity-theoretic results indicate that such approaches are inherently limited under standard input models. In particular, Schmidhuber and Lloyd~\cite{schmidhuber2022complexity} showed that Betti number estimation is NP-hard for general graphs. For clique complexes, stronger hardness results are known: Crichigno and Kohler~\cite{crichigno2024clique} proved that the counting version is \#BQP-complete, while the corresponding decision problem with exponentially small promise gap is QMA$_1$-hard. Together with subsequent work by Rudolph~\cite{rudolph2024towards}, this further implies PSPACE-completeness for clique complexes. Consequently, under standard complexity-theoretic assumptions, efficient quantum algorithms for Betti number estimation are unlikely to exist when the input is provided solely via pairwise connectivity.

These results suggest that the central issue is not homological computation itself, but rather the way the simplicial complex is specified. In particular, they leave open the possibility of efficient quantum algorithms under richer input models in which higher-order combinatorial structure is provided explicitly. Motivated by this perspective, we study a collection of topology-related computational problems under an alternative representation of the simplicial complex. Our goal is to identify settings in which quantum algorithms remain efficient and to understand the corresponding complexity tradeoffs.

\begin{figure*}
    \centering
    \includegraphics[width=\linewidth]{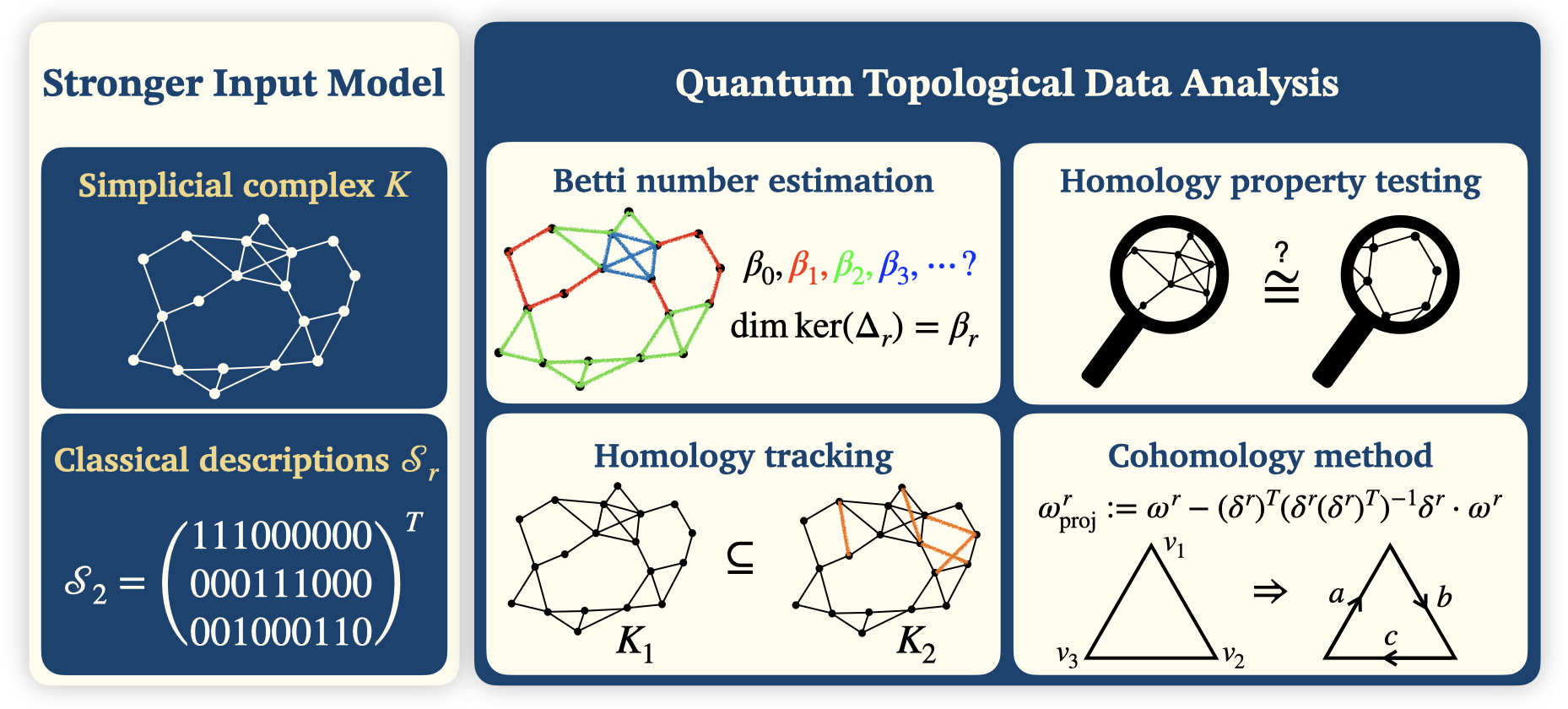}
    \caption{\justifying \textbf{Overview of our results.} We study quantum algorithms for topological data analysis under explicit classical access to a simplicial complex $K$ via its specification matrices $\{\mathcal{S}_r\}$. Our contributions include algorithms for estimating Betti numbers $\beta_r$ and persistent Betti numbers, as well as procedures for homology testing (triviality and equivalence). We also discuss applications to tracking homology classes and introduce a cohomological perspective based on differential operators.}
    \label{fig:overview}
\end{figure*}

\subsection{Motivation and problem statement}

A standard pipeline in topological data analysis begins with a finite set of data points equipped with a metric. From this, one constructs a simplicial complex by introducing edges between nearby points and extending to higher-dimensional simplices according to a chosen rule (e.g., via a threshold parameter). The resulting complex $K$ encodes multi-scale geometric structure of the data.

The central computational task is to extract topological invariants of $K$, most notably the Betti numbers $\{\beta_r\}$, which quantify the number of independent $r$-dimensional features (such as connected components, loops, and higher-dimensional voids). Formally, letting $S_r^K = \{\sigma_{r_i}\}_{i=1}^{|S_r^K|}$ denote the set of $r$-simplices, the space of $r$-chains is the vector space spanned by $S_r^K$, and the boundary operators $\{\partial_r\}$ define a chain complex. The associated combinatorial Laplacian is given by
\begin{equation}\label{eq: r_th_comb_laplacian}
    \Delta_r = \partial_{r+1}\partial_{r+1}^\dagger + \partial_r^\dagger \partial_r,
\end{equation}
and the $r$-th Betti number satisfies $\beta_r = \dim \ker(\Delta_r)$. A natural question is therefore:

\begin{center}
    \emph{Given a simplicial complex $K$, compute its Betti numbers $\{\beta_r\}$.}
\end{center}

Several quantum algorithms have been proposed for this problem~\cite{lloyd2016quantum, berry2024analyzing, hayakawa2022quantum, mcardle2022streamlined}, typically assuming oracle access to pairwise connectivity of the underlying data. Under this model, one can construct block-encodings of $\Delta_r$ and estimate its kernel dimension using techniques such as quantum phase estimation.

However, recent complexity-theoretic results place strong limitations on this approach. These hardness results suggest that efficient quantum algorithms are unlikely to exist in this setting without additional structural assumptions.

This motivates the study of alternative input models in which richer information about the simplicial complex is available. In particular, we consider settings where the complex is specified explicitly through its higher-order combinatorial structure, rather than implicitly through pairwise connectivity alone. Our goal is to understand whether such representations admit efficient quantum algorithms for topological inference.

\subsection{Main results and technical overview}\label{sec: overview}

We now present a high-level overview of our model, techniques, and main results. We also situate our contributions within the broader landscape of quantum topological data analysis. Formal statements and detailed proofs are deferred to subsequent sections. A schematic summary is given in~\cref{fig:overview}.

\subsubsection{Input model for the simplicial complex}

Motivated by the limitations of oracle-based input models, we consider an alternative representation in which the simplicial complex is specified explicitly. Let $S_r^K$ denote the set of $r$-simplices in $K$, and write $K = \{ S_r^K \}_{r=0}^{n-1}$, where $n$ is the number of vertices.

We assume access to a \emph{specification} of $K$ in the following form. For each $r \ge 1$, we are given a matrix
\begin{equation}
    \mathcal{S}_r \in \{0,1\}^{|S_{r-1}^K| \times |S_r^K|},
\end{equation}
where the $i$-th column corresponds to an $r$-simplex $\sigma_{r_i} \in S_r^K$, and its nonzero entries indicate the $(r-1)$-simplices that are faces of $\sigma_{r_i}$. Equivalently, each column encodes the incidence structure between an $r$-simplex and its $(r-1)$-faces.

Under this representation, we consider the following computational problem.

\begin{problem}[Betti number estimation with matrix specification]\label{prob: betti_classical}
Given a simplicial complex $K$ specified by matrices $\{\mathcal{S}_r\}$, estimate its Betti numbers $\{\beta_r\}$.
\end{problem}

This formulation contrasts with prior quantum TDA models~\cite{lloyd2016quantum, berry2024analyzing, schmidhuber2022complexity, ubaru2021quantum}, where the input is given implicitly through oracle access to pairwise connectivity. In our setting, the higher-order combinatorial structure of the complex is explicitly available via $\{\mathcal{S}_r\}$.

We briefly outline two natural scenarios in which such an input model may arise.
\begin{itemize}
    \item \textbf{Classical enumeration.}
    Suppose the underlying graph of $K$ is given explicitly, including its vertex and edge sets. Classical algorithms~\cite{chiba1985arboricity, eppstein2010listing} can then be used to enumerate higher-dimensional simplices. This procedure yields the matrices $\{\mathcal{S}_r\}$, with complexity depending on structural properties of the graph, such as bounded degree. Recent work~\cite{nghiem2025hybrid} adopts a similar approach in a hybrid classical--quantum framework.

    \item \textbf{Direct specification.}
    In some applications, one may directly construct a simplicial complex of interest by specifying a collection of simplices. In this case, the matrices $\{\mathcal{S}_r\}$ form part of the input. More generally, one may assign entries in $\mathcal{S}_r$ (e.g., from $\{0,\pm1\}$) to encode a chosen combinatorial structure, and subsequently analyze its topology using the algorithms developed here.
\end{itemize}

\subsubsection{Block-encoding of $\{\partial_r^\dagger \partial_r\}$}

Given the specification matrices $\{\mathcal{S}_r\}$ of the simplicial complex $K$, our goal is to construct block-encodings of the operators $\{\partial_r^\dagger \partial_r\}$ (see~\cref{sec: reviewalgebraictopology} for the definition of $\partial_r$). To achieve this, we require access to the entries of $\{\mathcal{S}_r\}$. We consider two standard access models.

\begin{itemize}
    \item \textbf{Classical access.}
    The entries of $\{\mathcal{S}_r\}$ are explicitly given, and hence the boundary operators $\{\partial_r\}$ are directly determined. We obtain the following.

    \begin{lemma}
    \label{lemma: entrycomputablematrix}
    Given classical access to $\mathcal{S}_r$, and hence to $\partial_r$, there exists a quantum procedure using
    \begin{equation}
        \mathcal{O}\!\left(\log(|S_{r-1}^K||S_r^K|)\right)
    \end{equation}
    qubits and circuit complexity
    \begin{equation}
        \mathcal{O}\!\left(\log(|S_{r-1}^K||S_r^K|)\right)
    \end{equation}
    that block-encodes
    \begin{equation}
        \frac{\partial_r^\dagger \partial_r}{\|\partial_r\|_F^2}
        =
        \frac{\partial_r^\dagger \partial_r}{r|S_r^K|},
    \end{equation}
    using $\mathcal{O}(1)$ ancilla qubits and $\mathcal{O}(1)$ classical preprocessing. Here $\|\partial_r\|_F = \sqrt{r|S_r^K|}$.
    \end{lemma}
    \begin{proof}
    The proof is given in~\cref{sec: proofoflemmaentrycomputablematrix}.    
    \end{proof}

    \item \textbf{Sparse access.}
    We assume that $\mathcal{S}_r$ has sparsity $r$ and is accessed via the standard oracle model:
    \begin{align}
        O_r \ket{i}\ket{k} &= \ket{i}\ket{s_{ik}}, \\
        O_A \ket{i}\ket{j}\ket{\mathbf{0}} &= \ket{i}\ket{j}\ket{(\mathcal{S}_r)_{ij}},
    \end{align}
    where $1 \le i \le |S_r^K|$, $1 \le j \le |S_{r-1}^K|$, and $k \le r$. Since $\partial_r$ is determined by $\mathcal{S}_r$, this induces sparse access to $\partial_r$, which can be block-encoded via~\cref{lemma: blockencodesparseaccess}.

    \begin{lemma}[Based on~\cref{lemma: blockencodesparseaccess}]
    \label{lemma: sparsepartial}
    Given sparse access to $\mathcal{S}_r$, and hence to $\partial_r$, there exists a quantum circuit that $\epsilon$-approximately block-encodes
    \[
        \frac{\partial_r}{r\lambda_r},
    \]
    where $\lambda_r$ is the largest singular value of $\partial_r$, with complexity
    \begin{equation}
        \mathcal{O}\left(
            \log(|S_r^K|)
            +
            \log^{2.5}\!\left(\frac{1}{\epsilon}\right)
        \right).
    \end{equation}
    \end{lemma}
\end{itemize}

The normalization factors in the two access models differ. We unify them via rescaling. Let
\begin{equation}
    \Lambda := \max_r \lambda_r .
\end{equation}
In the classical-access model, we amplify the block-encoding of $\partial_r^\dagger \partial_r/(r|S_r^K|)$ by a factor $r|S_r^K|/\Lambda^2$, thereby obtaining a block-encoding of $\partial_r^\dagger \partial_r/\Lambda^2$. In the sparse-access model, we rescale $\partial_r/(r\lambda_r)$ to obtain $\partial_r/\Lambda$, and then use standard block-encoding composition to construct $\partial_r^\dagger \partial_r/\Lambda^2$.

Combining the constructions for $r$ and $r+1$, we obtain a block-encoding of the combinatorial Laplacian
\begin{equation}
    \frac{\Delta_r}{2\Lambda^2}
    =
    \frac{1}{2}
    \left(
        \frac{\partial_r^\dagger \partial_r}{\Lambda^2}
        +
        \frac{\partial_{r+1}^\dagger \partial_{r+1}}{\Lambda^2}
    \right).
\end{equation}

\begin{lemma}
\label{lemma: blockencodingpartialr}
For every $r$, the following hold.

\begin{itemize}
    \item \textbf{\textnormal{(Classical access)}} There exists a block-encoding of $\partial_r^\dagger \partial_r/\Lambda^2$ with complexity
    \begin{equation}
        \mathcal{O}\left(
            r|S_r^K| \log(|S_r^K||S_{r-1}^K|)
        \right).
    \end{equation}
    Moreover, $\Delta_r/(2\Lambda^2)$ can be block-encoded with complexity
    \begin{equation}
        \mathcal{O}\!\left(
            r(|S_r^K|+|S_{r+1}^K|)
            \log(|S_{r+1}^K||S_r^K||S_{r-1}^K|)
        \right).
    \end{equation}

    \item \textnormal{(Sparse access)} There exists an $\epsilon$-approximate block-encoding of $\partial_r^\dagger \partial_r/\Lambda^2$ with complexity
    \begin{equation}
        \mathcal{O}\left(
            r\left(
                \log(|S_r^K|)
                +
                \log^{2.5}\!\left(\frac{1}{\epsilon}\right)
            \right)
        \right).
    \end{equation}
    Moreover, $\Delta_r/(2\Lambda^2)$ can be block-encoded with complexity
    \begin{equation}
        \mathcal{O}\left(
            r\left(
                \log(|S_r^K||S_{r+1}^K|)
                +
                \log^{2.5}\!\left(\frac{1}{\epsilon}\right)
            \right)
        \right).
    \end{equation}
\end{itemize}
\end{lemma}

We will use these constructions as building blocks for estimating Betti numbers and persistent Betti numbers. To this end, we rely on the following standard subroutine.

\begin{lemma}[Rank estimation]
\label{lemma: traceestimation}
Let $A \in \mathbb{C}^{N \times N}$ be given via a block-encoding with cost $T_A$, and suppose $\|A\| \in (1/\kappa_A,1)$. Then there exists a quantum algorithm that estimates $\mathrm{rank}(A)/N$ to additive error $\epsilon$ with success probability at least $1-\xi$ using
\begin{equation}
    \mathcal{O}\!\left(
        \frac{T_A \kappa_A}{\sqrt{\epsilon}}
        \log\!\left(\frac{1}{\epsilon}\right)
        \log\!\left(\frac{1}{\xi}\right)
    \right)
\end{equation}
time.
\end{lemma}
\begin{proof}
The proof is given in~\cref{sec: rankestimation}. 
\end{proof}

\subsubsection{Estimating Betti numbers}
The overall procedure for our quantum algorithm for estimating Betti numbers is summarized in~\cref{fig: betti_workflow}.

\begin{algorithm}[t]
\DontPrintSemicolon
\textbf{Input:} Matrix specification $\{\mathcal{S}_r\}$ of a simplicial complex $K$.

Construct the boundary maps $\partial_r$ from $\mathcal{S}_r$.

Construct a block-encoding of $\frac{1}{\Lambda^2}\partial_r^\dagger \partial_r$.

Construct a block-encoding of the combinatorial Laplacian $\frac{1}{\Lambda^2}\Delta_r$.

Estimate $\operatorname{rank}(\Delta_r)/|S_r^K|$ using the rank estimation procedure
(\cref{lemma: traceestimation}).

Compute
\[
\frac{\beta_r}{|S_r^K|}
= 1 - \frac{\operatorname{rank}(\Delta_r)}{|S_r^K|}.
\]

\textbf{Output:} Normalized Betti number $\beta_r / |S_r^K|$.

\caption{Quantum algorithm for estimating normalized Betti numbers}
\label{fig: betti_workflow}
\end{algorithm}

We define the quantity ${\dim\ker(\Delta_r)}/{|S_r^K|} = {\beta_r}/{|{S_r^K}|}$
as the \emph{$r$-th normalized Betti number}. The algorithm described above yields the following result:

\begin{theorem}[Time complexity of estimating (normalized) Betti numbers, see~\cref{sec: estimatingBettinumbers}]
\label{thm: estimatingbettinumbers}
Let $\{ \mathcal{S}_r \}$ be the specification of a simplicial complex $K$, and let $\Lambda = \max_r \lambda_r$ denote the maximum singular value of the boundary operators $\{ \partial_r \}$. Then the following hold.

\begin{itemize}
    \item \textnormal{(Classical access)} 
    The normalized Betti number $\beta_r / |S_r^K|$ can be estimated to additive error $\varepsilon$ in time
    \begin{equation}
        \mathcal{O}\!\left(
        \Lambda^2 \frac{1}{\sqrt{\varepsilon}} \,
        r (|S_r^K| + |S_{r+1}^K|)\,
        \log\!\big( |S_{r+1}^K|\,|S_r^K|\,|S_{r-1}^K| \big)
        \right).
    \end{equation}
    Moreover, $\beta_r$ can be estimated to multiplicative error $\delta$ in time
    \begin{equation}
        \mathcal{O}\!\left(
        \Lambda^2 \frac{1}{\sqrt{\delta}} \,
        \sqrt{\frac{|S_r^K|}{\beta_r}}\,
        r (|S_r^K| + |S_{r+1}^K|)\,
        \log\!\big( |S_{r+1}^K|\,|S_r^K|\,|S_{r-1}^K| \big)
        \right).
    \end{equation}

    \item \textnormal{(Sparse access)} 
    The normalized Betti number $\beta_r / |S_r^K|$ can be estimated to additive error $\varepsilon$ in time
    \begin{equation}
        \mathcal{O}\!\left(
        r \Lambda^2 \frac{1}{\sqrt{\varepsilon}}\,
        \left(
        \log(|S_r^K|\,|S_{r+1}^K|)
        + \log^{2.5}\!\frac{1}{\varepsilon}
        \right)
        \right).
    \end{equation}
    Moreover, $\beta_r$ can be estimated to multiplicative error $\delta$ in time
    \begin{equation}
        \mathcal{O}\!\left(
        r \Lambda^2 \frac{1}{\sqrt{\delta}}\,
        \sqrt{\frac{|S_r^K|}{\beta_r}}\,
        \left(
        \log(|S_r^K|\,|S_{r+1}^K|)
        + \log^{2.5}\!\frac{1}{\varepsilon}
        \right)
        \right).
    \end{equation}
\end{itemize}
\end{theorem}

As we summarize in~\cref{sec: TDA}, to the best of our knowledge, the state-of-the-art quantum algorithm (referred to as the LGZ algorithm) achieves the following time complexities for estimating normalized Betti numbers (to additive precision $\varepsilon$) and unnormalized Betti numbers (to multiplicative precision $\delta$), respectively:
\begin{equation}
    \mathcal{O}\! \left(\frac{1}{\varepsilon}\cdot\left(n^2 \sqrt{\frac{\binom{n}{r+1}}{|{S_r^K}|}} + n\kappa\right)\right), \quad
    \mathcal{O}\! \left(\frac{1}{\delta} \cdot \left( n^2 \sqrt{\frac{\binom{n}{r+1}}{\beta_r}} + n\kappa \sqrt{\frac{|{S_r^K}|}{\beta_r}} \right) \right),
\end{equation}
where $\kappa$ denotes the condition number of the combinatorial Laplacian $\Delta_r$.

\paragraph{Comparison to existing quantum algorithms.}
For normalized Betti numbers $\beta_r / |S_r^K|$, the LGZ algorithm is efficient primarily in the dense regime, where $|S_r^K|$ is large. In particular, when
\[
    \frac{\binom{n}{r+1}}{|S_r^K|} = \mathcal{O}(1),
    \quad \text{i.e.,} \quad
    |S_r^K| \sim \binom{n}{r+1},
\]
the LGZ algorithm achieves time complexity $\mathcal{O}(n^2 + n\kappa)$. 

In contrast, as long as $|S_r^K| \in \mathcal{O}(\mathrm{poly}(n))$, our algorithm achieves time complexity $\mathcal{O}(\mathrm{poly}(\log n))$ in the sparse-access model, yielding a \emph{superpolynomial} speed-up over LGZ. This advantage is further amplified when
\[
    \frac{\binom{n}{r+1}}{|S_r^K|} \in \mathcal{O}(\mathrm{poly}(n)).
\]

In the dense regime, the classical-access model of our algorithm runs in time $\mathcal{O}(|S_r^K|)$, and thus matches LGZ-type complexities only when $|S_r^K| \sim n^2$ (e.g., for $r \leq 2$). For larger $r$, it is polynomially slower in $n$.

In the \emph{sparse regime}, where $|S_r^K| \ll \binom{n}{r+1}$, we have
\[
    \frac{\binom{n}{r+1}}{|S_r^K|} \approx \binom{n}{r+1},
\]
and the LGZ algorithm incurs significant overhead. If $\binom{n}{r+1} \in \mathcal{O}(\mathrm{poly}(n))$, our sparse-access algorithm achieves a superpolynomial speed-up; if $\binom{n}{r+1} \in \exp(n)$, the speed-up becomes \emph{nearly exponential}. 

Within the classical-access model, the degree of speed-up depends on the ratio $\binom{n}{r+1}/|S_r^K|$: it is polynomial when this ratio is $\mathrm{poly}(n)$, and exponential when it is $\exp(n)$.

For estimating the (unnormalized) Betti numbers $\beta_r$, both algorithms perform best when $\beta_r = \Theta(|S_r^K|)$, i.e., when the normalized Betti number is constant. In this regime, our algorithm runs in time
\[
    \mathcal{O}\!\left(
    \log(|S_{r-1}^K|\,|S_r^K|)
    \cdot
    \log(|S_r^K|\,|S_{r+1}^K|)
    \right),
\]
whereas the LGZ algorithm requires
\begin{equation}
    \mathcal{O}\!\left(
    n^2 \sqrt{\frac{\binom{n}{r+1}}{\beta_r}}
    \right)
    \approx
    \mathcal{O}\!\left(
    n^2 \sqrt{\frac{\binom{n}{r+1}}{|S_r^K|}}
    \right).
\end{equation}
Thus, the qualitative comparison remains the same: our algorithm achieves a superpolynomial speed-up in the dense regime and a nearly exponential speed-up in the sparse regime.

\paragraph{Comparison to classical algorithms.}
A straightforward classical approach to estimating $\beta_r$ is to diagonalize the combinatorial Laplacian $\Delta_r$, e.g., via Gaussian elimination or exact diagonalization, which requires time $\mathcal{O}(|S_r^K|^3)$. In comparison, our algorithm achieves exponential speed-up in the sparse-access model and polynomial speed-up in the classical-access model when $\beta_r = \Theta(|S_r^K|)$. Notably, this is also the regime where LGZ-type algorithms perform optimally~\cite{lloyd2016quantum, berry2024analyzing, schmidhuber2022complexity, ubaru2021quantum}.

More recently,~\cite{memoli2022persistent} proposed a classical algorithm with runtime $\mathcal{O}(|S_r^K|^\omega)$, where $\omega \approx 2.807$ is the matrix multiplication exponent. Even in this case, our algorithm achieves a superpolynomial improvement in the regime $\beta_r = \Theta(|S_r^K|)$.

\begin{algorithm}[t]
\DontPrintSemicolon
\textbf{Input:} Matrix specification of two simplicial complexes $K_1 \subseteq K_2$.

Construct the boundary maps $\partial_r^{K_1}$ and $\partial_r^{K_2}$ from the classical input.

Construct block-encodings of $\frac{1}{\Lambda^2}(\partial_r^{K_1})^\dagger \partial_r^{K_1}$ and 
$\frac{1}{\Lambda^2}(\partial_r^{K_2})^\dagger \partial_r^{K_2}$.

Construct a block-encoding of the $r$-th persistent combinatorial Laplacian
\[
\Delta_r^{K_1,K_2}
=
\partial_{r+1}^{K_1,K_2}(\partial_{r+1}^{K_1,K_2})^\dagger
+
(\partial_r^{K_1})^\dagger \partial_r^{K_1}.
\]

Estimate $\operatorname{rank}(\Delta_r^{K_1,K_2}) / |S_r^K|$ using the rank estimation procedure 
(\cref{lemma: traceestimation}).

Compute
\[
\frac{\beta_r^{\mathrm{pers}}}{|S_r^K|}
=
1 - \frac{\operatorname{rank}(\Delta_r^{K_1,K_2})}{|S_r^K|}.
\]

\textbf{Output:} Normalized persistent Betti number $\beta_r^{\mathrm{pers}} / |S_r^K|$.

\caption{Quantum algorithm for estimating normalized persistent Betti numbers}
\label{fig: persistentbetti_workflow}
\end{algorithm}

\subsubsection{Estimating persistent Betti numbers}
We remark that the above result assumes a fixed simplicial complex $K$, denoted hereafter by $K_1$, constructed from pairwise connectivity under a given threshold (or length scale). As the threshold increases, additional connections are formed among the data points, yielding a denser simplicial complex $K_2$. It is straightforward to verify that $K_1 \subseteq K_2$, since all connections present in $K_1$ are preserved in $K_2$, while $K_2$ may contain additional pairwise connections. While Betti numbers quantify the ``holes'' within a given complex (e.g., $K_1$), the \emph{persistent Betti numbers} capture the topological features that persist from the earlier complex (corresponding to a smaller threshold) to the later one (with a larger threshold). Our objective is thus to compute the persistent Betti numbers associated with the inclusion $K_1 \subseteq K_2$. This problem was previously considered in \cite{hayakawa2022quantum}, where the input model assumes oracle access encoding pairwise connectivity of the underlying data points at two distinct thresholds—essentially requiring two separate oracles for $K_1$ and $K_2$. As before, we depart from the oracle-based model and instead assume that classical descriptions of $K_1$ and $K_2$ are available. This leads us to the following reformulation of the computational task:

\begin{problem}[Persistent Betti number estimation with classical description, see~\cref{sec: estimatingpersistentbettinumbers}]\label{prob: persistent_betti}
Given two simplicial complexes $K_1 \subseteq K_2$, specified via classical boundary matrix descriptions $\{ \mathcal{S}_r^{K_1} \}$ and $\{ \mathcal{S}_r^{K_2} \}$, estimate the persistent Betti numbers associated with the inclusion $K_1 \subseteq K_2$. 
\end{problem}

Our algorithm builds on the method proposed in \cite{hayakawa2022quantum}. To estimate persistent Betti numbers, one must consider the \emph{persistent combinatorial Laplacian}. Recall that the standard $r$-th combinatorial Laplacian is defined by~\cref{eq: r_th_comb_laplacian}. The $r$-th persistent combinatorial Laplacian is given by
\begin{align}
    \Delta_r^{K_1,K_2} = \partial_{r+1}^{K_1,K_2} (\partial_{r+1}^{K_1,K_2})^\dagger + (\partial_r^{K_1})^\dagger \partial_r^{K_1}
\end{align}
where $\partial_{r+1}^{K_1,K_2}$ denotes the restriction of $\partial_{r+1}^{K_2}$ to an appropriate subspace. A more detailed discussion of this operator, including the construction of the first term, appears in~\cref{sec: estimatingpersistentbettinumbers}. We emphasize here that the spectrum of $\Delta_r^{K_1,K_2}$ encodes the persistent Betti numbers; specifically, $\beta_r^{\text{persistent}} = \dim \ker(\Delta_r^{K_1,K_2})$. While the full algorithmic procedure is deferred to~\cref{sec: estimatingpersistentbettinumbers}, we summarize the essential steps in~\cref{fig: persistentbetti_workflow} and yields the following complexity result:

\begin{theorem}[Time complexity of estimating (normalized) persistent Betti numbers, see~\cref{sec: estimatingpersistentbettinumbers}]
\label{thm: estimatingpersistentbettinumbers}
Let $K_1 \subseteq K_2$ be simplicial complexes with specifications $\{\mathcal{S}_r^{K_1}\}$ and $\{\mathcal{S}_r^{K_2}\}$. Let $\Lambda = \max_r \lambda_r$ denote the maximum singular value among $\{\partial_r^{K_1}\}$. Then the normalized persistent Betti number $\beta_r^{\mathrm{persistent}} / |S_r^{K_1}|$ can be estimated to additive error $\varepsilon$ with the following complexities:

\begin{itemize}
    \item \textnormal{(Classical access)} with time
    \begin{equation}
        \mathcal{O}\left(
        \Lambda^2 \frac{1}{\sqrt{\varepsilon}} \log\left(\frac{1}{\varepsilon}\right)
        \cdot \mathcal{C}_r^{\mathrm{cl}}
        \right),
    \end{equation}
    where
    \begin{align}
        \mathcal{C}_r^{\mathrm{cl}}
        :=
        &\, r |S_r^{K_1}|
        \log\!\big(|S_{r-1}^{K_1}|\,|S_r^{K_1}|\,|S_{r+1}^{K_1}|\big) \notag \\
        &+ r (|S_{r+1}^{K_2}| - |S_{r+1}^{K_1}|)
        \log\!\big(|S_r^{K_1}| (|S_{r+1}^{K_2}| - |S_{r+1}^{K_1}|)\big) \notag \\
        &+ (|S_r^{K_2}| - |S_r^{K_1}|)(|S_{r+1}^{K_2}| - |S_{r+1}^{K_1}|)
        \log\!\big((|S_r^{K_2}| - |S_r^{K_1}|)(|S_{r+1}^{K_2}| - |S_{r+1}^{K_1}|)\big).
    \end{align}

    \item \textnormal{(Sparse access)} with time
    \begin{equation}
        \mathcal{O}\left(
        \Lambda^2 \frac{1}{\sqrt{\varepsilon}} \log\left(\frac{1}{\varepsilon}\right)
        \cdot \mathcal{C}_r^{\mathrm{sp}}
        \right),
    \end{equation}
    where
    \begin{align}
        \mathcal{C}_r^{\mathrm{sp}}
        :=
        &\, r \log\!\big(|S_{r-1}^{K_1}|\,|S_r^{K_1}|\,|S_{r+1}^{K_1}|\big) \notag \\
        &+ r \log\!\big(|S_r^{K_1}| (|S_{r+1}^{K_2}| - |S_{r+1}^{K_1}|)\big) \notag \\
        &+ \log\!\big((|S_r^{K_2}| - |S_r^{K_1}|)(|S_{r+1}^{K_2}| - |S_{r+1}^{K_1}|)\big).
    \end{align}
\end{itemize}
\end{theorem}

\paragraph{Comparison to existing work.}
For comparison, we recall the bounds from~\cite{hayakawa2022quantum} for estimating the $r$-th normalized persistent Betti number to additive error $\varepsilon$, and the persistent Betti number to multiplicative error $\delta$:
\begin{equation}
    \mathcal{O}\left(
        \frac{\log(1/\varepsilon)}{\varepsilon^2}
        \left(
            r n^2 \sqrt{ \frac{ \binom{n}{r+1} }{ |S_r^{K_1}| } }
            + r^4 n^8 \log(1/\varepsilon)
        \right)
    \right),
\end{equation}
\begin{equation}
    \mathcal{O}\left(
        \frac{|S_r^{K_1}|^{3/2}}{\delta^2 (\beta_r^{\mathrm{persistent}})^2}
        \left(
            r n^2 \sqrt{ \binom{n}{r+1} }
            + r^4 n^8 |S_r^{K_1}|^{1/2}
            \log\left(\frac{|S_r^{K_1}|}{\delta \beta_r^{\mathrm{persistent}}}\right)
        \right)
        \log\left(\frac{|S_r^{K_1}|}{\delta \beta_r^{\mathrm{persistent}}}\right)
    \right).
\end{equation}

If the increment $|S_{r+1}^{K_2}| - |S_{r+1}^{K_1}|$ is small (e.g., $\mathcal{O}(1)$), the comparison reduces to that for ordinary Betti number estimation. In this regime, our algorithm achieves a \emph{superpolynomial} speed-up in the sparse regime and a \emph{near-exponential} speed-up when $\binom{n}{r+1}$ is exponential.

\paragraph{Comparison to classical algorithms.}
The best-known classical algorithm for persistent Betti numbers~\cite{memoli2022persistent} runs in time $\mathcal{O}(|S_r^K|^\omega)$, where $\omega \approx 2.807$. As in the non-persistent case, our algorithm achieves a superpolynomial improvement in the regime $\beta_r = \Theta(|S_r^{K_1}|)$.

\subsubsection{Testing for trivial homology classes}
Thus far, we have considered the estimation of (normalized) Betti numbers and persistent Betti numbers, quantities that encapsulate topological invariants of the underlying simplicial complex. Motivated by this, we now turn to a related foundational task in algebraic topology that extends beyond the computation of Betti numbers. Subsequently, we also discuss how it can be applied to TDA. Specifically, we consider the following:
\begin{problem}[Homology triviality testing with matrix specification, see~\cref{sec: zerohomologyclass}]\label{prob: homology_detection}
    Given the classical description $\{ \mathcal{S}_r \}$ of a simplicial complex $K$, and an $r$-cycle $c_r$, determine whether $c_r$ is homologous to zero.
\end{problem}

Our solution to~\cref{prob: homology_detection} relies on the algebraic structure of chain complexes: $r$-simplices serve as a basis for a vector space, and an $r$-cycle $c_r$ is a (formal) linear combination of $r$-simplices, which can be represented as a vector in this space. A necessary and sufficient condition for $c_r$ to be homologous to zero is that the following linear equation admits a solution: $\partial_{r+1} c_{r+1} = c_r$, which directly follows from the definition: an $r$-cycle is homologous to zero if it is the boundary of some $(r{+}1)$-chain (see~\cref{sec: reviewalgebraictopology} for further details).  To determine whether a solution exists, it suffices to check whether the matrices $\partial_{r+1}$ and $[\partial_{r+1}|c_r]$ have the same rank. A more detailed discussion is provided in~\cref{sec: zerohomologyclass}, and we summarize our algorithmic approach in~\cref{fig: homologytesting}.

\begin{algorithm}[t]
\DontPrintSemicolon
\textbf{Input:} Simplicial complex $K = \{ S_r^K \}_{r=1}^n$ with specification matrices $\{ \mathcal{S}_r\}$, and an input $r$-cycle $c_r$.

Construct the boundary map $\partial_{r+1}$ from the classical input.

Construct a block-encoding of $\frac{1}{\Lambda^2}\partial_{r+1}^\dagger \partial_{r+1}$.

Estimate
\[
\frac{\operatorname{rank}(\partial_{r+1})}{|S_{r+1}^K|}
\quad\text{and}\quad
\frac{\operatorname{rank}([\partial_{r+1}\mid c_r])}{|S_{r+1}^K|+1}
\]
using the rank estimation procedure (\cref{lemma: traceestimation}).

\uIf{$\operatorname{rank}(\partial_{r+1}) = \operatorname{rank}([\partial_{r+1}\mid c_r])$}{
    \Return ``$c_r$ is homologous to zero''
}
\Else{
    \Return ``$c_r$ is not homologous to zero''
}

\textbf{Output:} Decision of whether $c_r$ is homologous to zero.

\caption{Quantum algorithm for homology triviality testing}
\label{fig: homologytesting}
\end{algorithm}

\begin{theorem}[Time complexity of testing homology triviality, see~\cref{sec: zerohomologyclass}]
\label{thm: thm3}
Let $K$ be a simplicial complex specified by $\{\mathcal{S}_r\}$. Given an $r$-cycle $c_r$, one can determine whether it is homologous to zero with high probability as follows:
\begin{itemize}
    \item \textnormal{(Classical access)} using a quantum circuit of complexity
    \begin{equation}
        \mathcal{O}\!\left(
        (r|S_{r+1}^K| + L)\,
        \Lambda^2\,
        \sqrt{\frac{\mathrm{rank}(\partial_{r+1})}{|S_{r+1}^K|}}\,
        \log\!\big(|S_r^K|\,|S_{r+1}^K|\big)
        \right),
    \end{equation}
    where $L$ is the bit-length required to describe $c_r$. This assumes that both $\mathrm{rank}(\partial_{r+1})$ and $\mathrm{rank}([\partial_{r+1} \mid c_r])$ can be estimated to constant multiplicative accuracy.

    \item \textnormal{(Sparse access)} using a quantum circuit of complexity
    \begin{equation}
        \mathcal{O}\!\left(
        r \Lambda^2\,
        \sqrt{\frac{\mathrm{rank}(\partial_{r+1})}{|S_{r+1}^K|}}\,
        \log |S_{r+1}^K|
        \right).
    \end{equation}
\end{itemize}
\end{theorem}

\paragraph{Comparison to prior quantum algorithms.}
A related quantum algorithm for testing homology triviality was proposed in~\cite{nghiem2023constant}. That work focuses on 2-dimensional triangulated manifolds and assumes access to a precomputed cohomology basis, along with an oracle specifying the input cycle. Its complexity is independent of $L$, but scales as $\beta_1 \sqrt{E}$, where $E$ is the number of edges and $\beta_1$ is the first Betti number.

In the 2-dimensional setting, taking $r=1$ gives $|S_1^K| = E$. When $\mathrm{rank}(\partial_2)$ is large (i.e., $\mathrm{rank}(\partial_2) = \Theta(|S_2^K|)$, equivalently $\beta_2$ is small), our algorithm achieves a superpolynomial speed-up in $E$ compared to~\cite{nghiem2023constant}. We emphasize that the input models differ.

\paragraph{Comparison to classical algorithms.}
A straightforward classical approach computes $\mathrm{rank}(\partial_{r+1})$ and $\mathrm{rank}([\partial_{r+1} \mid c_r])$ via Gaussian elimination, with time complexity $\mathcal{O}(|S_{r+1}^K|^3)$. Our quantum algorithm can therefore achieve an exponential speed-up when $\mathrm{rank}(\partial_{r+1}) = \Theta(|S_{r+1}^K|)$, i.e., when $\beta_{r+1} \ll |S_{r+1}^K|$. This is precisely the regime where our method performs best. Notably, this contrasts with prior quantum algorithms for estimating Betti numbers (see~\cref{thm: estimatingbettinumbers,thm: estimatingpersistentbettinumbers}), which are most effective when the Betti numbers are large.

\begin{algorithm}[t]
\DontPrintSemicolon
\textbf{Input:} Simplicial complex $K = \{ S_r^K \}_{r=1}^n$ with specification matrices $\{\mathcal{S}_r\}$, and two $r$-cycles $c_{r_1}, c_{r_2}$.

Construct the boundary map $\partial_{r+1}$ from the classical input.

Construct a block-encoding of $\frac{1}{\Lambda^2}\partial_{r+1}^\dagger \partial_{r+1}$.

Estimate
\[
\frac{\operatorname{rank}(\partial_{r+1})}{|S_{r+1}^K|}
\quad\text{and}\quad
\frac{\operatorname{rank}\!\left([\partial_{r+1} \mid (c_{r_1}-c_{r_2})]\right)}{|S_{r+1}^K|+1}
\]
using the rank estimation procedure (\cref{lemma: traceestimation}).

Classically compute $\operatorname{rank}(\partial_r)$ and 
$\operatorname{rank}\!\left([\partial_r \mid (c_{r_1}-c_{r_2})]\right)$.

\uIf{$\operatorname{rank}(\partial_r)=\operatorname{rank}\!\left([\partial_r \mid (c_{r_1}-c_{r_2})]\right)$}{
    \Return ``$c_{r_1} \sim c_{r_2}$ (homology equivalent)''
}
\Else{
    \Return ``$c_{r_1} \not\sim c_{r_2}$ (not homologous)''
}

\textbf{Output:} Decision of whether $c_{r_1} \sim c_{r_2}$.

\caption{Quantum algorithm for homology equivalence testing}
\label{fig: nontrivialhomologytesting}
\end{algorithm}

\subsubsection{Testing homology equivalence}
Beyond checking whether a single cycle is homologous to zero, the above algorithm can be naturally extended to test whether two given $r$-cycles are homologous to each other.

\begin{problem}[Homology equivalence testing with classical description, see~\cref{sec: nonzerohomologyclass}]\label{prob: homology_testing_nontrivial}
Given a simplicial complex $K$, specified via a classical description of its boundary matrices $\{ \mathcal{S}_r \}$, and two $r$-cycles $c_1$ and $c_2$ represented as explicit vectors in the $r$-chain space, determine whether $c_1$ and $c_2$ are homologous.
\end{problem}

We first recall the basic translative property of homology: if $c_{r_1} \sim c_{r_2}$ and $c_{r_2} \sim c_{r_3}$, then $c_{r_1} \sim c_{r_3}$. To test the relation $c_{r_1} \sim c_{r_2}$, our algorithm again relies on two key facts: (i) $c_{r_1}$ and $c_{r_2}$ are known vectors, and (ii) a sufficient condition for their homology is that the linear system
\begin{align}
    \partial_{r+1} c_{r+1} = c_{r_1} - c_{r_2}
    \label{eqn: 11}
\end{align}
has a solution. This follows directly from the definition that two $r$-cycles are homologous if their difference is the boundary of an $(r{+}1)$-chain. Since the classical descriptions of $c_{r_1}$ and $c_{r_2}$ are given, we can construct the difference vector $c_{r_1} - c_{r_2}$ efficiently. Then, testing whether the system admits a solution reduces to checking whether $\operatorname{rank}(\partial_{r+1}) = \operatorname{rank}([\partial_{r+1} | (c_{r_1} {-} c_{r_2})])$. This allows us to reuse the same quantum procedure outlined in~\cref{fig: homologytesting}, with essentially the same time complexity:

\begin{theorem}[Time complexity of testing homology equivalence, see~\cref{sec: nonzerohomologyclass}]
\label{thm: thm4}
Let $K$ be a simplicial complex specified by $\{\mathcal{S}_r\}$. Given two $r$-cycles $c_{r_1}$ and $c_{r_2}$, one can determine whether they are homologous with high probability as follows:
\begin{itemize}
    \item \textnormal{(Classical access)} using a quantum circuit of complexity
    \begin{equation}
        \mathcal{O}\left(
        (r|S_{r+1}^K| + L)\,
        \Lambda^2\,
        \sqrt{\frac{\mathrm{rank}(\partial_{r+1})}{|S_{r+1}^K|}}\,
        \log\!\big(|S_r^K|\,|S_{r+1}^K|\big)
        \right),
    \end{equation}
    where $L = \max\{L(c_{r_1}), L(c_{r_2})\}$ is the larger bit-length of the two cycles. This assumes that both $\mathrm{rank}(\partial_{r+1})$ and $\mathrm{rank}([\partial_{r+1} \mid c_r])$ can be estimated to constant multiplicative accuracy.

    \item \textnormal{(Sparse access)} using a quantum circuit of complexity
    \begin{equation}
        \mathcal{O}\left(
        r \Lambda^2
        \frac{1}{\sqrt{\delta}}\,
        \sqrt{\frac{\mathrm{rank}(\partial_{r+1})}{|S_{r+1}^K|}}\,
        \log |S_{r+1}^K|
        \right).
    \end{equation}
\end{itemize}
\end{theorem}

As with the zero-homology case, if the rank of $\partial_{r+1}$ is large (e.g., $\approx |S_{r+1}^K|$), our quantum algorithm achieves an exponential speed-up over the classical method, which again requires Gaussian elimination to compute the ranks of $\partial_{r+1}$ and $[\partial_{r+1} | (c_{r_1} {-} c_{r_2})]$. 

Beside, to our knowledge, there has not been a quantum algorithm for testing homology equivalence between curves. Therefore, our work marks a new step toward quantum computing application along this direction.

\begin{algorithm}[t]
\DontPrintSemicolon
\textbf{Input:} Simplicial complex $K$, candidate $r$-cycles $c_{r_1},\dots,c_{r_s}$.

Initialize the representative set $\mathcal{H} = \{c_{r_1}\}$.

\For{$i = 2$ \KwTo $s$}{
    \ForEach{$c_{r_j}^h \in \mathcal{H}$}{
        Test homology equivalence between $c_{r_j}^h$ and $c_{r_i}$ using Algorithm~\ref{fig: nontrivialhomologytesting}\;
        \If{equivalent}{
            \textbf{break} the inner loop
        }
    }
    \If{no equivalent representative is found}{
        Add $c_{r_i}$ to $\mathcal{H}$\;
    }
}

Construct the matrix
\[
M = [c_{r_1}^h, c_{r_2}^h, \dots, c_{r_p}^h]
\]
from the representatives in $\mathcal{H}$.

Estimate $\operatorname{rank}(M)$ using stochastic rank estimation (\cref{lemma: traceestimation}).

Compute
\[
\beta_r = \operatorname{rank}(M).
\]

\textbf{Output:} Estimated Betti number $\beta_r$.

\caption{Quantum algorithm for Betti number estimation via homology property testing}
\label{fig: trackingBettinumbers}
\end{algorithm}

\subsubsection{An alternative approach to Betti number estimation}

In our earlier discussion, as well as in several related works~\cite{schmidhuber2022complexity, lloyd2016quantum, hayakawa2022quantum, berry2024analyzing, mcardle2022streamlined}, the quantum algorithm for estimating (normalized) Betti numbers relies on the basic identity $\beta_r = \dim \ker (\Delta_r)$. Accordingly, the goal of these approaches (see, e.g.,~\cref{fig: betti_workflow}) is to construct the combinatorial Laplacian $\Delta_r$ and then estimate the dimension of its kernel. This remains one of the most standard methods for computing Betti numbers of a given complex. As stated in~\cref{thm: estimatingbettinumbers}, the complexity of estimating the $r$-th Betti number of a complex $K$ scales approximately as $\sim {|S_r^K|}/{\beta_r}$. Thus, for the algorithm to be efficient, it is necessary that $\beta_r$ be comparable to $|S_r^K|$—that is, the complex must exhibit \emph{large} Betti numbers. 

At first glance, this may appear counterintuitive: higher Betti numbers correspond to more intricate topological structures, characterized by a larger number of ``holes.'' Motivated by this observation, we investigate the opposite regime—when $\beta_r$ is small—and explore whether an efficient algorithm can still be devised. Interestingly, such an algorithm does exist, and it naturally emerges from our earlier procedure for testing whether two given cycles are homologous.

To obtain elements of $H_r$, we observe that the quotient space $\ker \partial_r / \operatorname{im} \partial_{r+1}$ induces an equivalence relation on $\ker \partial_r$, the space of $r$-cycles. Two $r$-cycles are homologous if and only if their difference lies in $\operatorname{im} \partial_{r+1}$. This relation underlies our homology testing procedure (see~\cref{eqn: 11}).

Our approach is as follows. We sample a collection of $r$-cycles, determine their pairwise homology relations using the procedure in~\cref{fig: nontrivialhomologytesting}, and group homologous cycles together. From each equivalence class, we retain one representative. We then estimate the number of linearly independent representatives, which equals the number of independent homology classes, i.e., the Betti number. A schematic overview is given in~\cref{fig: trackingBettinumbers}, and a full analysis appears in~\cref{sec: trackingbettinumber}.

The complexity depends on the access model for $\{\mathcal{S}_r\}$.

\begin{itemize}
    \item \textbf{(Classical access)} 
    The complexity of estimating the normalized rank
    \begin{equation}
        \frac{1}{p}\,\mathrm{rank}[c_{r_1}^h, \dots, c_{r_p}^h]
    \end{equation}
    to additive error $\varepsilon$ is
    \begin{equation}
        \mathcal{O}\left(
        \mathcal{T}_{\mathrm{hom}}^{\mathrm{cl}}
        +
        \log(p|S_r^K|)\,
        \frac{\|\mathcal{C}\|_F^2}{\sqrt{\varepsilon}}
        \log\left(\frac{1}{\varepsilon}\right)
        \right),
    \end{equation}
    where
    \begin{equation}
        \mathcal{T}_{\mathrm{hom}}^{\mathrm{cl}}
        :=
        (r|S_{r+1}^K| + L)\,
        \Lambda^2\,
        \sqrt{\frac{\mathrm{rank}(\partial_{r+1})}{|S_{r+1}^K|}}\,
        \log\!\big(|S_r^K|\,|S_{r+1}^K|\big),
    \end{equation}
    and $\|\mathcal{C}\|_F^2 = \sum_{i=1}^p \|c_{r_i}^h\|^2$.

    To estimate
    \begin{equation}\label{eq:rankrank}
        \mathrm{rank}[c_{r_1}^h, \dots, c_{r_p}^h]
    \end{equation}
    to multiplicative error $\delta$, the complexity becomes
    \begin{equation}
        \mathcal{O}\left(
        \mathcal{T}_{\mathrm{hom}}^{\mathrm{cl}}
        +
        \log(p|S_r^K|)\,
        \frac{\|\mathcal{C}\|_F^2}{\sqrt{\delta}}\,
        \sqrt{\frac{p}{\beta_r}}\,
        \log\left(\frac{1}{\varepsilon}\right)
        \right).
    \end{equation}

    \item \textbf{(Sparse access)} 
    Replacing $\mathcal{T}_{\mathrm{hom}}^{\mathrm{cl}}$ with its sparse-access counterpart from~\cref{thm: thm4}, we obtain the additive-error complexity
    \begin{equation}
        \mathcal{O}\left(
        \mathcal{T}_{\mathrm{hom}}^{\mathrm{sp}}
        +
        \log(p|S_r^K|)\,
        \frac{\|\mathcal{C}\|_F^2}{\sqrt{\varepsilon}}
        \log\left(\frac{1}{\varepsilon}\right)
        \right),
    \end{equation}
    where
    \begin{equation}
        \mathcal{T}_{\mathrm{hom}}^{\mathrm{sp}}
        :=
        r \Lambda^2\,
        \sqrt{\frac{\mathrm{rank}(\partial_{r+1})}{|S_{r+1}^K|}}\,
        \log |S_{r+1}^K|.
    \end{equation}

    For multiplicative error $\delta$, the complexity becomes
    \begin{equation}
        \mathcal{O}\left(
        \mathcal{T}_{\mathrm{hom}}^{\mathrm{sp}}
        +
        \log(p|S_r^K|)\,
        \frac{\|\mathcal{C}\|_F^2}{\sqrt{\delta}}\,
        \sqrt{\frac{p}{\beta_r}}\,
        \log\left(\frac{1}{\varepsilon}\right)
        \right).
    \end{equation}
\end{itemize}

\paragraph{Comparison to classical algorithms.}
A classical approach computes $\mathrm{rank}(\partial_{r+1})$ and~\cref{eq:rankrank} via Gaussian elimination, with total complexity $\mathcal{O}(|S_{r+1}^K|^3 + p^3)$. 

Our quantum algorithm achieves an exponential speed-up in both $|S_{r+1}^K|$ and $p$ provided that
\[
    \mathrm{rank}(\partial_{r+1}) = \Theta(|S_{r+1}^K|)
    \quad \text{and} \quad
    \beta_r = \Theta(p).
\]
The first condition holds when $\beta_{r+1} \ll |S_{r+1}^K|$, while the second holds when only a small number of representatives is needed to span $H_r$. In this regime, the quantum algorithm attains optimal performance.

\begin{algorithm}[t]
\DontPrintSemicolon
\textbf{Input:} Simplicial complex $K = \{ S_r^K \}_{r=1}^n$, two $r$-cycles $c_{r_1}, c_{r_2}$.

Construct the coboundary map $\delta^r$ from $K$.

Sample a random $r$-cochain $\omega^r$.

Project $\omega^r$ onto the cocycle space:
\[
\omega_{\mathrm{proj}}^r
=
\omega^r
-
(\delta^r)^\top
\big(\delta^r (\delta^r)^\top\big)^{-1}
\delta^r\,\omega^r .
\]

Evaluate $\omega_{\mathrm{proj}}^r(c_{r_1})$ and $\omega_{\mathrm{proj}}^r(c_{r_2})$.

\If{$\omega_{\mathrm{proj}}^r(c_{r_1})=\omega_{\mathrm{proj}}^r(c_{r_2})$}{
    \Return ``$c_{r_1}\sim c_{r_2}$ (homologous)''
}
\Else{
    \Return ``$c_{r_1}\not\sim c_{r_2}$ (not homologous)''
}

\textbf{Output:} Decision of whether $c_{r_1}\sim c_{r_2}$.

\caption{Cohomological algorithm for homology equivalence testing}
\label{fig: cohomologytesting}
\end{algorithm}

\subsubsection{A cohomology-based algorithm for homology equivalence testing}

Previously, our algorithms were primarily based on homology theory. In this section, we explore alternative solutions to the same problem using \textit{cohomology theory}. Roughly speaking, homology theory builds upon simplices and the linear mappings between them (i.e., boundary maps), whereas cohomology theory deals with linear functionals that assign real numbers to simplices. In this sense, cohomology can be viewed as a ``dual'' theory to homology.

A formal introduction to cohomology is provided in~\cref{sec: overviewcohomology}, but we briefly summarize the essential concepts relevant to our algorithm for testing homology equivalence:
\begin{enumerate}[label=(\roman*)]
    \item An $r$-cochain is a linear functional that assigns a real value to any $r$-chain.
    \item The dual operator of the boundary map $\partial_r$ is called the \emph{coboundary map} $\delta^r$.
    \item An $r$-cochain $\omega^r$ satisfying $\delta^r \omega^r = 0$ is called an $r$-cocycle.
\end{enumerate}

Our cohomological algorithm is based on the following key property:
\begin{center}
    \textit{If two cycles $c_{r_1}$ and $c_{r_2}$ are homologous, then $\omega^r(c_{r_1}) = \omega^r(c_{r_2})$ for all $r$-cocycles $\omega^r$. \\ Otherwise, there exists some $\omega^r$ such that $\omega^r(c_{r_1}) \neq \omega^r(c_{r_2})$.}
\end{center}
We will provide a proof of this property in~\cref{sec: overviewcohomology,sec: cohomologydetectinghomology}. Our quantum algorithm based on this idea is summarized in~\cref{fig: cohomologytesting}. This leads to the following performance guarantee:
\begin{theorem}[Time complexity of homology equivalence testing via cohomology, see~\cref{sec: hom_equiv_test_cohom}]
    Given a simplicial complex $K$ with classical description $\{\mathcal{S}_r\}$, determining whether two given $r$-cycles $c_{r_1}, c_{r_2}$ are homologous requires different resources depending on the construction method of the $r$-cocycle. If block encodings are constructed via projection onto $\ker(\delta_r)$, the quantum time complexity is $\mathcal{O}(\log |S_r||S_{r+1}|)$. Alternatively, in manual construction via explicit representatives, the quantum time complexity is $\mathcal{O}(\log |S_r||S_{r+1}|)$.
\end{theorem}

In comparison, the homology-based approach to this problem has the complexity given in Thm~\ref{thm: thm4}. As discussed earlier, this homology-based method is effective only when $\operatorname{rank}(\partial_{r+1})$ is large. In contrast, the cohomological approach does not depend on this rank and thus performs robustly regardless of the topological structure of the complex. This highlights the surprising power and generality of cohomology compared to homology in this context.

\subsection{Conclusion, discussion and open problems}

In this work, we explored several new directions in quantum topological data analysis. 
Our main contribution is the introduction of a refined input model for estimating Betti numbers and persistent Betti numbers, which allows quantum algorithms to operate under more structured access to simplicial complexes. 
Within this framework, we proposed a homology tracking method that avoids explicitly computing the dimension of the kernel of combinatorial Laplacians, a step that was central to previous approaches. 
This leads to a new algorithmic framework for Betti number estimation that significantly improves the performance of prior quantum methods, particularly in regimes where earlier algorithms were efficient only for complexes with large Betti numbers. 
Under this model, our results show that quantum algorithms can achieve substantial—and in certain parameter regimes, exponential—speedups over both the best-known classical algorithms and previous quantum approaches.

Beyond Betti number estimation, we introduced the \emph{homology property testing} problem. 
While closely related to Betti number computation, this task captures finer structural features of simplicial complexes. 
We also developed a cohomological formulation of the problem and proposed quantum algorithms for testing homology classes via cohomology. 
This dual viewpoint often leads to simpler or more efficient procedures. 
In particular, we showed that both homology triviality testing and homology class distinction admit efficient quantum algorithms, yielding exponential advantages under suitable assumptions. 
Together, these results strengthen the connection between quantum computation and computational topology and suggest that homological and cohomological invariants provide a promising domain for demonstrating quantum advantage.

Several natural directions remain open. 
A fundamental operation in cohomology is the \emph{cup product}, which endows the cohomology groups $H^\ast(K)$ with the structure of a graded-commutative ring. 
Whether quantum algorithms can efficiently compute cup products, or more generally perform cohomological ring operations, remains largely unexplored. 
Progress in this direction could lead to new quantum algorithms for richer topological invariants, including invariants of manifolds and higher-order interactions in data.

More broadly, our results suggest that cohomology may provide algorithmic advantages over homology in certain settings. 
It would therefore be interesting to identify additional computational tasks in which cohomological methods yield improved quantum algorithms. 
Developing new primitives based on cohomological constructions may open algorithmic avenues that are inaccessible to purely homological techniques.

Another promising direction concerns connections with categorified invariants. 
Recent work such as~\cite{schmidhuber2025quantum} suggests possible links between quantum algorithms and categorified topological invariants such as Khovanov homology and its persistent analogue, \emph{Persistent Khovanov Homology}. 
Understanding how such invariants interact with the cohomological framework introduced here may lead to new perspectives on quantum algorithms for topological structures.

Finally, the complexity-theoretic status of Betti number computation remains only partially understood. 
Existing hardness results show that deciding whether the $k$-th Betti number exceeds a given threshold is PSPACE-complete for clique complexes~\cite{rudolph2024towards}, and that counting versions of this problem are \#BQP-complete~\cite{crichigno2024clique}. 
However, these results do not address the complexity of \emph{approximate} Betti number computation. 
This gap is particularly important, since most applications in topological data analysis require only approximate estimates. 
Currently, the only indication of potential quantum advantage for approximate computation comes from the DQC1-hardness of generalized chain complexes~\cite{cade2024complexity}, although this result applies to structures more general than those typically used in applied settings. 
Determining the complexity of approximate Betti number estimation therefore remains a central open problem at the intersection of quantum algorithms and computational topology.

\newpage
\section{Preliminaries and related work}\label{sec: preliminaries}
In this section, we provide a self-contained summary of the quantum algorithms and the mathematical background required for our approach.

\subsection{Block-encoding and quantum singular value transformation}
\label{sec: summaryofnecessarytechniques}

We summarize the main quantum primitives used in our algorithm. We state only the key results and refer to~\cite{gilyen2019quantum} for details.

\begin{definition}[Block encoding unitary, see e.g.~\cite{low2017optimal, low2019hamiltonian, gilyen2019quantum}]
\label{def: blockencode} 
Let $A$ be a Hermitian matrix of size $N \times N$ with $\|A\| < 1$. A unitary $U$ is said to be an \emph{exact block encoding} of $A$ if
\begin{equation}
    U = \ket{\mathbf{0}}\bra{\mathbf{0}} \otimes A + U_\perp,
\end{equation}
where $U_\perp$ denotes the remaining component orthogonal to $\ket{\mathbf{0}}\bra{\mathbf{0}} \otimes A$. If instead $U$ satisfies
\begin{equation}
    U = \ket{\mathbf{0}}\bra{\mathbf{0}} \otimes \tilde{A} + U_\perp,
\end{equation}
for some $\tilde{A}$ such that $\|\tilde{A} - A\| \le \varepsilon$, then $U$ is called an $\varepsilon$-approximate block encoding of $A$. Moreover,
\begin{equation}
\label{eqn: action}
    U \ket{\mathbf{0}}\ket{\phi} = \ket{\mathbf{0}} A\ket{\phi} + \ket{\mathrm{garbage}},
\end{equation}
where $\ket{\mathrm{garbage}}$ is orthogonal to $\ket{\mathbf{0}}A\ket{\phi}$.
\end{definition}

\begin{remark}[Properties of block-encoding unitary]
The block-encoding framework has the following
immediate consequences:
\begin{enumerate}[label=(\roman*)]
    \item Any unitary $U$ is trivially an exact block encoding of itself.
    \item If $U$ is a block encoding of $A$, then so is $I_m \otimes U$ for any $m \geq 1$.
    \item The identity matrix $I_m$ can be trivially block encoded.
\end{enumerate}
\end{remark}

\paragraph{Algebra of block encodings.} Given a set of block-encoded operators, a variety of arithmetic operations can be performed on them. In the following, we present several operations that are particularly relevant and important to
our algorithm. 

\begin{lemma}[Product, see e.g.~\cite{gilyen2019quantum}]
\label{lemma: product}
Given unitary block encodings of $A_1$ and $A_2$ with complexities $T_1$ and $T_2$, there exists a procedure that constructs a block encoding of $A_1 A_2$ with complexity $T_1 + T_2$.
\end{lemma}

\begin{lemma}[Tensor product, see e.g.~{\cite[Theorem 1]{camps2020approximate}}]
\label{lemma: tensorproduct}
Given unitary block encodings $\{U_i\}_{i=1}^m$ of $\{M_i\}_{i=1}^m$, one can construct a block encoding of $\bigotimes_{i=1}^m M_i$ using one call to each $U_i$ and $\mathcal{O}(1)$ additional gates.
\end{lemma}

\begin{lemma}[Linear combination, see e.g.~{\cite[Theorem 52]{gilyen2019quantum}}]
\label{lemma: sumencoding}
Given unitary block encodings of $\{A_i\}_{i=1}^m$, one can construct a block encoding of $\sum_{i=1}^m \pm (A_i/m)$ using $\mathcal{O}(m)$ calls to the inputs.
\end{lemma}

\begin{lemma}[Scaling multiplication of block-encoded operators] 
\label{lemma: scale}
Given a block encoding of $A$, the block encoding of $A/p$ for $p > 1$ can be prepared with $\mathcal{O}(1)$ overhead.
\end{lemma}

To see this, note that the matrix representation of the $R_Y$ rotation gate is
\begin{equation}
    R_Y(\theta) =
    \begin{pmatrix}
        \cos(\theta/2) & -\sin(\theta/2) \\
        \sin(\theta/2) & \cos(\theta/2)
    \end{pmatrix}.
\end{equation}

Choosing $\theta = 2\cos^{-1}(1/p)$, one can construct a block encoding of $R_Y(\theta) \otimes I_{\dim(A)}$, which yields a diagonal operator with entries $1/p$. By applying~\cref{lemma: product}, we obtain a block encoding of
\begin{equation}
    (1/p)\, I_{\dim(A)} \cdot A = A/p.
\end{equation}

For generality, we define
\begin{equation}
\mathrm{size}(A) := (\text{number of rows of } A) \times (\text{number of columns of } A),
\end{equation}
which denotes the total number of entries of $A$.

\begin{lemma}[Matrix inversion and pseudo-inversion, see e.g.~\cite{gilyen2019quantum, childs2017quantum}]
\label{lemma: matrixinversion}
Given a block encoding of $A$ with $\|A\| \leq 1$ and complexity $T_A$, there exists a quantum circuit that produces an $\varepsilon$-approximate block encoding of $A^{-1}/\kappa$ (if invertible) or $A^+/\kappa$, with complexity
\begin{equation}
    \mathcal{O}\left(
    \kappa T_A \log \frac{1}{\varepsilon}
    \right).
\end{equation}
\end{lemma}


\begin{lemma}[Amplification, see e.g.~{\cite[Theorem 30]{gilyen2019quantum}}]
\label{lemma: amp_amp}
Let $U,\Pi,\widetilde{\Pi} \in \mathrm{End}(\mathcal{H}_U)$, where $U$ is unitary and $\Pi,\widetilde{\Pi}$ are orthogonal projectors. Let $\gamma > 1$ and $\delta,\epsilon \in (0,1/2)$. Suppose
\begin{equation}
    \widetilde{\Pi} U \Pi = \sum_i \varsigma_i \ket{w_i}\bra{v_i}.
\end{equation}
Then there exists
\begin{equation}
    m = \mathcal{O}\left(
    \frac{\gamma}{\delta}
    \log \left(\frac{\gamma}{\epsilon}\right)
    \right),
\end{equation}
and an efficiently computable $\Phi \in \mathbb{R}^m$ such that
\begin{equation}
\left(
\bra{+} \otimes \widetilde{\Pi}_{\le (1-\delta)/\gamma}
\right)
U_\Phi
\left(
\ket{+} \otimes \Pi_{\le (1-\delta)/\gamma}
\right)
=
\sum_{i:\,\varsigma_i \le (1-\delta)/\gamma}
\tilde{\varsigma}_i \ket{w_i}\bra{v_i},
\end{equation}
where
\begin{equation}
    \left|
    \frac{\tilde{\varsigma}_i}{\gamma \varsigma_i} - 1
    \right|
    \le \epsilon.
\end{equation}
\end{lemma}

\begin{lemma}[{\cite[Lemma 48]{gilyen2019quantum}}]
\label{lemma: blockencodesparseaccess}
Let $A \in \mathbb{C}^{n \times n}$ be an $s$-sparse matrix with operator norm $\|A\| \le 1$
(otherwise we consider the rescaled matrix $A/\lambda_{\max}$, where $\lambda_{\max}$ denotes
the largest eigenvalue of $A$, or an upper bound on it).
Given oracle access to $A$, there exists an $\epsilon$-approximate unitary block-encoding of
$A/s$ that can be implemented with gate complexity
\[
\mathcal{O}\!\left(\log n + \log^{2.5}\!\left(\frac{1}{\epsilon}\right)\right).
\]
\end{lemma}

\begin{lemma}[{\cite[Theorem 56]{gilyen2019quantum}}]
\label{lemma: qsvt}
Suppose that $U$ is an $(\alpha, a, \epsilon)$-block-encoding of a Hermitian matrix $A$
(see~{\cite[Definition 43]{gilyen2019quantum}} for the definition).
Let $P \in \mathbb{R}[x]$ be a polynomial of degree $d$ satisfying
\begin{itemize}
\item for all $x \in [-1,1]$, $|P(x)| \le \frac{1}{2}$.
\end{itemize}
Then there exists a quantum circuit $\tilde{U}$ that is a $(1, a+2, 4d\sqrt{\epsilon/\alpha})$-block-encoding
of $P(A/\alpha)$ and uses
\begin{itemize}
\item $d$ applications of $U$ and $U^\dagger$,
\item one controlled-$U$ gate, and
\item $\mathcal{O}((a+1)d)$ additional one- and two-qubit gates.
\end{itemize}
\end{lemma}

\subsection{Algebraic topology}\label{sec: reviewalgebraictopology}
This section provides a self-contained introduction to algebraic topology, with an emphasis on key concepts from homology theory and their application to the emerging field of topological data analysis. For a more comprehensive treatment, we refer the reader to standard texts such as \cite{hatcher2005algebraic, nakahara2018geometry}.

We begin by introducing the discrete geometric objects known as \textit{simplexes}. A simplex is a set of \textit{geometrically independent} points. More precisely, a collection of $(r{+}1)$ points forms an $r$-simplex if no $(r{-}1)$-dimensional affine subspace contains all of them. Let $v_0, v_1, \dots, v_r$ be $(r{+}1)$ points in $\mathbb{R}^m$ (with $m \geq r$); then the corresponding $r$-simplex is denoted by $\sigma_r = [v_0, v_1, \dots, v_r]$. The~\cref{fig: simplex} illustrates examples of simplexes of various dimensions.

\begin{figure}[ht]
\centering
\begin{tikzpicture}[scale = 1.7, every node/.style={font=\small}]
    \filldraw (-1,1) circle (1pt);
    \node[above left] at (-1,1) {$v_0$}; 
    \node[below right] at (-1.5,0.8) {$0$-simplex}; 
    
    \draw (1,1) -- (2,1);
    \filldraw (1,1) circle (1pt);
    \filldraw (2,1) circle (1pt);
    \node[above] at (1,1) {$v_0$};
    \node[above] at (2,1) {$v_1$}; 
    \node[below] at (1.5,0.8) {$1$-simplex}; 
    
    \coordinate (v0) at (-1,-0.5); 
    \coordinate (v1) at (-1.6, -1.5); 
    \coordinate (v2) at (-0.4, -1.5); 
    \draw[fill=gray!10] (v0)--(v1)--(v2)--cycle; 
    \filldraw (v0) circle (1pt);
    \filldraw (v1) circle (1pt);
    \filldraw (v2) circle (1pt);
    \node[above] at (v0) {$v_0$};
    \node[below left] at (v1) {$v_1$};
    \node[below right] at (v2) {$v_2$};
    \node[below] at (-1.0, -1.8) {$2$-simplex}; 
    
    \coordinate (P0) at (1.5,-0.4);
    \coordinate (P1) at (0.7,-1.4);
    \coordinate (P2) at (1.9,-1.6);
    \coordinate (P3) at (1.9,-1.0);
    \draw[fill=gray!10] (P0) -- (P1) -- (P2) -- cycle; 
    \draw[thick] (P0) -- (P1) -- (P2) -- (P0);
    \draw[thick] (P0) -- (P3) -- (P2);
    \draw[dashed] (P1) -- (P3);
    \filldraw (P0) circle (1pt);
    \filldraw (P1) circle (1pt);
    \filldraw (P2) circle (1pt);
    \filldraw (P3) circle (1pt);
    \node[above] at (P0) {$v_0$};
    \node[below left] at (P1) {$v_1$};
    \node[below right] at (P2) {$v_2$};
    \node[right] at (P3) {$v_3$}; 
    \node[below] at (1.3, -1.8) {$3$-simplex}; 
\end{tikzpicture}
\caption{\justifying \textbf{Illustration of standard simplexes.} Top left: a point ($0$-simplex); top right: a line segment ($1$-simplex); bottom left: a filled triangle ($2$-simplex); bottom right: a filled tetrahedron ($3$-simplex). Each $r$-simplex is formed by $(r{+}1)$-geometrically independent vertices in Euclidean space.}
\label{fig: simplex}
\end{figure}

Higher-dimensional simplices arise as natural generalizations of the basic examples. 
Intuitively, a simplex is a collection of geometrically independent points that are mutually connected. 
For example, a $3$-simplex $\sigma_3 = [v_0,v_1,v_2,v_3]$ contains several $2$-simplex faces, including 
$[v_0,v_1,v_2]$, $[v_0,v_2,v_3]$, $[v_1,v_2,v_3]$, and $[v_0,v_1,v_3]$. 
In general, an $r$-simplex has exactly $(r+1)$ \emph{faces}, each obtained by removing one vertex and thus forming an $(r-1)$-simplex.

\begin{remark}[Simplicial complex]
A \emph{simplicial complex} $K$ is a collection of simplices satisfying the following conditions:
\begin{enumerate}[label=(\roman*)]
\item Every face of a simplex in $K$ also belongs to $K$.
\item The intersection of any two simplices in $K$, if nonempty, is a common face of both.
\end{enumerate}
\end{remark}

Next we define the chain groups and boundary maps. 
Let $S_r^K$ denote the set of $r$-simplices in $K$, i.e.,
\begin{equation}
    S_r^K := \{\sigma_{r_i}\}_{i=1}^{|S_r^K|}.    
\end{equation}
The $r$-th chain group $C_r^K$ is the free Abelian group generated by the simplices in $S_r^K$. 
An element $c_r \in C_r^K$, called an $r$-chain, has the form
\begin{equation}
c_r = \sum_{i=1}^{|S_r^K|} a_i \sigma_{r_i},
\end{equation}
where the coefficients $a_i$ lie in a coefficient ring, typically $\mathbb{Z}$, though in many applications one also considers coefficients in $\mathbb{R}$.

The boundary map $\partial_r$ is a group homomorphism
\begin{equation}
    \partial_r : C_r^K \rightarrow C_{r-1}^K,    
\end{equation}
defined on basis elements as follows. 
For an $r$-simplex $\sigma_r=[v_0,v_1,\dots,v_r]$,
\begin{equation}
\partial_r [v_0,v_1,\dots,v_r]
=
\sum_{i=0}^r (-1)^i [v_0,\dots,\hat{v}_i,\dots,v_r],
\end{equation}
where $\hat{v}_i$ indicates that the vertex $v_i$ is omitted. 
Thus $\partial_r$ decomposes an $r$-simplex into the alternating sum of its $(r-1)$-dimensional faces. 
The map extends linearly to arbitrary chains:
\begin{equation}
\partial_r c_r
=
\sum_{i=1}^{|S_r^K|} a_i\,\partial_r \sigma_{r_i}.
\end{equation}

An $r$-chain $c_r$ satisfying $\partial_r c_r = 0$ is called an \emph{$r$-cycle}. 
If there exists an $(r+1)$-chain $c_{r+1}$ such that $\partial_{r+1}c_{r+1}=c_r$, then $c_r$ is called an \emph{$r$-boundary}. 
A fundamental identity is
\begin{equation}
    \partial_r \partial_{r+1} = 0,    
\end{equation}
which states that the boundary of a boundary is zero.

Let $Z_r^K$ denote the group of $r$-cycles and $B_r^K$ the group of $r$-boundaries. 
Since $\partial_r\partial_{r+1}=0$, every boundary is a cycle, and hence
\begin{equation}
    B_r^K \subseteq Z_r^K .    
\end{equation}
The $r$-th homology group is defined as the quotient
\begin{equation}
    H_r^K := Z_r^K / B_r^K .    
\end{equation}
Its rank is called the \emph{$r$-th Betti number} $\beta_r$, which counts the number of independent $r$-dimensional holes in $K$. 
For example, $\beta_1$ corresponds to loops and $\beta_2$ to voids. 
Betti numbers are topological invariants: they remain unchanged under homeomorphisms. 
Consequently, they provide a fundamental tool for distinguishing topological spaces represented by simplicial complexes.

\subsection{Topological data analysis}
\label{sec: TDA}

Topological data analysis (TDA) is an emerging framework in data science that applies tools from algebraic topology, particularly homology theory, to the analysis of high-dimensional data~\cite{wasserman2016topological,bubenik2015statistical}. 
The central idea is that large and high-dimensional datasets often exhibit geometric and topological structures that are difficult to capture using traditional statistical or machine-learning methods. 
TDA provides a principled framework for extracting such structural information in a mathematically robust and computationally tractable manner.

A common task in TDA is to infer the ``shape'' of a dataset given as a point cloud. 
Suppose we are given $n$ points in $\mathbb{R}^m$. 
By equipping the space with a metric (e.g., the Euclidean metric) and fixing a threshold $\bar{\varepsilon}$, we connect two points by an edge whenever their distance is less than $\bar{\varepsilon}$. 
This produces a graph whose cliques can be interpreted as simplices: a clique of $(r+1)$ vertices corresponds to an $r$-simplex. 
In this way we obtain a simplicial complex $K$.

Betti numbers, defined as the ranks of homology groups, are topological invariants of $K$. 
They capture fundamental topological features such as connected components, loops, and voids. 
The parameter $\bar{\varepsilon}$ acts as a length scale, and analyzing the complex over different scales reveals the persistence of these features. 
Features that appear only within a narrow range of scales are typically regarded as noise, while persistent features reflect intrinsic geometric structure.

Formally, let $S_r^K$ denote the set of $r$-simplices in $K$. 
The $r$-th chain group is generated by these simplices. 
Since homology theory is Abelian, it is convenient to treat the chain group as a vector space. 
We therefore define the $r$-th chain space
\begin{equation}
    C_r^K = \mathrm{span}\{\ket{\sigma_{r_i}}\}_{i=1}^{|S_r^K|},    
\end{equation}
whose dimension is $|S_r^K|$.

The boundary map $\partial_r$ is a linear operator
\begin{equation}
    \partial_r : C_r^K \rightarrow C_{r-1}^K .    
\end{equation}
Its matrix representation is determined with respect to the bases
$\{\ket{\sigma_{r_i}}\}$ for $C_r^K$ and $\{\ket{\sigma_{(r-1)_i}}\}$ for $C_{r-1}^K$. 
Applying $\partial_r$ to a basis vector yields a linear combination of $(r-1)$-simplices.

The homology group is defined as
\begin{equation}
    H_r^K = Z_r^K / B_r^K ,    
\end{equation}
where $Z_r^K$ denotes the cycle space and $B_r^K$ the boundary space. 
The rank of this group is the $r$-th Betti number $\beta_r$. 
A classical approach to computing $\beta_r$ is to analyze the spectrum of the $r$-th combinatorial Laplacian
\begin{equation}
    \Delta_r = \partial_{r+1}\partial_{r+1}^\dagger + \partial_r^\dagger\partial_r ,    
\end{equation}
for which a standard result gives
\begin{equation}
    \dim \ker(\Delta_r) = \beta_r .    
\end{equation}
Computing the kernel dimension via Gaussian elimination requires time $\mathcal{O}(|S_r^K|^3)$~\cite{wasserman2016topological}.

\paragraph{Quantum algorithms.}

Quantum algorithms for estimating Betti numbers were first proposed by Lloyd, Garnerone, and Zanardi (LGZ)~\cite{lloyd2016quantum}. 
The LGZ algorithm encodes simplices as computational basis states of an $n$-qubit system. 
Each $r$-simplex $\sigma_r$ is represented as a binary string $\ket{\sigma_r}\in\mathbb{C}^{2^n}$ with Hamming weight $(r+1)$, where the positions of the ones correspond to the vertices of the simplex.

The algorithm assumes access to a membership oracle that determines whether a candidate simplex belongs to the complex:
\begin{equation}
    O_r^K\ket{\sigma_{r_i}}\ket{0}
=
\ket{\sigma_{r_i}}\ket{0\ \text{or}\ 1}.    
\end{equation}
Using this encoding, the LGZ algorithm estimates Betti numbers through the following procedure. 
Starting from the point cloud, a simplicial complex $K$ is constructed and its simplices are encoded as quantum basis states. 
A uniform mixture over these states is prepared, and the combinatorial Laplacian $\Delta_r$ is constructed from the boundary operators. 
Quantum phase estimation applied to the simulated dynamics $\exp(-i\Delta_r)$ then yields an estimate of the fraction of zero eigenvalues of $\Delta_r$, which corresponds to the normalized Betti number. 

The algorithm outputs an estimate of the normalized Betti number
\begin{equation}
    \frac{\dim\ker(\Delta_r)}{|S_r^K|}.    
\end{equation}
Subsequent works have refined and analyzed this approach~\cite{ubaru2016fast,berry2024analyzing,mcardle2022streamlined,schmidhuber2022complexity,hayakawa2022quantum}. 
Two aspects are particularly relevant for the present work.

\begin{enumerate}[label=(\roman*)]
\item Computing Betti numbers exactly is \#P-hard, and even estimating them is NP-hard under standard oracle access models~\cite{schmidhuber2022complexity}.
\item The best-known quantum complexities for approximating normalized and unnormalized Betti numbers are
\begin{align}
\mathcal{O}\!\left(
\frac{1}{\varepsilon}
\left(
n^2\sqrt{\frac{\binom{n}{r+1}}{|S_r^K|}} + n\kappa
\right)
\right),
\quad
\mathcal{O}\!\left(
\frac{1}{\delta}
\left(
n^2\sqrt{\frac{\binom{n}{r+1}}{\beta_r}} + n\kappa\sqrt{\frac{|S_r^K|}{\beta_r}}
\right)
\right),
\end{align}
where $\kappa$ denotes the condition number of $\Delta_r$.
\end{enumerate}

These results reveal that quantum speedups arise primarily in the \emph{simplex-dense} regime, where $|S_r^K|$ approaches the combinatorial upper bound $\binom{n}{r+1}$ and $\beta_r$ is large. 
However, such instances are rare in practical applications~\cite{schmidhuber2022complexity,berry2024analyzing}. 
Consequently, whether quantum algorithms can provide meaningful advantages for realistic topological data analysis remains an open question.

A key technical bottleneck of the LGZ approach is the preparation of the uniform mixture
\begin{equation}
    \frac{1}{|S_r^K|}\sum_{\sigma_r\in K}\ket{\sigma_r}\bra{\sigma_r},    
\end{equation}
which is constructed using a multi-solution variant of Grover search with query complexity
\begin{equation}
    \mathcal{O}\!\left(
\sqrt{\frac{\binom{n}{r+1}}{|S_r^K|}}
\right).    
\end{equation}
This cost becomes prohibitive in the simplex-sparse regime where $|S_r^K|\ll\binom{n}{r+1}$.

Recent work on quantum cohomology~\cite{nghiem2023quantum} suggests that better performance may be achievable in sparse regimes, aligning more closely with classical intuition. 
These observations motivate a re-examination of the algorithmic structure underlying quantum approaches to homology.

Another important insight from~\cite{schmidhuber2022complexity} is that quantum speedups are unlikely when the simplicial complex is specified only implicitly via vertex and edge lists. 
Instead, the complex must be provided with additional structural information to avoid the cost of Grover search. 
Examples include datasets where higher-order relations are explicitly known.

Motivated by this observation, we consider a refined input model in which the simplicial complex is explicitly specified. 
Under this model, we show that quantum algorithms can estimate normalized Betti numbers with exponential speedup over known classical approaches, particularly in simplex-sparse regimes.

\section{Alternative quantum algorithm for estimating Betti numbers}\label{sec: qalgorithm}
In this section, we propose a new quantum algorithm for estimating Betti numbers under a new input model, which differs from those considered in prior works. We also analyze the time complexity of the proposed algorithm.

\subsection{Simplicial complex specification}
\label{sec: complex specification}
Our approach is based on homology theory, where the central computational task is to determine the dimension of the kernel of the combinatorial Laplacian $\Delta_r$. 
We therefore begin by specifying the simplicial complex $K$ and providing access to its simplices. 
Let $S_r^K$ denote the set of $r$-simplices in $K$, where $K$ is a simplicial complex defined over a set of $n$ vertices.

Instead of encoding simplices as binary strings with Hamming weight $(r+1)$, as in the LGZ algorithm, we index them using integers. 
Specifically, we label the simplices as
\[
S_r^K = \{\sigma_{r_i}\}_{i=1}^{|S_r^K|},
\qquad
[|S_r^K|] := \{1,2,\dots,|S_r^K|\}.
\]
The specification of $K$ consists of classical knowledge of all simplices together with their face relations. 
In particular, for each $r$-simplex $\sigma_{r_i}$ and $(r-1)$-simplex $\sigma_{(r-1)_j}$, the specification determines whether
\begin{equation}
    \sigma_{(r-1)_j} \subseteq \sigma_{r_i},
\end{equation}
for all $i \in [|S_r^K|]$ and $j \in [|S_{r-1}^K|]$.

To encode this information, we define the matrix
\begin{equation}
\mathcal{S}_r \in \{0,1\}^{|S_{r-1}^K| \times |S_r^K|},
\end{equation}
where the $(j,i)$-th entry is $1$ if $\sigma_{(r-1)_j}$ is a face of $\sigma_{r_i}$ and $0$ otherwise. 
Each column corresponds to an $r$-simplex and contains exactly $(r+1)$ nonzero entries, corresponding to its $(r-1)$-faces. 
Since two distinct $r$-simplices share at most one $(r-1)$-face, any two columns of $\mathcal{S}_r$ overlap in at most one position. 
The collection $\{\mathcal{S}_r\}_{r=1}^n$ therefore provides a compact, face-based specification of the simplicial complex $K$.

To illustrate this representation, consider five vertices $\{v_i\}_{i=0}^4$. 
Suppose the set of $2$-simplices is
\begin{equation}
S_2^K = \{[v_0,v_1,v_2],\; [v_0,v_3,v_4],\; [v_1,v_2,v_3]\},
\end{equation}
which we label as $1,2,3$. 
Let the $1$-simplices be
\begin{align}
S_1^K = \{&[v_0,v_1], [v_0,v_2], [v_1,v_2], [v_0,v_3], [v_3,v_4],\nonumber \\
&[v_0,v_4], [v_1,v_3], [v_2,v_3], [v_2,v_4]\},
\end{align}
labeled $1$ through $9$.

The matrix $\mathcal{S}_2 \in \{0,1\}^{9 \times 3}$ then encodes the face relations between $1$- and $2$-simplices. 
Each column corresponds to a $2$-simplex and each row to a $1$-simplex, with entry $(i,j)=1$ if the $i$-th $1$-simplex is a face of the $j$-th $2$-simplex. 
In this example,
\begin{align}
\mathcal{S}_2 =
\begin{pmatrix}
1 & 1 & 1 & 0 & 0 & 0 & 0 & 0 & 0 \\
0 & 0 & 0 & 1 & 1 & 1 & 0 & 0 & 0 \\
0 & 0 & 1 & 0 & 0 & 0 & 1 & 1 & 0
\end{pmatrix}^\top .
\end{align}

A similar construction defines $\mathcal{S}_1$ for the relations between $0$- and $1$-simplices. 
The labeling of simplices does not affect the topology of the complex, and thus the specification above is without loss of generality. 
Indeed, simplicial complexes are commonly represented in classical settings by explicitly listing simplices and their inclusion relations.

Given the matrices $\{\mathcal{S}_r\}$, we can explicitly construct the boundary operator $\partial_r$, which is a matrix of size $|S_{r-1}^K| \times |S_r^K|$. 
The $j$-th column of $\partial_r$, corresponding to the simplex $\sigma_{r_j}$, has nonzero entries at the rows corresponding to the $(r-1)$-faces of $\sigma_{r_j}$. 
These entries take values $\pm1$, determined by the orientation convention. 
Thus, starting from $\mathcal{S}_r$, the boundary operator $\partial_r$ can be obtained by assigning appropriate signs to the nonzero entries. 
The block-encoding of $\partial_r^\dagger \partial_r$ then follows directly from~\cref{lemma: blockencodingpartialr}, depending on the access model assumed for the matrices $\{\mathcal{S}_r\}$ (and hence for $\{\partial_r\}$).

\subsection{Estimating (normalized) Betti numbers}
\label{sec: estimatingBettinumbers}
In the following, we elaborate in detail the procedure that starts from~\cref{lemma: entrycomputablematrix}, proceeds to the first part of~\cref{lemma: blockencodingpartialr}, and culminates in the main result stated in~\cref{thm: estimatingbettinumbers}. We describe the construction under the classical-access model; however, the procedure for the sparse-access model is essentially identical.

Applying~\cref{lemma: entrycomputablematrix} to $\partial_r$, we obtain a block-encoding of the normalized matrix
\begin{equation}\label{eq: block_todo_1}
    \frac{\partial_r^\dagger \partial_r}{ r  |{S_r^K}|}.
\end{equation}
Analogously, applying the same construction to $\partial_{r+1}$ yields a block-encoding of
\begin{equation}\label{eq: block_todo_2}
    \frac{\partial_{r+1} \partial_{r+1}^\dagger}{ (r+1) |{S_{r+1}^K}|}.
\end{equation}

Our next objective is to construct a block-encoding of $\Delta_r$, up to a proportional constant. To this end, we apply~\cref{lemma: amp_amp} to rescale the previously obtained block-encoded operators \cref{eq: block_todo_1} and \cref{eq: block_todo_2}. First, applying the lemma to the block-encoding in~\cref{eq: block_todo_1} yields
\begin{align}
     \frac{\partial_r^\dagger \partial_r}{ r  |{S_r^K}|} \longrightarrow \frac{\partial_r^\dagger \partial_r}{\Lambda^2}. 
\end{align}
Similarly, applying the same rescaling to~\cref{eq: block_todo_2} yields
\begin{equation}\label{eq: block_encoded_2}
    \frac{\partial_{r+1} \partial_{r+1}^\dagger}{ \Lambda^2}.
\end{equation}
We then apply~\cref{lemma: sumencoding} to obtain a block-encoding of their sum:
\begin{equation}\label{eq: block_encoded_3}
    \frac{\Delta_r}{2 \Lambda^2}.
\end{equation}

The final step is to estimate the dimension of the kernel of $\Delta_r$. In principle, several approaches are available for this task. For example, one may apply quantum phase estimation, as in the LGZ algorithm, or use the block measurement technique introduced in~\cite{hayakawa2022quantum}. Both approaches incur complexity that is polynomial in the inverse spectral gap (the difference between the zero eigenvalue and the smallest nonzero eigenvalue) and linear in the inverse of the desired precision. Recently, \cite{ubaru2021quantum} introduced the stochastic rank estimation method, which requires shallower quantum circuits than previous methods (although it requires more repetitions). In our work, we propose an alternative approach, namely~\cref{lemma: traceestimation}, which builds on the ideas of \cite{ubaru2021quantum}. The detailed proof of~\cref{lemma: traceestimation} can be found in~\cref{sec: rankestimation}.

To analyze the overall complexity, we summarize the key steps of the construction.
\begin{enumerate}[label=(\roman*)]
    \item \textbf{Block-encoding of boundary operators:} We first apply~\cref{lemma: entrycomputablematrix} to construct block-encodings of
    \begin{equation}
        \frac{\partial_r^\dagger \partial_r}{  r |{S_r^K}|}, \quad   \frac{\partial_{r+1} \partial_{r+1}^\dagger }{  (r+1) |{S_{r+1}^K}|}
    \end{equation}
    and their transposes. According to~\cref{lemma: entrycomputablematrix}, since the matrix $\partial_r$ has size $|S_{r-1}^K| \times |S_{r}^K|$ and sparsity $r+1$, this step has circuit complexity $\mathcal{O}\!\left(\log(|S_{r-1}^K||{S_r^K}|)\right)$.
    
    \item \textbf{Rescaling of block-encoded operators:} Using~\cref{lemma: amp_amp}, we obtain the rescaled block-encoding of $ \frac{\partial_r^\dagger \partial_r}{  \Lambda^2 } $. This requires $\mathcal{O}\left(  r|S_r^K| \right)$ uses of the block-encoding of $\frac{\partial_r^\dagger \partial_r}{  r |{S_r^K}|}$, leading to complexity
    $$\mathcal{O}\!\left(r|S_r^K| \log(|S_{r-1}^K||{S_r^K}|)\right).$$
    Likewise, the complexity for constructing the block-encoding of $   \frac{\partial_{r+1}^\dagger \partial_{r+1}}{  \Lambda^2 }$ is
    $$ \mathcal{O}\!\left(r|S_{r+1}^K| \log(|S_{r}^K||{S_{r+1}^K}|)\right).$$

    \item \textbf{Summation of rescaled operators:} By~\cref{lemma: sumencoding}, we construct the block-encoding of their sum, which uses each block-encoding once. The resulting complexity is therefore the sum of the above two complexities:
    $$  \mathcal{O}\!\left(r (|S_r^K| + |S_{r+1}^K|) \log( |S_{r-1}^K| |S_{r}^K||{S_{r+1}^K}|)\right).$$

    \item \textbf{Rank estimation:} Finally, we apply the rank estimation method in Lemma \ref{lemma: traceestimation} to approximate the normalized rank ${\mathrm{rank}(\Delta_r)}/{|{S_r^K}|}$, from which the normalized kernel dimension follows as
    \begin{equation}
        \frac{\dim\ker(\Delta_r)}{|{S_r^K}|} = 1 - \frac{\mathrm{rank}(\Delta_r)}{|{S_r^K}|}.
    \end{equation}
    The complexity of this estimation, up to additive error $\varepsilon$, is the product of the complexity of preparing the block-encoding in~\cref{eq: block_encoded_3} and the complexity of estimating the above ratio via rank estimation. In particular, we apply Lemma \ref{lemma: traceestimation} with accuracy $\varepsilon$ and 
    \begin{equation}
        \lambda_{\rm min}\!\left(  \frac{\Delta_r}{2\Lambda^2}\right) \sim \frac{1}{ \Lambda^2}.
    \end{equation}
    Therefore, the total complexity is
    \begin{equation}
         \mathcal{O}\!\left( \frac{1}{\sqrt{\epsilon}} \log(\frac{1}{\epsilon}) \Lambda^2 r (|S_r^K| + |S_{r+1}^K|) \log( |S_{r-1}^K| |S_{r}^K||{S_{r+1}^K}|)\right).
    \end{equation}
\end{enumerate}

We remark that the outcome of the above procedure is an estimate of the quantity
\begin{equation}
    1 - \frac{\mathrm{rank}(\Delta_r)}{|{S_r^K}|} = \frac{\beta_r}{|{S_r^K}|},
\end{equation}
which is referred to as the normalized $r$-th Betti number. To estimate the (unnormalized) $r$-th Betti number $\beta_r$ to multiplicative accuracy $\delta$, it suffices to estimate the normalized Betti number to additive error
\begin{equation}
    \varepsilon = \delta \cdot \frac{\beta_r}{|{S_r^K}|}.
\end{equation}
Substituting this value into the complexity bound of the stochastic rank estimation yields the total complexity
\begin{equation}
      \mathcal{O}\!\left( \frac{1}{\sqrt{\delta}} \sqrt{\frac{|S_r^K|}{\beta_r}} \log(\frac{1}{\delta})  \Lambda^2 r (|S_r^K| + |S_{r+1}^K|) \log( |S_{r-1}^K| |S_{r}^K||{S_{r+1}^K}|)\right).
\end{equation}

The construction above assumes the classical-access model. As emphasized earlier, the procedure for the sparse-access model is identical. The only difference is that, in the sparse-access model, the complexity of obtaining the block-encoding of $ \partial_r^\dagger\partial_r/\Lambda^2$ differs by approximately a factor of $|S_r^K|$. Following the same analysis therefore yields the complexity stated in Thm~\ref{thm: estimatingbettinumbers}.

\subsection{Estimating (normalized) persistent Betti numbers}
\label{sec: estimatingpersistentbettinumbers}

We now extend our discussion to the estimation of \emph{persistent Betti numbers}, which generalize ordinary Betti numbers. In typical applications of TDA, the connectivity between data points is governed by a scale parameter $\bar{\varepsilon}$. For a fixed threshold $\bar{\varepsilon}$, one constructs a simplicial complex whose Betti numbers quantify topological features such as connected components, loops, and voids. In particular, the $r$-th Betti number counts the number of $r$-dimensional holes; for example, the first Betti number measures one-dimensional loops, while the second Betti number corresponds to two-dimensional voids.

As the scale parameter increases from $\bar{\varepsilon}_1$ to $\bar{\varepsilon}_2$ with $\bar{\varepsilon}_2 \geq \bar{\varepsilon}_1$, additional simplices are introduced, producing a new simplicial complex that is typically denser and topologically distinct. The goal is therefore to identify topological features that persist across this range of scales. Persistent Betti numbers quantify the number of such features that survive from the first complex to the second.

Quantum algorithms for estimating Betti numbers in this persistent setting were proposed in~\cite{hayakawa2022quantum} and~\cite{mcardle2022streamlined}, under the assumption of oracle access to pairwise connectivity information at two different scales. In contrast, our approach assumes classical descriptions of the simplicial complexes. Specifically, let $K_1$ and $K_2$ be simplicial complexes satisfying $K_1 \subseteq K_2$, meaning that every simplex in $K_1$ also appears in $K_2$. Let $\{\mathcal{S}_r^1\}$ and $\{\mathcal{S}_r^2\}$ denote the specification matrices associated with $K_1$ and $K_2$, respectively, as defined in~\cref{sec: complex specification}.

Given these classical specifications, we can apply the previously described procedures to construct block-encodings of the normalized combinatorial Laplacians
\[
\frac{\Delta_r^{K_1}}{2\Lambda_{K_1}^2},
\qquad
\frac{\Delta_r^{K_2}}{2\Lambda_{K_2}^2},
\]
where $\Lambda_{K_1}$ and $\Lambda_{K_2}$ denote the maximum eigenvalues of the boundary operators $\{\partial_r^{K_1}\}$ and $\{\partial_r^{K_2}\}$, respectively. These constructions form the basis for estimating persistent Betti numbers when the simplicial complexes are explicitly known.

We now formalize the notion of persistent Betti numbers. Let $K_1 \subseteq K_2$ correspond to two length scales $\bar{\varepsilon}_1 \leq \bar{\varepsilon}_2$ in the filtration of a simplicial complex. For each $r \geq 0$, denote by $\Delta_r^{K_1}$ and $\Delta_r^{K_2}$ the $r$-th combinatorial Laplacians of $K_1$ and $K_2$, respectively. These are square matrices of dimensions
\[
|S_r^{K_1}| \times |S_r^{K_1}|
\qquad \text{and} \qquad
|S_r^{K_2}| \times |S_r^{K_2}|.
\]
Since $K_1 \subseteq K_2$, we have $S_r^{K_1} \subseteq S_r^{K_2}$ and hence
\begin{equation}
    |S_r^{K_1}| \leq |S_r^{K_2}|    
\end{equation}
for all $r$.

Although the previously described quantum algorithm can estimate the normalized Betti numbers of $K_1$ and $K_2$ individually, the difference of these quantities does not, in general, equal the $r$-th persistent Betti number. As observed in~\cite{hayakawa2022quantum}, a more refined construction is required. We therefore adopt the formalism developed in~\cite{hayakawa2022quantum}.

Let $\{\partial_r^{K_1}\}$ and $\{\partial_r^{K_2}\}$ denote the boundary operators of $K_1$ and $K_2$. Since $K_1 \subseteq K_2$, the domain of $\partial_r^{K_1}$ is naturally a subspace of the domain of $\partial_r^{K_2}$. The $r$-th persistent homology group is defined as
\begin{equation}
H_r^{K_1,K_2}
=
\frac{\ker(\partial_r^{K_1})}
{\operatorname{im}(\partial_r^{K_2}) \cap \ker(\partial_r^{K_1})}.
\end{equation}
The corresponding $r$-th persistent Betti number is the rank of this quotient space.

Recall that the $r$-th chain group $C_r^K$ is the vector space spanned by the $r$-simplices of $K$. The boundary operator $\partial_r^K$ acts linearly from $C_r^K$ to $C_{r-1}^K$. For two complexes $K_1 \subseteq K_2$, define the subspace
\begin{equation}
    C_r^{K_1,K_2}
:=
\{\, c \in C_r^{K_2} \;:\; \partial_r^{K_2}(c) \in C_{r-1}^{K_1} \,\}
\subseteq C_r^{K_2}.    
\end{equation}
This subspace consists of $r$-chains in $K_2$ whose boundaries lie in $K_1$. Let $\partial_r^{K_1,K_2}$ denote the restriction of $\partial_r^{K_2}$ to the domain $C_r^{K_1,K_2}$. By construction, this operator maps
\begin{equation}
    C_r^{K_1,K_2} \rightarrow C_{r-1}^{K_1}.    
\end{equation}

Using this operator, we define the $r$-th \emph{persistent combinatorial Laplacian}
\begin{equation}
\Delta_r^{K_1,K_2}
:=
\partial_{r+1}^{K_1,K_2}
(\partial_{r+1}^{K_1,K_2})^\dagger
+
(\partial_r^{K_1})^\dagger \partial_r^{K_1}.
\end{equation}
As shown in~\cite{lim2020hodge}, the dimension of the kernel of this operator equals the $r$-th persistent Betti number.

Consequently, our objective reduces to constructing a block-encoding of $\Delta_r^{K_1,K_2}$. Among its two terms, the second term
\begin{equation}
    (\partial_r^{K_1})^\dagger \partial_r^{K_1}
\end{equation}
can be efficiently block-encoded using the techniques developed in~\cref{sec: summaryofnecessarytechniques}. The main challenge lies in constructing a block-encoding of
\begin{equation}
    \partial_{r+1}^{K_1,K_2}
(\partial_{r+1}^{K_1,K_2})^\dagger    
\end{equation}
from the classical descriptions of $K_1$ and $K_2$.

To construct this operator, we use the \emph{Schur complement}. Let $M \in \mathbb{R}^{N \times N}$ be a real square matrix. For index sets $I,J \subseteq [N] := \{1,2,\dots,N\}$, denote by $M(I,J)$ the submatrix of $M$ formed by rows indexed by $I$ and columns indexed by $J$. The Schur complement of the principal submatrix $M(I,I)$ in $M$ is defined as
\begin{align}
M / M(I,I)
=
M(\bar I,\bar I)
-
M(\bar I,I)\,
M(I,I)^+\,
M(I,\bar I),
\end{align}
where $\bar I := [N] \setminus I$ and $M(I,I)^+$ denotes the Moore–Penrose pseudoinverse.

Returning to our setting, let $S_r^{K_1}$ and $S_r^{K_2}$ denote the sets of $r$-simplices of $K_1$ and $K_2$, respectively, with $S_r^{K_1} \subseteq S_r^{K_2}$. Define
\[
[|S_r^{K_1}|] := \{1,2,\dots,|S_r^{K_1}|\},
\qquad
I_{K_1}^{K_2} := [|S_r^{K_2}|] \setminus [|S_r^{K_1}|].
\]
The index set $I_{K_1}^{K_2}$ therefore identifies the coordinates in $S_r^{K_2}$ that do not belong to $S_r^{K_1}$.

As shown in Hodge theory~\cite{lim2020hodge}, the operator
\begin{equation}
    \partial_{r+1}^{K_1,K_2}
(\partial_{r+1}^{K_1,K_2})^\dagger    
\end{equation}
can be expressed as the Schur complement of the operator
\begin{equation}
    \partial_{r+1}^{K_2}
(\partial_{r+1}^{K_2})^\dagger.    
\end{equation}
Explicitly,
\begin{align}
\label{eqn:schur_laplacian}
\partial_{r+1}^{K_1,K_2}
(\partial_{r+1}^{K_1,K_2})^\dagger
&=
\bigl(\partial_{r+1}^{K_2}(\partial_{r+1}^{K_2})^\dagger\bigr)
/
\bigl(
\partial_{r+1}^{K_2}(\partial_{r+1}^{K_2})^\dagger(I_{K_1}^{K_2},I_{K_1}^{K_2})
\bigr)
\nonumber \\
&=
\partial_{r+1}^{K_2}(\partial_{r+1}^{K_2})^\dagger(\bar I,\bar I)
-
\partial_{r+1}^{K_2}(\partial_{r+1}^{K_2})^\dagger(\bar I,I)
\bigl(
\partial_{r+1}^{K_2}(\partial_{r+1}^{K_2})^\dagger(I,I)
\bigr)^+
\partial_{r+1}^{K_2}(\partial_{r+1}^{K_2})^\dagger(I,\bar I),
\end{align}
where $I := I_{K_1}^{K_2}$ and $\bar I := [|S_r^{K_2}|] \setminus I$.

This representation shows that the persistent Laplacian can be constructed using only the combinatorial Laplacian of $K_2$ together with the inclusion structure between $K_1$ and $K_2$. In particular, it implies that
\begin{equation}
    \partial_{r+1}^{K_1,K_2}
(\partial_{r+1}^{K_1,K_2})^\dagger    
\end{equation}
can be obtained as a Schur complement of the full Laplacian. Consequently, assuming access to a block-encoding of
\begin{equation}
    \partial_{r+1}^{K_2}
(\partial_{r+1}^{K_2})^\dagger,    
\end{equation}
we can construct a block-encoding of the persistent operator.

\begin{figure*}
\begin{center}
    \scalebox{1.2}
    {\begin{tikzpicture}
        \node at (-2.4,0) {$\partial_{r+1}^{K_2} = ~~~~~$};
        \matrix[matrix of math nodes,
        left delimiter={[}, right delimiter={]},
        nodes={minimum height=1.5em, text depth=0.5ex, text height=1.5ex, anchor=center},
        column sep=0.8em, row sep=0.8em,
        ampersand replacement=\&,
        ] (m) {
        \partial_{r+1}^{K_1} \& * \& * \& * \\
        \mathbf{0} \& * \& * \& * \\
        \mathbf{0} \& * \& * \& * \\
        \mathbf{0} \& * \& * \& * \\
        };
        
        \begin{scope}[on background layer]
        \node[fill=blue!20, draw=blue, thick, rounded corners=4pt, inner sep=1pt, fit=(m-1-1), opacity=0.7] (Bbox) {};
        \node[fill=red!20, draw=red, thick, rounded corners=4pt, inner sep=1pt, fit=(m-1-2)(m-1-4), opacity=0.7] (Rbox) {};
        \node[fill=green!15, draw=black!40!green, thick, rounded corners=4pt, inner sep=1pt, fit=(m-2-2)(m-4-4), opacity=0.7] (Pbox) {};
        \end{scope}

        \node[above=2pt of Bbox, blue] {\large $\mathcal{B}$};
        \node[above=2pt of Rbox, red] {\large $\mathcal{R}$};
        \node[below=2pt of Pbox, black!40!green] {\large $\mathcal{G}$};
    \end{tikzpicture}}
    \scalebox{1.2}
    {\begin{tikzpicture}
        \node at (-3.5,0) {and,$\quad (\partial_{r+1}^{K_2})^\dagger = ~~~~~$};
        \matrix[matrix of math nodes,
        left delimiter={[}, right delimiter={]},
        nodes={minimum height=1.5em, text depth=0.5ex, text height=1.5ex, anchor=center},
        column sep=0.8em, row sep=0.8em,
        ampersand replacement=\&,
        ] (m) {
        (\partial_{r+1}^{K_1})^\dagger \& \mathbf{0} \& \mathbf{0} \& \mathbf{0} \\
        ~~~~*~~~~ \& * \& * \& * \\
        * \& * \& * \& * \\
        * \& * \& * \& * \\
        };
        
        \begin{scope}[on background layer]
        \node[fill=blue!20, draw=blue, thick, rounded corners=4pt, inner sep=1pt, fit=(m-1-1), opacity=0.7, dash pattern=on 4pt off 2pt] (Bbox) {};
        \node[fill=red!20, draw=red, thick, rounded corners=4pt, inner sep=1pt, fit=(m-2-1)(m-4-1), opacity=0.7, dash pattern=on 4pt off 2pt] (Rbox) {};
        \node[fill=green!15, draw=black!40!green, thick, rounded corners=4pt, inner sep=1pt, fit=(m-2-2)(m-4-4), opacity=0.7, dash pattern=on 4pt off 2pt] (Pbox) {};
        \end{scope}

        \node[above=2pt of Bbox, blue] {\large $\mathcal{B}^\dagger$};
        \node[below=2pt of Rbox, red] {\large $\mathcal{R}^\dagger$};
        \node[below=2pt of Pbox, black!40!green] {\large $\mathcal{G}^\dagger$};
    \end{tikzpicture}}
\end{center}
\caption{\justifying \textbf{Block structure of the boundary matrix $\partial_{r+1}^{K_2}$ and its adjoint $(\partial_{r+1}^{K_2})^\dagger$.} The blue block $\mathcal{B}$ corresponds to the original boundary operator $\partial_{r+1}^{K_1}$, while the red block $\mathcal{R}$ represents the interaction between simplices in $K_1$ and those newly added in $K_2 \setminus K_1$. The green block $\mathcal{G}$ encodes the internal structure among new simplices. Dashed outlines in the adjoint matrix highlight the transpose-like dual roles of each sub-block.}
\label{fig: block_matrix}
\end{figure*}

We now examine the structure of the boundary matrix $\partial_{r+1}^{K_2}$ in greater detail. Recall that this matrix represents the boundary operator
\begin{equation}
    \partial_{r+1}^{K_2} : C_{r+1}^{K_2} \rightarrow C_r^{K_2},
\end{equation}
and therefore has dimensions $|S_r^{K_2}| \times |S_{r+1}^{K_2}|$. 
Since the simplices satisfy
\[
S_{r+1}^{K_1} \subseteq S_{r+1}^{K_2}, \qquad
S_r^{K_1} \subseteq S_r^{K_2},
\]
the matrix $\partial_{r+1}^{K_2}$ contains $\partial_{r+1}^{K_1}$ as a submatrix in its top-left block (see~\cref{fig: block_matrix}, left).

We now describe the block structure of $\partial_{r+1}^{K_2}$.  
The \emph{blue block}, denoted by $\mathcal{B}$, corresponds exactly to $\partial_{r+1}^{K_1}$ and has size
\begin{equation}
    \mathrm{size}(\mathcal{B})
=
|S_r^{K_1}| \times |S_{r+1}^{K_1}|.    
\end{equation}

The \emph{red block}, denoted by $\mathcal{R}$, consists of the columns indexed by the new $(r+1)$-simplices in $K_2$, i.e.
\begin{equation}
    S_{r+1}^{K_2} \setminus S_{r+1}^{K_1},    
\end{equation}
and rows indexed by the $r$-simplices in $K_1$ that are faces of these simplices. Its size is therefore
\begin{equation}
    \mathrm{size}(\mathcal{R})
=
|S_r^{K_1}|
\times
\bigl(|S_{r+1}^{K_2}| - |S_{r+1}^{K_1}|\bigr).    
\end{equation}

The \emph{green block}, denoted by $\mathcal{G}$, corresponds to both new rows and new columns, namely
\begin{equation}
    S_r^{K_2} \setminus S_r^{K_1}
\quad\text{and}\quad
S_{r+1}^{K_2} \setminus S_{r+1}^{K_1}.    
\end{equation}
Hence
\begin{equation}
    \mathrm{size}(\mathcal{G})
=
\bigl(|S_r^{K_2}| - |S_r^{K_1}|\bigr)
\times
\bigl(|S_{r+1}^{K_2}| - |S_{r+1}^{K_1}|\bigr).    
\end{equation}

This decomposition clarifies how the boundary matrix grows when new simplices are added to the filtration.  
The matrix representation of $(\partial_{r+1}^{K_2})^\dagger$ is shown in~\cref{fig: block_matrix} (right) and contains the blocks
\[
\mathcal{B}^\dagger, \qquad
\mathcal{R}^\dagger, \qquad
\mathcal{G}^\dagger.
\]

We now analyze the matrix product
\begin{equation}
    \partial_{r+1}^{K_2}(\partial_{r+1}^{K_2})^\dagger,    
\end{equation}
which has size $|S_r^{K_2}| \times |S_r^{K_2}|$.

Define the index set
\begin{equation}
    I_{K_1}^{K_2}
:=
[|S_r^{K_2}|] \setminus [|S_r^{K_1}|],    
\end{equation}
which corresponds to the $r$-simplices appearing in $K_2$ but not in $K_1$.

Then the submatrix
\begin{equation}
    \partial_{r+1}^{K_2}
(\partial_{r+1}^{K_2})^\dagger
(I_{K_1}^{K_2}, I_{K_1}^{K_2})    
\end{equation}
corresponds to the product
\begin{equation}
    \mathcal{G}\mathcal{G}^\dagger.    
\end{equation}

Similarly,
\begin{equation}
    \partial_{r+1}^{K_2}
(\partial_{r+1}^{K_2})^\dagger
([|S_r^{K_2}|]\setminus I_{K_1}^{K_2}, I_{K_1}^{K_2})
=
\mathcal{R}\mathcal{G}^\dagger,    
\end{equation}
while its Hermitian transpose equals
\begin{equation}
    \mathcal{G}\mathcal{R}^\dagger.    
\end{equation}

Finally, the principal block
\begin{equation}
    \partial_{r+1}^{K_2}
(\partial_{r+1}^{K_2})^\dagger
([|S_r^{K_2}|]\setminus I_{K_1}^{K_2},
[|S_r^{K_2}|]\setminus I_{K_1}^{K_2})    
\end{equation}
is given by
\begin{equation}
    \mathcal{B}\mathcal{B}^\dagger
+
\mathcal{R}\mathcal{R}^\dagger.    
\end{equation}

Combining these terms and applying the Schur complement (cf.~\cref{eqn:schur_laplacian}), we obtain
\begin{align}
\partial_{r+1}^{K_1,K_2}
(\partial_{r+1}^{K_1,K_2})^\dagger
=
\mathcal{B}\mathcal{B}^\dagger
+
\mathcal{R}\mathcal{R}^\dagger
-
\mathcal{R}\mathcal{G}^\dagger
(\mathcal{G}\mathcal{G}^\dagger)^+
\mathcal{G}\mathcal{R}^\dagger.
\end{align}

Here $(\mathcal{G}\mathcal{G}^\dagger)^+$ denotes the Moore–Penrose pseudoinverse.

Now, let us return to the matrix representation of $\partial_{r+1}^{K_2}$. Given a classical description of $K_2$, the matrix $\partial_{r+1}^{K_2}$ is fully determined, and consequently all the block matrices $\mathcal{B}$, $\mathcal{R}$, and $\mathcal{G}$ are explicitly known. Therefore, by invoking~\cref{lemma: blockencodingpartialr}, we can construct block-encodings of the normalized operators
\begin{equation}
    \frac{\mathcal{B}\mathcal{B}^\dagger}{\Gamma^2},
    \qquad
    \frac{\mathcal{G}\mathcal{G}^\dagger}{\Gamma^2},
    \qquad
    \frac{\mathcal{R}\mathcal{R}^\dagger}{\Gamma^2},
\end{equation}
where $\Gamma$ denotes the maximum eigenvalue among these matrices. Recall that these block-encodings are obtained by first applying~\cref{lemma: entrycomputablematrix} to construct a block-encoding of
\begin{equation}
    \frac{\mathcal{B}\mathcal{B}^\dagger}{\|\mathcal{B}\|_F^2},
\end{equation}
followed by an application of~\cref{lemma: amp_amp} to obtain a block-encoding of
\begin{equation}
    \frac{\mathcal{B}\mathcal{B}^\dagger}{\Gamma^2}.
\end{equation}

Recall that the sizes of these matrices are
\begin{align}
    \mathrm{size}(\mathcal{B})
    &= |S_r^{K_1}| \times |S_{r+1}^{K_1}|, \\
    \mathrm{size}(\mathcal{R})
    &= |S_r^{K_1}| \times \bigl( |S_{r+1}^{K_2}| - |S_{r+1}^{K_1}| \bigr), \\
    \mathrm{size}(\mathcal{G})
    &= \bigl( |S_r^{K_2}| - |S_r^{K_1}| \bigr)
       \times
       \bigl( |S_{r+1}^{K_2}| - |S_{r+1}^{K_1}| \bigr).
\end{align}
Thus, their Frobenius norms are
\begin{align}
    \|\mathcal{B}\|_F^2
    &= (r+1)|S_{r+1}^{K_1}|, \\
    \|\mathcal{R}\|_F^2
    &= r \bigl( |S_{r+1}^{K_2}| - |S_{r+1}^{K_1}| \bigr), \\
    \|\mathcal{G}\|_F^2
    &= \bigl( |S_r^{K_2}| - |S_r^{K_1}| \bigr)
       \bigl( |S_{r+1}^{K_2}| - |S_{r+1}^{K_1}| \bigr).
\end{align}

The circuit complexities for block-encoding
\(
{\mathcal{B}\mathcal{B}^\dagger}/{\Gamma^2},
{\mathcal{G}\mathcal{G}^\dagger}/{\Gamma^2},
{\mathcal{R}\mathcal{R}^\dagger}/{\Gamma^2}
\)
are, respectively,
\begin{align}
    &\mathcal{O}\!\left(
        r |S_{r+1}^{K_1}|
        \log \bigl( |S_r^{K_1}|\,|S_{r+1}^{K_1}| \bigr)
    \right), \\
    &\mathcal{O}\!\left(
        r \bigl( |S_{r+1}^{K_2}| - |S_{r+1}^{K_1}| \bigr)
        \log \!\left(
            |S_r^{K_1}|
            \bigl( |S_{r+1}^{K_2}| - |S_{r+1}^{K_1}| \bigr)
        \right)
    \right), \\
    &\mathcal{O}\!\left(
        \bigl( |S_r^{K_2}| - |S_r^{K_1}| \bigr)
        \bigl( |S_{r+1}^{K_2}| - |S_{r+1}^{K_1}| \bigr)
        \log \!\left(
            \bigl( |S_r^{K_2}| - |S_r^{K_1}| \bigr)
            \bigl( |S_{r+1}^{K_2}| - |S_{r+1}^{K_1}| \bigr)
        \right)
    \right).
\end{align}

To proceed, we introduce the following key result.

\begin{lemma}[Positive power of a positive matrix, see e.g.~{\cite{gilyen2019quantum}}]
\label{lemma: positive}
Let $\mathcal{M}$ be a positive matrix with a block-encoding, satisfying
\begin{equation}
    \frac{\mathbb{I}}{\kappa_M} \leq \mathcal{M} \leq \mathbb{I}.
\end{equation}
Then, for any $c \in (0,1)$, one can implement an $\varepsilon$-approximate block-encoding of $\mathcal{M}^c/2$ with time complexity
\begin{equation}
    \mathcal{O}\!\left(
        \kappa_M T_M \log^2\!\left(\frac{\kappa_M}{\varepsilon}\right)
    \right),
\end{equation}
where $T_M$ denotes the complexity of the block-encoding of $\mathcal{M}$.
\end{lemma}

By applying this lemma, we obtain the transformations
\begin{align}
    \frac{\mathcal{R}\mathcal{R}^\dagger}{\Gamma^2}
    &\rightarrow
    \frac{\mathcal{R}^\dagger}{\Gamma},
    \\
    \frac{\mathcal{G}\mathcal{G}^\dagger}{\Gamma^2}
    &\rightarrow
    \frac{\mathcal{G}^\dagger}{\Gamma}.
\end{align}

Moreover,~\cref{lemma: matrixinversion} allows us to implement the pseudoinverse transformation
\begin{equation}
    \frac{\mathcal{G}\mathcal{G}^\dagger}{\Gamma^2}
    \rightarrow
    \frac{1}{\kappa} (\mathcal{G}\mathcal{G}^\dagger)^+,
\end{equation}
where $\kappa$ denotes the condition number of ${\mathcal{G}\mathcal{G}^\dagger}/{\Gamma^2}$, assumed to be known or upper bounded.

Next, by employing~\cref{lemma: product}, we can construct block-encodings of the matrix products
\begin{equation}
    \frac{\mathcal{G}\mathcal{R}^\dagger}{\Gamma^2},
    \qquad
    \frac{\mathcal{R}\mathcal{G}^\dagger}{\Gamma^2}.
\end{equation}

Finally, using~\cref{lemma: sumencoding}, we obtain a block-encoding of the full operator
\begin{equation}
    \frac{
        \mathcal{B}\mathcal{B}^\dagger
        + \mathcal{R}\mathcal{R}^\dagger
        - \mathcal{R}\mathcal{G}^\dagger
          (\mathcal{G}\mathcal{G}^\dagger)^+
          \mathcal{G}\mathcal{R}^\dagger
    }{
        4 \kappa \Gamma^4
    }
    =
    \frac{
        \partial_{r+1}^{K_1,K_2}
        (\partial_{r+1}^{K_1,K_2})^\dagger
    }{
        4 \kappa \Gamma^4
    }.
\end{equation}

On the other hand, by Lemma~\ref{lemma: blockencodingpartialr}, under the classical-access model we already have a block-encoding of
\begin{equation}
    \frac{(\partial_r^{K_1})^\dagger \partial_r^{K_1}}{\Lambda^2},
\end{equation}
where $\Lambda$ is the maximum eigenvalue of $\partial_r^{K_1}$. The circuit complexity of this block-encoding is
\begin{equation}
    \mathcal{O}\!\left(
        r |S_r^{K_1}|
        \log \bigl( |S_{r-1}^{K_1}|\,|S_r^{K_1}| \bigr)
    \right).
\end{equation}

Therefore, by another application of~\cref{lemma: sumencoding}, we can construct a block-encoding of
\begin{equation}
    \frac{
        \partial_{r+1}^{K_1,K_2}
        (\partial_{r+1}^{K_1,K_2})^\dagger
        +
        (\partial_r^{K_1})^\dagger \partial_r^{K_1}
    }{
        4 \kappa \Gamma^4 + \Lambda^2
    }.
\end{equation}

As in the final step of the previous section, we then apply the rank-estimation procedure from Lemma~\ref{lemma: traceestimation} to approximate the normalized $r$-th persistent Betti number,
\begin{equation}
    \frac{\beta_r^{\mathrm{persistent}}}{|S_r^{K_1}|}
    =
    1
    -
    \frac{
        \operatorname{rank}\!\left(
            \partial_{r+1}^{K_1,K_2}
            (\partial_{r+1}^{K_1,K_2})^\dagger
            +
            (\partial_r^{K_1})^\dagger \partial_r^{K_1}
        \right)
    }{
        |S_r^{K_1}|
    },
\end{equation}
to additive error $\varepsilon$.

To simplify the complexity expressions, define
\begin{align}
    N_r &:= |S_r^{K_1}|, &
    N_{r+1} &:= |S_{r+1}^{K_1}|, \\
    M_r &:= |S_r^{K_2}| - |S_r^{K_1}|, &
    M_{r+1} &:= |S_{r+1}^{K_2}| - |S_{r+1}^{K_1}|.
\end{align}

Then the cost of block-encoding
\(
\partial_{r+1}^{K_1,K_2}
(\partial_{r+1}^{K_1,K_2})^\dagger
\)
is dominated by the costs of block-encoding
\(
{\mathcal{B}\mathcal{B}^\dagger}/{\Gamma^2},
{\mathcal{G}\mathcal{G}^\dagger}/{\Gamma^2},
{\mathcal{R}\mathcal{R}^\dagger}/{\Gamma^2}
\),
whose total complexity is
\begin{equation}
    \mathcal{O}\!\left(
        r N_{r+1} \log(N_r N_{r+1})
        + r M_{r+1} \log(N_r M_{r+1})
        + M_r M_{r+1} \log(M_r M_{r+1})
    \right).
\end{equation}

The remaining component is the block-encoding of
\(
{(\partial_r^{K_1})^\dagger \partial_r^{K_1}}/{\Lambda^2}
\),
whose complexity is
\begin{equation}
    \mathcal{O}\!\left(
        r N_r \log(|S_{r-1}^{K_1}| N_r)
    \right).
\end{equation}

Hence, the total block-encoding complexity becomes
\begin{equation}
    \mathcal{O}\!\left(
        r N_r \log(|S_{r-1}^{K_1}| N_r)
        + r N_{r+1} \log(N_r N_{r+1})
        + r M_{r+1} \log(N_r M_{r+1})
        + M_r M_{r+1} \log(M_r M_{r+1})
    \right).
\end{equation}

Applying~\cref{lemma: traceestimation}, we obtain the total complexity for estimating the normalized persistent Betti number up to additive error $\epsilon$:
\begin{equation}
\begin{split}
    \mathcal{O}\!\Biggl(
        \Bigl[
            r N_r \log(|S_{r-1}^{K_1}| N_r)
            + r N_{r+1} \log(N_r N_{r+1})
            + r M_{r+1} \log(N_r M_{r+1}) \\
            + M_r M_{r+1} \log(M_r M_{r+1})
        \Bigr]
        \left(
            (4\kappa \Gamma^4 + \Lambda^2)
            \frac{1}{\sqrt{\epsilon}}
            \log \frac{1}{\epsilon}
        \right)
    \Biggr).
\end{split}
\end{equation}

The above analysis is for the classical-access setting. In the sparse-access setting, the overall procedure is the same, but the corresponding block-encoding costs become
\begin{align}
    \mathcal{O}\!\left( r \log(N_r N_{r+1}) \right), \qquad
    \mathcal{O}\!\left( r \log(N_r M_{r+1}) \right), \qquad
    \mathcal{O}\!\left( \log(M_r M_{r+1}) \right),
\end{align}
and
\begin{equation}
    \mathcal{O}\!\left( r \log N_r \right)
\end{equation}
for
\(
{(\partial_r^{K_1})^\dagger \partial_r^{K_1}}/{\Lambda^2}
\).

Thus, repeating the same analysis yields the total complexity
\begin{equation}
\begin{split}
    \mathcal{O}\!\Biggl(
        \Bigl[
            r \log(|S_{r-1}^{K_1}| N_r N_{r+1})
            + r \log(N_r M_{r+1})
            + \log(M_r M_{r+1})
        \Bigr]
        \times
        \left(
            (4\kappa \Gamma^4 + \Lambda^2)
            \frac{1}{\sqrt{\epsilon}}
            \log \frac{1}{\epsilon}
        \right)
    \Biggr).
\end{split}
\end{equation}

\section{Homology property testing}
\label{sec: detectinghomologyclass}

\subsection{Triviality testing}
\label{sec: zerohomologyclass}

While our proposed algorithm and previous efforts \cite{schmidhuber2022complexity, hayakawa2022quantum, berry2024analyzing, mcardle2022streamlined} primarily concentrate on the estimation of (normalized) Betti numbers, we now turn our attention to a closely related and fundamentally significant problem in algebraic topology. Specifically, we consider the problem of verifying whether a given $r$-cycle belongs to the trivial homology class.

Recall that the $r$-th homology group of a simplicial complex $K$ is defined as the quotient $H_r^K = Z_r^K / B_r^K$, where $Z_r^K$ is the group of $r$-cycles and $B_r^K$ is the group of $r$-boundaries. This quotient endows $Z_r^K$ with an equivalence relation: two $r$-cycles are homologous if their difference lies in $B_r^K$. Each equivalence class under this relation is referred to as a homology class. The dimension of $H_r^K$, namely the $r$-th Betti number $\beta_r$, counts the number of linearly independent homology classes. Consequently, identifying all such classes is a {\#P-hard} problem.

Instead of attempting to enumerate all homology classes, we focus on a restricted variant of this problem: determining whether a given $r$-cycle $c_r$ is homologous to zero, i.e., whether $c_r$ represents the trivial homology class. The trivial homology class consists of all cycles that are themselves boundaries of some $(r{+}1)$-chain. In formal terms, $c_r$ is null-homologous if there exists a chain $c_{r+1}$ such that
\begin{equation}\label{eq:homology-system}
    c_r = \partial_{r+1} c_{r+1},    
\end{equation}
where $\partial_{r+1}$ denotes the $(r{+}1)$-st boundary operator.

Let $S_r^K = \{ \sigma_{r_i} \}_{i=1}^{|{S_r^K}|}$ be the set of $r$-simplices in the complex $K$. We represent each simplex $\sigma_{r_i}$ by the computational basis state $\ket{i}$ in a Hilbert space of $\log |{S_r^K}|$ qubits. Let $C = \{ i_1, i_2, \dots, i_L \} \subseteq \{1, \dots, |{S_r^K}| \}$ be the index set of simplices comprising $c_r$, with $0 \leq L \leq |{S_r^K}|$. The $r$-cycle is then encoded as
\begin{equation}
    c_r = \sum_{i_j \in C} \ket{i_j}.    
\end{equation}
Given that the constituent simplices of $c_r$ are known, the vector representation of $c_r$ is sparse, with ones at the positions corresponding to the indices in $C$.

To determine whether $c_r$ is null-homologous, it suffices to check whether the linear system~\cref{eq:homology-system} admits a solution. This is equivalent to verifying whether $c_r$ lies in the column space (image) of $\partial_{r+1}$. In linear algebraic terms, this holds if and only if the augmented matrix $[\partial_{r+1} | c_r]$ has the same rank as $\partial_{r+1}$. We leverage the \cref{lemma: traceestimation} to estimate the rank of both $\partial_{r+1}$ and the augmented matrix $[\partial_{r+1}  |  c_r]$. If the estimated ranks match up to the allowed accuracy $\varepsilon$, we conclude that $c_r$ lies in the image of $\partial_{r+1}$ with high confidence. Hence, the quantum algorithm enables us to test null-homology for a given $r$-cycle efficiently, even when the problem is embedded in high-dimensional simplicial complexes.

In the previous section, we obtained the block encoding of (via classical-access model)
\begin{equation}\label{eq: eq_det_hom_1}
    \frac{\partial_{r+1}^\dagger \partial_{r+1}}{\Lambda^2}.
\end{equation}
A direct application of the~\cref{lemma: traceestimation} enables estimation of the ratio
\begin{equation}
    \frac{\mathrm{rank} (\partial_{r+1}^\dagger\partial_{r+1}/\Lambda^2)}{ |{S_{r+1}^K}|}=  \frac{\mathrm{rank} (\partial_{r+1}^\dagger\partial_{r+1})}{ |{S_{r+1}^K}|} = \frac{\mathrm{rank}\left(\partial_{r+1}\right)}{|{S_{r+1}^K}|}
\end{equation}
to additive precision $\varepsilon$, since $\mathrm{rank}\left(\partial_{r+1}\right) = \mathrm{rank}(\partial_{r+1}^\dagger \partial_{r+1})$.

Let $\lambda_1$ denote the smallest nonzero eigenvalue of~\cref{eq: eq_det_hom_1}. Owing to the factor $\Lambda^2$, it is reasonable to expect ${\lambda_1}^{-1} = \mathcal{O} (\Lambda^2)$. Since the block-encoding of~\cref{eq: eq_det_hom_1} can be implemented with complexity $\mathcal{O}( r |S_{r+1}^K| \log(|S_r^K||S_{r+1}^K|))$ by ~\cref{lemma: blockencodingpartialr}, the overall time complexity for estimating the ratio ${\mathrm{rank}(\partial_{r+1})}/{|S_{r+1}^K|}$ to precision $\varepsilon$ is
\begin{equation}
    \mathcal{O}\!\left(  \frac{1}{\sqrt{\epsilon}} \Lambda^2 r |S_{r+1}^K| \log(|S_r^K||S_{r+1}^K|)\right),
\end{equation}
as established in~\cref{lemma: traceestimation}.

Next, consider the matrix $\partial_* \coloneqq \left[\partial_{r+1} | c_r \right]$, formed by appending the column vector $c_r$ to $\partial_{r+1}$. The resulting matrix has dimension $|{S_r^K}| \times (|{S_{r+1}^K}| + 1)$. Once the cycle $c_r$ of interest is specified, that is, the set of $r$-simplices $C = \{ i_1, i_2, \dots, i_L \}$ comprising $c_r$ is given, its vector representation is known. The $\ell_2$-norm of $c_r$ is then $\sqrt{L}$, and the Frobenius norm of $\partial_*$ becomes $\norm{\partial_*}_F = ({ (r+1)|{S_{r+1}^K}| + L})^{1/2}$.

Within the classical-access model, applying~\cref{lemma: entrycomputablematrix} yields a block encoding of the normalized operator
\begin{equation}\label{eq: eq_hom_det_2}
    \frac{\partial_*^\dagger \partial_*}{ (r+1)|{S_{r+1}^K}| + L}
\end{equation}
with complexity $\mathcal{O}\left( \log \left( |S_r^K|(|{S_{r+1}^K}| + 1\right) \right) = \mathcal{O}( \log (|S_r^K| |S_{r+1}^K|))$. Then we use Lemma \ref{lemma: amp_amp} to multiply such operator with the factor $((r+1)|{S_{r+1}^K}| + L)/\Lambda^2 $, resulting in the block-encoding of:
$$ \frac{\partial_*^\dagger \partial_*}{ \Lambda^2} $$
incurring a total complexity 
$$ \mathcal{O}\left(  ( (r+1)|{S_{r+1}^K}| + L) \log (|S_r^K| |S_{r+1}^K|)   \right) = \mathcal{O}\left(  ( r|{S_{r+1}^K}| + L) \log (|S_r^K| |S_{r+1}^K|)   \right) $$

Let $\lambda_2$ denote the smallest nonzero eigenvalue the operator above. Then, ${\lambda_2}^{-1} = \mathcal{O}\left( \Lambda^2 \right)$. By~\cref{lemma: traceestimation}, the rank ratio
\begin{equation}
    \frac{\mathrm{rank}(\partial_*^\dagger \partial)}{ |S_{r+1}^K| +1} = \frac{\mathrm{rank}(\partial_*)}{ |S_{r+1}^K| +1}
\end{equation}
can be estimated to additive precision $\varepsilon$ with time complexity
\begin{equation}
    \mathcal{O}\left(  (r|S_{r+1}^K| + L ) \Lambda^2 \frac{1}{\sqrt{\epsilon}} \log (|S_r^K| |S_{r+1}^K|) \right).
\end{equation}

We remark that comparing the ratios ${\text{rank}(\partial_{r+1})}/{|{S_{r+1}^K}|}$ and ${\text{rank}(\partial_*)}/{(|{S_{r+1}^K}| + 1)}$ is insufficient to infer a potential change in the rank of the matrix upon appending the column vector $c_r$. To rigorously determine whether the rank has increased (e.g., by 1), it is necessary to estimate both $\text{rank} (\partial_{r+1})$ and $\text{rank} (\partial_*)$ to a multiplicative accuracy $\delta$. This necessitates setting the estimation precision parameter $\varepsilon$ from the previous procedures to
\[
    \delta \cdot \frac{\text{rank}(\partial_{r+1})}{|{S_{r+1}^K}|} \quad \text{and} \quad \delta \cdot \frac{\text{rank} (\partial_*)}{|{S_{r+1}^K}| + 1},
\]
respectively.

With access to such multiplicative approximations, one can reliably determine whether the addition of $c_r$ changes the rank of the boundary matrix $\partial_{r+1}$. The total complexity of this procedure is the sum of the complexities of estimating the rank of $\partial_{r+1}$ and of $\partial_*$. Hence, the overall time complexity is
\begin{equation}
     \mathcal{O}\left(  (r|S_{r+1}^K| + L ) \Lambda^2 \frac{1}{\sqrt{\delta}} \sqrt{\frac{ |S_{r+1}^K| }{\rm rank(\partial_{r+1})}}  \left( \log (|S_r^K| |S_{r+1}^K|)  + \log^{2.5} \frac{1}{\delta} \right)\right).
\end{equation}
Here, note that setting $\delta$ to be a constant still ensures that the testing algorithm succeeds with high probability.

Classically, one may directly compute the ranks of $\partial_{r+1}$ and $\partial_*$ using Gaussian elimination, which has a time complexity of $\mathcal{O}\left( |S_{r+1}^K|^3 \right)$. Thus, for the quantum algorithm to offer a genuine computational advantage, the rank of $\partial_{r+1}$ must be on the order of $|S_{r+1}^K|$.  This is in contrast with the setting considered in previous works (e.g., \cite{lloyd2016quantum, schmidhuber2022complexity, berry2024analyzing}), where the quantum advantage in estimating Betti numbers becomes manifest when the Betti number is large, i.e., when $\beta_r \sim |S_r^K|$.

We now investigate the condition under which $\text{rank} \left( \partial_{r+1} \right) \approx |S_{r+1}^K|$. Recall that the $(r{+}1)$-st combinatorial Laplacian is defined as $\Delta_{r+1} = \partial_{r+2} \partial_{r+2}^\dagger + \partial_{r+1}^\dagger \partial_{r+1}$, and that its kernel has dimension equal to the $(r{+}1)$-st Betti number, $\beta_{r+1}$. Since $\Delta_{r+1}$ is the sum of two positive semidefinite operators, its kernel is contained in the intersection of the kernels of these two terms, implying $\beta_{r+1} \leq \dim \ker (\partial_{r+1}^\dagger \partial_{r+1})$. Moreover, as $\partial_{r+1}^\dagger \partial_{r+1}$ is Hermitian and positive semidefinite, we have the decomposition
\begin{equation}
    \dim \ker (\partial_{r+1}^\dagger \partial_{r+1}) + \text{rank} (\partial_{r+1}^\dagger \partial_{r+1} ) = |{S_{r+1}^K}|.    
\end{equation}
Since $\text{rank} (\partial_{r+1}^\dagger \partial_{r+1}) = \text{rank} \left(\partial_{r+1}\right)$, it follows that if $\text{rank} \left( \partial_{r+1} \right) \approx |{S_{r+1}^K}|$, then
\begin{equation}
    \dim \ker ( \partial_{r+1}^\dagger \partial_{r+1} ) \ll |{S_{r+1}^K}| \Rightarrow \beta_{r+1} \ll |{S_{r+1}^K}|.    
\end{equation}
Consequently, the proposed quantum algorithm for testing the triviality of a homology class performs optimally in the regime where the Betti numbers are small—e.g., in configurations such as low-genus surfaces. This is in direct contrast to the quantum algorithms for Betti number estimation discussed in~\cref{sec: estimatingBettinumbers}, which exhibit optimal performance in the high Betti number regime.

Last, the above procedure was described within the classical-access model. The same procedure can be executed within the sparse-access model, and thus the analysis of complexity can be done similarly. The only difference is that the complexity for obtaining the block-encoding of $ \partial_*^\dagger \partial_*/\Lambda^2$ is $\mathcal{O}\left( r \log |S^K_{r+1}|  \right) $, which leads to a different final complexity.

\subsection{Equivalence testing}
\label{sec: nonzerohomologyclass}
As previously discussed, the zero homology class represents the simplest case, consisting solely of trivial $r$-cycles. The procedure outlined above enables us to determine whether a given cycle is homologous to zero. Specifically, if the linear system $\partial_{r+1} c_{r+1} = c_r$ has no solution, then $c_r$ is not a boundary, and hence not homologous to zero. In this case, $c_r$ must represent a non-trivial homology class.

The solution relies on the fact that two $r$-cycles are homologous if and only if their difference is a boundary. That is, $c_1 \sim c_2$ if and only if there exists a $(r{+}1)$-chain $c_{r+1} \in C_{r+1}^K$ such that $\partial_{r+1} c_{r+1} = c_1 - c_2$.  Suppose that classical descriptions of $c_1$ and $c_2$ are given. As in the previous case, let $C_1$ and $C_2$ denote the sets of $r$-simplices supporting $c_1$ and $c_2$, respectively. Then the entries of the vectors $c_1$ and $c_2$ are classically accessible, and hence so are the entries of $c_1 - c_2$. 

To determine whether $c_1 - c_2$ is a boundary, we check whether the linear system $\partial_{r+1} c_{r+1} = c_1 - c_2$ has a solution. As in the zero-class case, this can be done by comparing the ranks of $\partial_{r+1}$ and the augmented matrix $[\partial_{r+1} | (c_1 {-} c_2)]$. Therefore, the computational complexity of this procedure is identical to that of verifying zero-homology, as previously discussed.

\subsection{Tracking homology classes}\label{sec: potentialappTDA}
The preceding sections were devoted to the problem of testing whether a given cycle is homologous to zero, or whether two given cycles are homologous to each other. Meanwhile, as discussed in~\cref{sec: TDA}, the central problem in topological data analysis (TDA) is the estimation of (persistent) Betti numbers associated with a simplicial complex. This task has been shown to be {NP-hard}~\cite{schmidhuber2022complexity}, thereby ruling out the possibility of an exponential quantum speed-up for generic inputs. As a result, significant asymptotic improvements through quantum algorithms for Betti number estimation appear unlikely.

Motivated by the techniques developed in our preceding analysis, we now consider whether they can provide utility within the broader framework of TDA. In~\cref{sec: estimatingpersistentbettinumbers}, we introduced the concept of persistent Betti numbers and their topological significance. In particular, the $r$-th persistent homology group encodes information about $r$-dimensional features that persist between two filtration scales. The rank of this group, the persistent Betti number, quantifies the number of such features. Intuitively, features that appear only within a narrow range of filtration values are considered topological noise, whereas those that persist across a wide range are interpreted as robust, intrinsic structures of the underlying dataset.

Inspired by this perspective, we propose a cycle-centric approach: rather than estimating persistent Betti numbers directly, we track the homological behavior of a specific cycle across filtration scales. To this end, suppose we are given three simplicial complexes $K_1 \subseteq K_2 \subseteq K_3$ obtained at increasing filtration values. Let $c_r$ be an $r$-cycle in $K_1$—and hence also in $K_2$ and $K_3$, since the inclusion of simplices preserves cycles.

Note that the boundary maps at each filtration scale differ, and we denote them as $\partial_{r+1}^{K_1}$, $\partial_{r+1}^{K_2}$, and $\partial_{r+1}^{K_3}$, respectively. By applying the zero-homology testing algorithm from~\cref{sec: zerohomologyclass} to each of these maps, we can determine whether $c_r$ is homologous to zero at each scale. For instance, if $c_r$ is not a boundary in $K_1$ but becomes a boundary in $K_2$ and $K_3$, then we observe that the homology class containing $c_r$ appears at the first scale and disappears in the later ones—precisely the type of topological change captured by persistent homology.

Similarly, we may apply the algorithm from~\cref{sec: nonzerohomologyclass} to compare the homological relationship between two cycles $c_1$ and $c_2$ at various filtration levels. Suppose that $c_1 \sim c_2$ in $K_1$, but not in $K_2$. Then at least one of the cycles must transition into a different homology class, indicating a change in the topological structure of the complex. This method therefore provides a complementary approach to analyzing persistent topological features—not by computing Betti numbers directly, but by tracking individual cycles through the filtration. Such cycle-based methods may offer additional insights or computational advantages in scenarios where specific cycles are of interest or where the homology classes themselves carry semantic meaning.

\subsection{From testing homology class to estimating Betti numbers}
\label{sec: trackingbettinumber}
In the preceding sections, we have shown that quantum algorithms can determine whether two given cycles belong to the same homology class. Furthermore, we have argued that such algorithms can be leveraged to track homology classes across varying length scales, i.e., over a filtration of a simplicial complex of interest. The appearance or disappearance of a homology class may indicate a change in the underlying topological structure; hence, the ability to track individual homology classes can be interpreted as the capability to detect topological changes in the dataset. For instance, a homology class that appears at a particular length scale but disappears shortly thereafter may reasonably be regarded as topological noise.

Motivated by this observation, we extend our consideration to a more general problem: \textit{tracking Betti numbers}. At first glance, this seems closely related to the context of~\cref{sec: estimatingpersistentbettinumbers}, in which we considered the estimation of persistent Betti numbers. These quantities, by definition, count the number of homology classes that persist across a range of filtration values. However, persistent Betti numbers capture the topological features of the entire complex in a \emph{global} manner, which can be computationally demanding. In contrast, our approach focuses on \emph{local} structure, thereby narrowing the scope and potentially reducing the computational overhead. The underlying expectation is that local analysis of homological features can effectively reveal global topological changes with less effort.

Suppose that at a given filtration level, corresponding to a simplicial complex $ K_1 $, we are given a collection of $ r $-cycles $ c_{r_1}, c_{r_2}, \dots, c_{r_s} $. Using the algorithm presented in~\cref{sec: nonzerohomologyclass}, we can determine whether any pair among them belongs to the same homology class. By performing such comparisons, we group homologous cycles together and select one representative from each group. Without loss of generality, let these representatives (modulo boundaries) be denoted $ c_{r_1}^h, c_{r_2}^h, \dots, c_{r_p}^h $, where $ p \leq s $. These representatives are elements of the homology group $ H_r $, and the $ r $-th Betti number $ \beta_r $ is defined as the dimension of $ H_r $. By definition, the dimension of a vector space corresponds to the maximal number of linearly independent elements in it. Therefore, if we can determine the number of linearly independent vectors among the set $ \{c_{r_1}^h, c_{r_2}^h, \dots, c_{r_p}^h\} $, we can infer the dimension of a subspace of $ H_r $.

To that end, we organize these vectors into a matrix
\begin{align}
    \mathcal{C} = [ c_{r_1}^h, c_{r_2}^h, \dots, c_{r_p}^h ],
\end{align}
where the rank of $ \mathcal{C} $ is precisely the number of linearly independent cycles among the given representatives. Since these vectors are assumed to be classically known, we can apply~\cref{lemma: entrycomputablematrix} to construct a block-encoding of the normalized Gram matrix ${\mathcal{C}^\dagger \mathcal{C}}/{\|\mathcal{C}\|_F^2}$, where the normalization by the Frobenius norm ensures that the spectral norm is bounded. Because the dimension of each column of $\mathcal{C}$ is $\mathcal{O}( |S_r^K|)$, so the complexity of this block-encoding is 
$$ \mathcal{O}\left(  \log (p |S_r^K|) \right)$$

The matrix $\mathcal{C}$ is of dimension $ p \times p $ and shares the same rank as $ \mathcal{C} $. Hence, our goal reduces to estimating the rank of this matrix. As discussed in~\cref{sec: estimatingBettinumbers}, several techniques exist for estimating the rank of a Hermitian matrix. Among them, the rank estimation algorithm, e.g., Lemma \ref{lemma: traceestimation} is particularly well-suited for our setting. It allows us to estimate the ratio
\begin{equation}
    \frac{1}{p} \cdot \operatorname{rank}\left( \frac{\mathcal{C}^\dagger \mathcal{C}}{ \|\mathcal{C}\|_F^2}  \right)
\end{equation}
to within additive error $ \varepsilon $ in time complexity
\begin{equation}
    \mathcal{O}\left( \frac{\lambda_{\min}(\mathcal{C}^\dagger \mathcal{C} /   \|\mathcal{C}\|_F^2 ))}{ \sqrt{\varepsilon}} \log \frac{1}{\epsilon}\right) = \mathcal{O}\left( \frac{ \|\mathcal{C}\|_F^2}{\sqrt{\epsilon}}\log \frac{1}{\epsilon} \right),
\end{equation}
where $ \lambda_{\min}(\mathcal{C}^\dagger \mathcal{C} / \mathrm{size} (\mathcal{C}) ) $ denotes the smallest nonzero eigenvalue of the matrix $ \mathcal{C}^\dagger \mathcal{C} /   \|\mathcal{C}\|_F^2 $, which is of order $\mathcal{O}( \|\mathcal{C}\|_F^2)$. The total complexity of this estimation is 
\begin{align}
    \mathcal{O}\left( \log (p |S_r^K|)  \frac{ \|\mathcal{C}\|_F^2}{\sqrt{\epsilon}} \log \frac{1}{\epsilon}    \right) 
\end{align}

Consequently, we can directly infer the desired ratio $ \operatorname{rank}(\mathcal{C})/p $, which provides an estimate of the number of linearly independent homology representatives. 

We emphasize that the vectors $ c_{r_1}^h, c_{r_2}^h, \dots, c_{r_p}^h $ belong to the (sub)space $ H_r $, whose dimension is precisely the $ r $th Betti number $ \beta_r $. Thus, the number of linearly independent such vectors yields a lower bound on $ \beta_r $, and in favorable cases, may even yield its exact value. This strategy therefore offers an alternative route to computing Betti numbers, supplementing the conventional approach based on combinatorial Laplacians, as reviewed in~\cref{sec: reviewalgebraictopology}.

\subsection{Cycle detection}\label{sec: detectingcycle}
As in~\cref{sec: zerohomologyclass}, let the indices of the $r$-simplices involved in $c_r$ be denoted by the index set $C := \{ i_1, i_2, \dots, i_L \}$, so that the $r$-chain can be formally written as
\begin{align}
    c_r = \sum_{i \in C} \ket{i}.
\end{align}
By definition, $c_r$ is an $r$-cycle if and only if $\partial_r c_r = 0$. Hence, to verify whether $c_r$ is a cycle, it suffices to examine the action of the boundary operator $ \partial_r $ on $ c_r $.

As discussed in~\cref{lemma: entrycomputablematrix}, given a classical description of the simplicial complex $ \{ \mathcal{S}_r \} $, i.e., via classical-access model, one can block-encode the operator
\begin{align}
    \frac{\partial_r^\dagger \partial_r}{ r|{S_r^K}|}
\end{align}
into a unitary operator, which we denote as $ U_r $.

Moreover, since the elements of the index set $ C $ are classically known, we can apply the method of~\cite{zhang2022quantum} to prepare the quantum state
\begin{align}
    \ket{c_r} = \frac{1}{\sqrt{L}} \sum_{i \in C} \ket{i},
\end{align}
using a quantum circuit of depth $ \mathcal{O}(\log (L)) $.

We now apply the unitary $ U_r $ to the state $ \ket{\mathbf{0}} \ket{c_r} $, where $ \ket{\mathbf{0}} $ denotes the ancilla qubits required for the block-encoding construction. According to~\cref{def: blockencode} and~\cref{eqn: action}, we have
\begin{align}
    U_r \ket{\mathbf{0}} \ket{c_r} = \ket{\mathbf{0}} \cdot \frac{\partial_r^\dagger \partial_r}{r|{S_r^K}|}\ket{c_r} + \ket{\mathrm{Garbage}},
\end{align}
where $ \ket{\mathrm{Garbage}} $ is orthogonal to the $ \ket{\mathbf{0}} $-component and takes the form
\begin{align}
    \ket{\mathrm{Garbage}} = \sum_{i \neq \mathbf{0}} \ket{i} \ket{\mathrm{Redundant}_i},
\end{align}
with $ \ket{\mathrm{Redundant}_i} $ denoting unnormalized and irrelevant residual states. Now, if $ c_r $ is indeed a cycle, then $ \partial_r \ket{c_r} = 0 $, and thus $\partial_r^\dagger \partial_r \ket{c_r} = 0$. This implies that the entire amplitude of the $ \ket{\mathbf{0}} $ component vanishes, and the resulting state is orthogonal to $ \ket{\mathbf{0}} $ in the ancilla register. Therefore, if we measure the ancilla qubits and never observe the outcome $ \ket{\mathbf{0}} $, we can infer that $ c_r $ is a cycle.

To make this inference statistically meaningful, we repeat the process $ T $ times. If the outcome $ \ket{\mathbf{0}} $ is never observed, then we can conclude, with success probability $ 1 - \eta $, that $ c_r $ is a cycle. It suffices to take $T = \mathcal{O}\left({1}/{\eta}\right)$. Thus, we obtain a \emph{probabilistic quantum algorithm} for verifying whether a given chain is a cycle.

\section{Cohomology and applications}
\label{sec: cohomologyandcupproduct}

\subsection{An overview of cohomology}\label{sec: overviewcohomology}
As discussed in~\cref{sec: reviewalgebraictopology,sec: TDA}, we introduced several foundational concepts in algebraic topology, including the $r$-simplex $\sigma_r$ (for $r \in \mathbb{Z}_+$), $r$-chains $c_r$ (formal linear combinations of $r$-simplices), and the $r$-th chain group $C_r^K$ of a simplicial complex $K$. For brevity, we drop the superscript $K$ and denote the $r$-chain group simply by $C_r$. These objects form the backbone of homology theory.

Cohomology theory, in contrast, centers on \emph{cochains}, which are linear functionals mapping chains to real numbers. Formally, an $r$-th cochain $\omega^r$ is a map $\omega^r : C_r \to \mathbb{R}$. The set of all $r$-th cochains, denoted $C^r$, forms a vector space over $\mathbb{R}$ with the natural additive structure:
\begin{align}
    (\omega_1^r + \omega_2^r)(c_r) = \omega_1^r(c_r) + \omega_2^r(c_r), \quad \forall c_r \in C_r.
\end{align}

In analogy with the boundary operator in homology, $\partial_r : C_r \to C_{r-1}$, cohomology introduces the \emph{coboundary} operator, $\delta^r : C^r \to C^{r+1}$. This operator is defined via duality: for any $c_{r+1} \in C_{r+1}$ and $\omega^r \in C^r$, $\delta^r \omega^r (c_{r+1}) := \omega^r ( \partial_{r+1}(c_{r+1}) )$. This definition is well-posed since $\partial_{r+1}(c_{r+1}) \in C_r$, and $\omega^r$ acts on elements of $C_r$. A fundamental fact is that the matrix representation of $\delta^r$ is the transpose (or adjoint) of $\partial_{r+1}$. $\delta^r = \partial_{r+1}^\dagger$.

We now mirror the homological concepts in the cohomological setting. An $r$-chain $c_r$ is called a \emph{cycle} if $\partial_r c_r = 0$. Analogously, an $r$-cochain $\omega^r$ is called a \emph{cocycle} if $\delta^r \omega^r = 0$. The set of all $r$-cocycles forms the cocycle group $Z^r$. A cochain $\omega^r$ is called a \emph{coboundary} if there exists $\omega^{r-1} \in C^{r-1}$ such that $\omega^r = \delta^{r-1} \omega^{r-1}$. The set of all $r$-coboundaries forms the coboundary group $B^r$. It is a standard result that the coboundary operators satisfy $\delta^r \delta^{r-1} = 0$, implying $B^r \subseteq Z^r$. The $r$-th cohomology group is then defined as the quotient: $H^r := Z^r / B^r$. A central theorem in algebraic topology asserts that the cohomology and homology groups are isomorphic: $H^r \cong H_r$, which illustrates the duality between cohomology and homology. In particular, this isomorphism implies that Betti numbers can equivalently be computed via the spectrum of coboundary operators.

Let us now consider the space $C^r$ of $r$-cochains, which has dimension $|S_r^K|$, the number of $r$-simplices in $K$. Let $\{ e_i \}_{i=1}^{|S_r^K|}$ denote a basis of $C^r$ such that for the $j$-th $r$-simplex $\sigma_{r_j}$, the basis element $e_i$ satisfies $e_i(\sigma_{r_j}) = \delta_{ij}$. Any cochain $\omega^r \in C^r$ can then be expressed as:
\begin{align}
    \omega^r = \sum_{i=1}^{|S_r^K|} \omega^r_i e_i,
\end{align}
where $\omega^r_i \in \mathbb{R}$. For a general $r$-chain $c_r \in C_r$ written as
\begin{align}
    c_r = \sum_{j=1}^{|S_r^K|} (c_r)_j \sigma_{r_j},
\end{align}
the evaluation of $\omega^r$ on $c_r$ is:
\begin{align}
    \omega^r(c_r) &= \sum_{i=1}^{|S_r^K|} \omega^r_i e_i(c_r) = \sum_{i=1}^{|S_r^K|} \omega^r_i e_i\left( \sum_{j=1}^{|S_r^K|} (c_r)_j \sigma_{r_j} \right) \nonumber\\
    &= \sum_{i=1}^{|S_r^K|} \sum_{j=1}^{|S_r^K|} \omega^r_i (c_r)_j \delta_{ij} = \sum_{i=1}^{|S_r^K|} \omega^r_i (c_r)_i.
\end{align}
This inner product representation provides a concrete numerical interpretation of cochain action on chains under the canonical basis.

\subsection{Cohomological frameworks for constructing $r$-cocycles}
\label{sec: cohomologydetectinghomology}
In the previous section, we introduced the foundational notions of cohomology theory. In particular, we defined an $r$-cochain $\omega^r$ as a linear functional that maps an arbitrary $r$-chain to a real number. An $r$-cochain $\omega^r$ is called an \emph{$r$-cocycle} if it lies in the kernel of the coboundary operator $\delta^r$, i.e., $\delta^r \omega^r = 0$. Meanwhile, an $r$-cochain is an \emph{$r$-coboundary} if it lies in the image of the coboundary operator on $(r{-}1)$-cochains, i.e., $\omega^r = \delta^{r-1} \omega^{r-1}$ for some $(r{-}1)$-cochain $\omega^{r-1}$.

The $r$-th cohomology group is then defined as the quotient
\begin{equation}
    H^r := \frac{\ker (\delta^r)}{\operatorname{im} (\delta^{r-1})},    
\end{equation}
which imposes an equivalence relation: two $r$-cocycles are cohomologous if their difference is an $r$-coboundary. That is, $\omega_1^r \sim \omega_2^r$ if and only if $\omega_1^r - \omega_2^r = \delta^{r-1} \omega^{r-1}$ for some $\omega^{r-1}$. The cohomology group $H^r$ thus consists of equivalence classes of $r$-cocycles modulo coboundaries, analogous to how $r$-cycles modulo boundaries form homology groups, as discussed in~\cref{sec: reviewalgebraictopology,sec: zerohomologyclass,sec: nonzerohomologyclass}.

An important property of cohomology is that any $r$-cocycle $\omega^r$ maps homologous $r$-cycles to the same real value. To see this, let $c_{r_1}, c_{r_2}$ be two $r$-cycles such that they are homologous, i.e., there exists an $(r{+}1)$-chain $c_{r+1}$ with $c_{r_1} - c_{r_2} = \partial_{r+1} c_{r+1}$. Then for any $r$-cocycle $\omega^r$, we compute:
\begin{align}
    \omega^r (c_{r_1}) &= \omega^r (c_{r_2} + \partial_{r+1} c_{r+1} ) \\
    &= \omega^r (c_{r_2}) + \omega^r(\partial_{r+1} c_{r+1}) \\
    &= \omega^r (c_{r_2}) +  (\delta^{r} \omega^r)(c_{r+1}).
\end{align}
Since $\omega^r$ is a cocycle, $\delta^{r} \omega^r = 0$, hence $\omega^r (c_{r_1}) = \omega^r (c_{r_2})$. In other words, the value of $\omega^r$ on a cycle depends only on its homology class. This observation motivates us to ask whether a similar invariance holds for cohomologous cocycles.

Let $\omega_1^r$ and $\omega_2^r$ be two cohomologous $r$-cocycles, i.e., there exists an $(r{-}1)$-cochain $\omega^{r-1}$ such that
\begin{equation}
    \omega_1^r - \omega_2^r = \delta^{r-1} \omega^{r-1}.
\end{equation}
We evaluate both cocycles on an arbitrary $r$-cycle $c_r$:
\begin{align}
    \omega_1^r (c_r) &= \omega_2^r (c_r ) + (\delta^{r-1} \omega^{r-1})(c_r) \\
    &= \omega_2^r (c_r ) + \omega^{r-1} (\partial_r c_r) \\
    &= \omega_2^r (c_r),
\end{align}
since $c_r$ is a cycle and thus $\partial_r c_r = 0$. Therefore, cohomologous cocycles agree on all cycles. Consequently, we may regard cohomology classes as functionals on homology classes. This leads to the key insight: 
\begin{remark}
    A cohomology class defines a well-defined linear functional on homology classes.
\end{remark}

Such properties reveal the deep duality between homology and cohomology. Equivalence relations are imposed respectively on cycles and cocycles via boundaries and coboundaries, and the topological structure of the underlying complex is revealed through how these classes interact. The duality described above motivates a reformulation of the homology testing problem using cohomological language.


Thanks to the established duality, we know that for any $r$-cocycle $\omega^r$, the difference $\omega^r(c_{r_1}) - \omega^r(c_{r_2})$ depends only on the homology class of $c_{r_1} - c_{r_2}$. In particular, $c_{r_1} \sim c_{r_2}$ if and only if $\omega^r(c_{r_1}) = \omega^r(c_{r_2})$ for all cocycles $\omega^r$, or equivalently, for all cohomology classes $[\omega^r] \in H^r$. Therefore, we obtain the following criterion:
\begin{remark}
    Two $r$-cycles are homologous if and only if all $r$-cocycles evaluate them equally.
\end{remark}
This cohomological lens offers not only a theoretical foundation but also a potential algorithmic strategy, particularly in settings where cocycles can be efficiently computed or represented, such as in persistent cohomology or combinatorial Hodge theory.

As outlined in~\cref{sec: nonzerohomologyclass}, the homological approach to testing whether two $r$-cycles $c_{r_1}$ and $c_{r_2}$ are homologous relies on the definition that they differ by a boundary. That is, there exists a $(r{+}1)$-chain $c_{r+1}$ such that $c_{r_1} - c_{r_2} = \partial_{r+1} c_{r+1}$. This provides a constructive means of certification: finding such a $c_{r+1}$ is sufficient to conclude that $c_{r_1} \sim c_{r_2}$. In contrast, the cohomological perspective is built upon the dual statement we proved earlier, namely, that for any $r$-cocycle $\omega^r$, one has $\omega^r(c_{r_1}) = \omega^r(c_{r_2})$ if $c_{r_1} \sim c_{r_2}$. This leads to the natural question: 
\begin{center}
    \emph{How does one obtain an $r$-cocycle, or at least the values $\omega^r(c_{r_1})$ and $\omega^r(c_{r_2})$?}    
\end{center}
This question is crucial for utilizing the cohomological viewpoint as a computational tool.

Following the same approach as in the previous section, we consider an explicit representation of the $r$-cocycle $\omega^r$. Let $\{e_i\}_{i=1}^{|S_r^K|}$ be an orthonormal basis for the cochain group $C^r$ (the dual space of $r$-chains). Then $\omega^r$ can be written as:
\begin{align}
    \omega^r = \sum_{i=1}^{|S_r^K|} \omega^r_i e_i,
\end{align}
where each $\omega^r_i \in \mathbb{R}$ represents the coefficient corresponding to basis functional $e_i$. In vector notation, we express this as:
\begin{align}
    \omega^r =
    \begin{pmatrix}
        \omega^r_1 \\
        \omega^r_2 \\
        \vdots \\
        \omega^r_{|S_r^K|}
    \end{pmatrix} \in \mathbb{R}^{|S_r^K|}.
\end{align}

To evaluate $\omega^r$ on an $r$-cycle $c_r$, it suffices to express $c_r$ in the same basis $\{e_i\}$ (via the identification of chains and cochains under the inner product), allowing us to compute $\omega^r(c_r)$ as a dot product $\omega^r(c_r) = \langle \omega^r, c_r \rangle$. Therefore, the action of an $r$-cocycle on a cycle can be implemented via a linear functional. The key point is that if $c_{r_1}$ and $c_{r_2}$ are not homologous, then there exists some cocycle $\omega^r$ such that $\omega^r(c_{r_1}) \ne \omega^r(c_{r_2})$. Consequently, homology detection reduces to finding such a separating cocycle in $Z^r$ (the space of cocycles), or equivalently, testing whether $\omega^r(c_{r_1}) = \omega^r(c_{r_2})$ for all $\omega^r \in Z^r$. We present two methodologies for obtaining the desired $r$-cocycle using such cohomology functionals.

\subsubsection{Constructing block encodings via projection onto $\ker(\delta^r)$}
Since the space of $r$-cocycles is precisely the kernel of the coboundary operator $\delta^r$, i.e., $\ker(\delta^r) \subseteq C^r$, any cochain $\omega^r \in C^r$ can be uniquely decomposed as $\omega^r = \omega^r_{\text{cycle}} + \omega^r_{\text{non-cycle}}$, where $\omega^r_{\text{cycle}} \in \ker(\delta^r)$ and $\omega^r_{\text{non-cycle}} \in \operatorname{im}((\delta^r)^\top)$. To project a general cochain $\omega^r$ onto the kernel space of $\delta^r$, we define the following projection:
\begin{align}
    \omega^r_{\text{proj}} := \omega^r - (\delta^r)^\top (\delta^r (\delta^r)^\top )^{-1} \delta^r \cdot \omega^r.
\end{align}
We verify that this projected vector indeed lies in $\ker(\delta^r)$ by direct computation:
\begin{align}
    \delta^r \omega^r_{\text{proj}} 
    &= \delta^r ( \omega^r - (\delta^r)^\top ( \delta^r (\delta^r)^\top )^{-1} \delta^r \cdot \omega^r ) \\
    &= \delta^r \omega^r - \delta^r (\delta^r)^\top ( \delta^r (\delta^r)^\top )^{-1} \delta^r \cdot \omega^r \\
    &= \delta^r \omega^r - \delta^r \omega^r = 0.
\end{align}
Thus, $\omega^r_{\text{proj}} \in \ker(\delta^r)$ as required.

We now consider the matrix representation of $\delta^r$, which, as discussed earlier, is the transpose of the boundary operator $\partial_{r+1}$. Recall from~\cref{sec: estimatingBettinumbers} that the operator
\begin{equation}
    \frac{\partial_{r+1}^\dagger \partial_{r+1}}{ (r+1) |{S_{r+1}^K}|}
\end{equation}
can be block-encoded efficiently using~\cref{lemma: entrycomputablematrix} if the classical knowledge is provided.. Since $\delta^r = \partial_{r+1}^\top$, we automatically obtain a block encoding of
\begin{equation} \label{eq: block_todo_cohom_1}
    \frac{\delta^r (\delta^r)^\top}{ (r+1)  |{S_{r+1}^K}|}.
\end{equation}
A similar construction yields a block encoding of
\begin{equation} \label{eq: block_todo_cohom_2}
    \frac{(\delta^r)^\top \delta^r}{ (r+1)  |{S_{r+1}^K}|}.
\end{equation}

Next, we apply~\cref{lemma: positive}, which allows for computing the square root of a positive semidefinite block-encoded operator, to obtain a block encoding of
\begin{equation} \label{eq: block_todo_cohom_3}
    \left(\frac{{(\delta^r)^\top \delta^r}}{{ (r+1)  |{S_{r+1}^K}|}}\right)^{1/2}.
\end{equation}

Finally, we observe the following decomposition of the projection matrix:
\begin{align}
    (\delta^r)^\top ( \delta^r (\delta^r)^\top )^{-1} \delta^r = ({(\delta^r)^\top \delta^r})^{1/2} \cdot ( (\delta^r)^\top \delta^r )^{-1} \cdot ({(\delta^r)^\top \delta^r})^{1/2}.
    \label{eq: projection_identity}
\end{align}
This factorization highlights that the projection onto $\operatorname{im}((\delta^r)^\top)$ is symmetric and idempotent, and its complement yields a projection onto $\ker(\delta^r)$, which corresponds to the subspace of cocycles.

Now, starting from the block encoding of~\cref{eq: block_todo_cohom_2}, we apply~\cref{lemma: matrixinversion} to obtain a block encoding of its inverse:
\begin{equation}\label{eq: big_condition_number}
    \frac{1}{\kappa_\delta} ((\delta^r)^\top \delta^r)^{-1},
\end{equation}
where $\kappa_\delta$ denotes the condition number of $(\delta^r)^\top \delta^r$. In general, this condition number may be large, making matrix inversion costly. To mitigate this, we invoke the preconditioning technique from~\cite{clader2013preconditioned} (see~\cref{sec: invertingillconditionmatrix}), which constructs a matrix $A$ such that $AM$ has bounded condition number, allowing one to invert $M$ efficiently via:
\begin{equation}
    M^{-1} = (AM)^{-1} A.
\end{equation}
Assuming $A$ can be constructed as in~\cite{clader2013preconditioned}, we first use~\cref{lemma: entrycomputablematrix} to block-encode
\begin{equation}
    \frac{A^\dagger A}{ ||A||_F^2}.
\end{equation}
Using~\cref{lemma: product}, we obtain a block encoding of
\begin{equation}
    \frac{A^\dagger A}{ ||A||_F^2} \cdot \frac{(\delta^r)^\top \delta^r}{ (r+1) |S_{r+1}^K|}.
\end{equation}
Then, applying~\cref{lemma: matrixinversion}, we get:
\begin{align}
    \frac{1}{\kappa} ((\delta^r)^\top \delta^r)^{-1} (A^\dagger A)^{-1},
\end{align}
where $\kappa$ is the condition number of $A^\dagger A \cdot  (\delta^r)^\top \delta^r$, which is guaranteed to be small by construction. We then multiply with the block encoding of $A^\dagger A / \mathrm{size}(A)$, resulting in:
\begin{align}
    \frac{((\delta^r)^\top \delta^r)^{-1}}{||A||_F^2}.
\end{align}
Finally, using~\cref{lemma: product} again with the block encoding of~\cref{eq: block_todo_cohom_3}, we obtain:
\begin{align}
    \left(\frac{{(\delta^r)^\top \delta^r}}{{ (r+1) |S_{r+1}^K|}}\right)^{1/2} \cdot \frac{((\delta^r)^\top \delta^r)^{-1}}{\kappa \cdot ||A||_F^2} \cdot \left(\frac{{(\delta^r)^\top \delta^r}}{{ (r+1) |S_{r+1}^K|}}\right)^{1/2} = \frac{(\delta^r)^\top (\delta^r (\delta^r)^\top)^{-1} \delta^r}{\kappa \cdot  ||A||_F^2 \cdot  (r+1) |S_{r+1}^K|}. \label{eq: block_todo_cohom_4}
\end{align}
This gives us a block encoding of the projection matrix onto $\operatorname{im}((\delta^r)^\top)$.

We now turn to computing the projected component of an arbitrary $r$-cochain $\omega^r$. We randomly generate a unitary $U_r$ of size $|S_r^K| \times |S_r^K|$ and use its first column as $\omega^r$. Using~\cref{lemma: product}, we compute a block encoding of:
\begin{equation} \label{eq: block_todo_cohom_5}
    \frac{(\delta^r)^\top (\delta^r (\delta^r)^\top)^{-1} \delta^r U_r}{\kappa \cdot ||A||_F^2 \cdot  (r+1)|S_{r+1}^K|}.
\end{equation}
The first column of this matrix is:
\begin{equation}
    \frac{(\delta^r)^\top (\delta^r (\delta^r)^\top)^{-1} \delta^r \omega^r}{\kappa \cdot ||A||_F^2 \cdot  (r+1)|S_{r+1}^K|}.
\end{equation}
Let us define the rescaled cochain
\begin{equation}
    (\omega^r)' := \frac{\omega^r}{\kappa \cdot ||A||_F^2 \cdot  (r+1)|S_{r+1}^K|}.
\end{equation}
Then the above vector equals
\begin{equation}
    (\delta^r)^\top (\delta^r (\delta^r)^\top)^{-1} \delta^r (\omega^r)'.
\end{equation}
We now use~\cref{lemma: scale} to block-encode 
\begin{equation}
    \frac{U_r}{\kappa \cdot ||A||_F^2 \cdot  (r+1)|S_{r+1}^K|},
\end{equation}
which contains $(\omega^r)'$ in the first column. Finally, using~\cref{lemma: sumencoding}, we compute the block encoding of:
\begin{align}
    \frac{U_r - (\delta^r)^\top (\delta^r (\delta^r)^\top)^{-1} \delta^r U_r}{2 \kappa \cdot ||A||_F^2 \cdot  (r+1)|S_{r+1}^K|},
\end{align}
whose first column equals
\begin{equation}
    \frac{1}{2} ( (\omega^r)' - (\delta^r)^\top (\delta^r (\delta^r)^\top)^{-1} \delta^r (\omega^r)' ).
\end{equation}
As shown earlier, this vector lies in $\ker(\delta^r)$, and hence is an $r$-cocycle. We summarize the above construction as follows:

\begin{lemma}[Efficient block-encoding of an $r$-cocycle]
    There exists a quantum procedure with circuit complexity $\mathcal{O}(\log ( |S_{r-1}^K| |S_r^K|))$ that returns the unitary block encoding $U_c$ of a matrix whose first column is an $r$-cocycle $\omega^r_c \in \ker(\delta^r)$.
\end{lemma}

\subsubsection{Manual construction via explicit representatives}
Recall that an $r$-cocycle $\omega^r_c$ belongs to the kernel of $\delta^r$. In other words, for any $(r{+}1)$-cochain $c_{r+1}$, we have
\begin{equation}\label{eq:cocycle_property}
    \delta^r \omega^r_c (c_{r+1}) = \omega^r_c (\partial_{r+1} c_{r+1}).
\end{equation}
Let $S_{r+1}^K = \{\sigma_{|S_r^K|_i}\}_{i=1}^{|S_{r+1}^K|}$ denote the collection of all $(r{+}1)$-simplices. Hence, any $(r{+}1)$-cochain can be expressed as
\begin{equation}
    c_{r+1} = \sum_{i=1}^{|S_{r+1}^K|} c_i\, \sigma_{|S_r^K|_i}.
\end{equation}
Substituting this expansion into \cref{eq:cocycle_property}, we obtain
\begin{align}
    \delta^r \omega^r_c(c_{r+1}) &= \omega^r_c \left(\partial_{r+1} \sum_{i=1}^{|S_{r+1}^K|} c_i\, \sigma_{|S_r^K|_i} \right) \\
    &= \sum_{i=1}^{|S_{r+1}^K|} c_i\, \omega^r_c\bigl(\partial_{r+1} \sigma_{|S_r^K|_i}\bigr).
\end{align}
If $\omega^r_c$ is indeed a cocycle, then the left-hand side of the above equation is zero. Consequently, one must have
\begin{equation}\label{eq:cocycle_boundary}
    \omega^r_c(\partial_{r+1} \sigma_{|S_r^K|_i}) = 0, \quad \forall\, i=1,2,\dots,|S_{r+1}^K|.
\end{equation}
Conversely, if an $r$-cochain $\omega^r_c$ satisfies \cref{eq:cocycle_boundary} for every $(r{+}1)$-simplex in $S_{r+1}^K$, then $\delta^r \omega^r_c (c_{r+1}) = 0$ for all $c_{r+1}$, so that $\omega^r_c \in \ker(\delta^r)$ and hence is an $r$-cocycle. The practical challenge is thus: 
\begin{center}
    \emph{how can one manually choose an $r$-cochain $\omega^r$ satisfying \cref{eq:cocycle_boundary}?}    
\end{center}
A straightforward strategy is as follows. One iterates over all $(r{+}1)$-simplices and assigns real values to the $r$-simplices lying on their boundaries in such a way that the conditions in \cref{eq:cocycle_boundary} hold. Specifically, one may proceed according to the following steps:
\begin{enumerate}[label=(\roman*)]
    \item Start with the first $(r{+}1)$-simplex $\sigma_{|S_r^K|_1}$ and consider its boundary $\partial_{r+1}\sigma_{|S_r^K|_1}$, whose constituent $r$-simplices are known from the classical description of the boundary operator $\partial_{r+1}$.
    \item Assign arbitrary (e.g., random) real values $x_1,x_2,\dots,x_{r+1}$ to the first $r$-simplices in the boundary, and then determine the value on the remaining $r$-simplex by requiring that the sum of the assigned values on the boundary is zero. For example, if the boundary of $\sigma_{|S_r^K|_1}$ has $(r{+}2)$ faces, one may set the value for the last face as $1 - (x_1 + x_2 + \cdots + x_{r+1})$, so that the boundary condition is met.
    \item Next, consider a second $(r{+}1)$-simplex $\sigma_{|S_r^K|_2}$ and similarly assign values on its boundary. When a face is shared with a previously processed simplex, the previously assigned value is re-used.
    \item Repeat this procedure until real values have been assigned to all $r$-simplices in such a way that each $(r{+}1)$-simplex satisfies the cocycle condition \cref{eq:cocycle_boundary}.
\end{enumerate}

This method yields an explicit construction of the desired $r$-cocycle $\omega^r_c$. Since every $(r{+}1)$-simplex is processed and the boundary operator $\partial_{r+1}$ has dimension $|S_r^K| \times |S_{r+1}^K|$, the overall classical time complexity for this manual selection process is $\mathcal{O}( |S_r^K| |S_{r+1}^K|)$. The above selection procedure has complexity $\mathcal{O}( |S_r^K||S_{r+1}^K|)$. Thus, in practice, it is only efficient when $|S_{r+1}^K|$ is not large. It turns out that things can be simpler under certain circumstances, which can further optimize the classical preproccessing time. We recall that we are seeking for $r$-cochain that satisfies the condition: 
\begin{equation}
    \omega^r_c(\partial_{r+1} \sigma_{|S_r^K|_i}) = 0, \quad \forall\, i=1,2,\dots,|S_{r+1}^K|.
\end{equation}

Suppose there exist two $r$-simplices that are both faces of a common $(r{+}1)$-simplex, and furthermore, are not faces of any other $(r{+}1)$-simplices. In this case, the matrix $\mathcal{S}_{r+1}$ (and equivalently, the boundary matrix $\partial_{r+1}$) contains two rows—indexed by $p$ and $q$ with $p,q \leq |S_r^K|$—each having a single nonzero entry located in the same column, indexed by $k \leq |S_{r+1}^K|$. The rows indexed by $p$ and $q$ correspond to the $r$-simplices $\sigma_{r_p}$ and $\sigma_{r_q}$, respectively, both of which are faces of the $(r{+}1)$-simplex $\sigma_{|S_r^K|_k}$. Define a cochain $\omega^r_c \in C^r$ supported only on $\sigma_{r_p}$ and $\sigma_{r_q}$ according to the following rule:
\begin{enumerate}[label=(\roman*)]
    \item If $(\partial_{r+1})_{p,k} = (\partial_{r+1})_{q,k}$, then set
    \begin{align}
        & \omega^r_c(\sigma_{r_p}) = -\omega^r_c(\sigma_{r_q}) = 1, \\    
        & \omega^r_c(\sigma_{r_i}) = 0 \quad \text{for all } i \notin \{p, q\}.
    \end{align}
    \item If $(\partial_{r+1})_{p,k} = -(\partial_{r+1})_{q,k}$, then set
    \begin{align}
        & \omega^r_c(\sigma_{r_p}) = \omega^r_c(\sigma_{r_q}) = 1, \\
        & \omega^r_c(\sigma_{r_i}) = 0 \quad \text{for all } i \notin \{p, q\}.
    \end{align}
\end{enumerate}
The above construction is elementary and can be implemented in nearly constant time, provided such a pair of $r$-simplices exists.

Once the values $\omega^r_c(\sigma_{r_i})$ on the $r$-simplices are determined for all $i=1,2,\dots,|S_r^K|$, we can employ the method described in \cite{zhang2022quantum} to prepare a quantum state corresponding to the normalized $r$-cocycle:
\begin{equation}
    \ket{\omega^r_c} = \frac{1}{\omega}\sum_{i=1}^{|S_r^K|} \omega^r_c(\sigma_{r_i}) \ket{i-1},
\end{equation}
where the normalization constant is given by
\begin{equation}
    \omega = \left({\sum_{i=1}^{|S_r^K|} \omega^r_c(\sigma_{r_i})^2}\right)^{1/2}.
\end{equation}
Thus, the classical procedure of manual selection can be efficiently interfaced with the quantum state preparation technique for further quantum processing.

\subsection{Homology equivalence testing with cohomological frameworks}\label{sec: hom_equiv_test_cohom}
We now return to our main objective: determining whether two given $r$-cycles $c_{r_1}$ and $c_{r_2}$ are homologous (\cref{prob: homology_testing_nontrivial}; homology equivalence testing). 

\subsubsection{Constructing block encodings via projection onto $\ker(\delta^r)$}
As in~\cref{sec: zerohomologyclass} and~\cref{sec: nonzerohomologyclass}, we denote the indices of the $r$-simplices contained in $c_{r_1}$ by $C_1 := \{i^1_1, i^1_2, \dots, i^1_{L_1} \}$, and those contained in $c_{r_2}$ by $C_2 := \{i^2_1, i^2_2, \dots, i^2_{L_2} \}$, where $i^1_j, i^2_j \in \{1, 2, \dots, |S_r^K|\}$, and $L_1, L_2 \leq |S_r^K|$ denote the lengths of the cycles $c_{r_1}$ and $c_{r_2}$, respectively.

Having obtained a block encoding of the $r$-cocycle $\omega^r_c$, we now seek to evaluate $\omega^r(c_{r_1})$ and $\omega^r(c_{r_2})$. Since $c_{r_1} = \sum_{i \in C_1} \sigma_{r_i}$ and $c_{r_2} = \sum_{i \in C_2} \sigma_{r_i}$, we have
\begin{align}
    \omega^r(c_{r_1}) = \sum_{i \in C_1} \omega^r(\sigma_{r_i}), \quad \omega^r(c_{r_2}) = \sum_{i \in C_2} \omega^r(\sigma_{r_i}).
\end{align}

Let $U_c$ be a unitary that block-encodes the vector $\omega^r_c$ in its first column. Applying $U_c$ to the initial state $\ket{\mathbf{0}} \ket{0}_{|S_r^K|}$, where $\ket{\mathbf{0}}$ is an ancilla register used for block encoding and $\ket{0}_{|S_r^K|}$ denotes the first computational basis state in the $|S_r^K|$-dimensional Hilbert space, we obtain
\begin{equation}
    U_c \ket{\mathbf{0}} \ket{0}_{|S_r^K|} = \ket{\mathbf{0}} \ket{\omega^r_c} + \ket{\mathrm{Garbage}}.
\end{equation}
Knowing the index set $C_1 = \{i^1_1, i^1_2, \dots, i^1_{L_1}\}$, we can use the method proposed in~\cite{zhang2022quantum} to construct the quantum state
\begin{equation}
    \ket{\phi_1} = \frac{1}{L_1} \sum_{i \in C_1} \ket{i}.
\end{equation}
By appending the ancilla register, we obtain the state $\ket{\mathbf{0}} \ket{\phi_1}$. Then, the inner product
\begin{align}
    \bra{\mathbf{0}} \bra{\phi_1} \left(\ket{\mathbf{0}} \ket{\omega^r_c} + \ket{\mathrm{Garbage}}\right)
    & = \braket{\phi_1 | \omega^r_c} \\
    & = \frac{1}{L_1} \sum_{i \in C_1} \omega^r(\sigma_{r_i}) = \frac{1}{L_1} \omega^r(c_{r_1})
\end{align}
can be estimated to additive accuracy $\varepsilon$ using either the Hadamard test or the SWAP test, with circuit complexity $\mathcal{O}(1/\varepsilon)$.  A similar procedure can be applied to estimate the ratio $\omega^r(c_{r_2}) / L_2$. By comparing the two values $\omega^r(c_{r_1})$ and $\omega^r(c_{r_2})$, one can determine whether the corresponding cycles are homologous.

The complexity for preparing the unitary $U_c$ is $\mathcal{O}(\log(r|S_r^K|))$. Preparing the states $\ket{\phi_1}$ and $\ket{\phi_2}$ requires complexity $\mathcal{O}(\log (L_1))$ and $\mathcal{O}(\log (L_2))$, respectively. The overlap estimation step has complexity $\mathcal{O}(1/\varepsilon)$. Hence, the total complexity is given by
\begin{align}
    \mathcal{O}\left( \frac{\left( \log(r|S_r^K|) + \log L_1 + \log L_2 \right)}{\varepsilon} \right) = \mathcal{O}\left( \frac{\log \left( r|S_r^K|\cdot L_1 L_2 \right)}{\varepsilon} \right).
\end{align}
By choosing $\varepsilon = \mathcal{O}(1)$, we obtain constant-additive-error estimates of the target quantities, which suffice for the purpose of comparison. Since the maximal values of $L_1$ and $L_2$ are at most $|S_r^K|$, the overall complexity simplifies to $\mathcal{O}(\log (r|S_r^K|))$.

\subsubsection{Manual construction via explicit representatives}
Alternatively, by manual selection, the $r$-cocycle $\omega_c^r$ is classically known. Then we use the method in \cite{zhang2022quantum} to prepare the state
\begin{equation}
    \ket{\omega^r_c} = \frac{1}{\omega} \sum_{i=1}^{|S_r^K|} \omega^r_c(\sigma_{r_i}) \ket{i-1},
\end{equation}
with normalization factor
\begin{equation}
    \omega = \left({ \sum_{i=1}^{|S_r^K|} \omega^r_c(\sigma_{r_i})^2 }\right)^{1/2},
\end{equation}
Then we can apply essentially the same procedure to evaluate the action of $\omega$ on the two $r$-cycles $c_{r_1}, c_{r_2}$. Specifically, the inner product
\begin{align}
    \braket{\phi_1 | \omega^r_c} = \frac{1}{\omega L_1} \sum_{i \in C_1} \omega^r(\sigma_{r_i}) = \frac{1}{\omega L_1} \omega^r(c_{r_1})
\end{align}
can again be estimated via the Hadamard or SWAP test. The same applies for the estimate of ${\omega^r(c_{r_2})}/{\omega L_2}$, after which we compare the results to decide homology equivalence.  The complexity of preparing the state $\ket{w^r_c}$ is $\mathcal{O}(|S_r^K|)$. Preparing the states $\ket{\phi_1}$ and $\ket{\phi_2}$ requires complexities $\mathcal{O}(\log (L_1))$ and $\mathcal{O}(\log (L_2))$, respectively. The final overlap estimation step incurs a complexity of $\mathcal{O}(1/\varepsilon)$. Therefore, the total complexity is given by
\begin{align}
    \mathcal{O}\left( \frac{\left( \log |S_r^K| + \log L_1 + \log L_2 \right)}{\varepsilon} \right) = \mathcal{O}\left( \frac{\log (|S_r^K| \cdot L_1 L_2)}{\varepsilon} \right).    
\end{align}
Assuming $\varepsilon = \mathcal{O}(1)$, this simplifies to $\mathcal{O}\left( \log (|S_r^K| \cdot L_1 L_2) \right)$. Since $L_1, L_2 \leq |S_r^K|$, we conclude that the total complexity is $ \mathcal{O}\left( \log ( |S_r^K|)\right)$. The time complexity of the classical processing for manual selection is considered separately.



\section*{Acknowledgements}
N.A.N. thanks Junseo Lee for helpful discussions. N.A.N. acknowledges support from the Center for Distributed Quantum Processing. Part of this work was completed during N.A.N.'s internship at QuEra Computing Inc.




\bibliographystyle{alpha}
\bibliography{ref.bib}

\newpage
\appendix
\section{Proof of~\cref{lemma: entrycomputablematrix}}
\label{sec: proofoflemmaentrycomputablematrix}

\subsection{General framework}
To begin, we note that~\cref{lemma: entrycomputablematrix} can be regarded as a corollary of a recent result~\cite{nghiem2025refined}. Here, we recapitulate and, to some extent, simplify and refine the proof in~\cite{nghiem2025refined} for our purposes.  
First, we point out the following property. Let $A$ be a matrix of size $M \times N$, and let $A^i$ denote the $i$-th column of $A$. Then, consider the following (not necessarily normalized) vector:
\begin{align}
     \frac{1}{\sqrt{MN}}\sum_{i=1}^N \underbrace{A^i}_{\rm register~\mathsf{1}}\ket{i},
\end{align}
which is of dimension $MN$. If we trace out the first register of the above vector, we obtain the following operator:
\begin{align}
  \frac{A^\dagger A}{MN}.
\end{align}
Our goal is to construct a block encoding of the above operator, assuming classical access to the entries of $A$.  
Before describing the procedure, we introduce a helpful lemma:
\begin{lemma}[Block encoding of density matrix, see e.g.~\cite{gilyen2019quantum}]
\label{lemma: improveddme}
Let $\rho = \Tr_A \ket{\Phi}\bra{\Phi}$, where $\rho \in \mathbb{H}_B$ and $\ket{\Phi} \in  \mathbb{H}_A \otimes \mathbb{H}_B$. Given a unitary $U$ that prepares $\ket{\Phi}$ from $\ket{\bf 0}_A \otimes \ket{\bf 0}_B$, there exists a highly efficient procedure that constructs an exact unitary block encoding of $\rho$ using $U$ and $U^\dagger$ once each.
\end{lemma}

\paragraph{Approach 1 (based on \cite{mcardle2022quantum}).}
Next, we note that a recently introduced state preparation method~\cite{mcardle2022quantum} achieves efficient circuit depth while using only a modest number of ancilla qubits. According to the description below Theorem~1 in~\cite{mcardle2022quantum}, let $f:[-a,a] \rightarrow \mathbb{R}$ be a (preferably smooth) function of arbitrary parity, and define  
\begin{equation}
    \ket{\Phi_f} = \frac{1}{\mathcal{N}_f}\sum_{ i=-{N}/{2}}^{{N}/{2} -1} f \left( \frac{2a i}{N} \right)  \ket{i},    
\end{equation}
where $\mathcal{N}_f = \sqrt{ \sum_{i} f(\cdot)^2}$. Then, there exists a deterministic procedure (see~{\cite[Figure~1]{mcardle2022quantum}}) that prepares the state
\begin{align}
    \frac{1}{\sqrt{N}}\ket{00}\sum_{i=1}^{N} f \left( \frac{2a i}{N}\right) \ket{i}+ \ket{\rm Garbage},
\end{align}
where $\ket{\rm Garbage}$ refers to the redundant part that is completely orthogonal to the first term, $\ket{00}\sum_{i=1}^{N} f \left( \frac{2a i}{N}\right) \ket{i}$. The quantum circuit complexity of the above procedure is $\mathcal{O}\!\left( \deg(f) \log (N) \right)$ (where $\deg(f)$ denotes the degree of $f$), requiring an additional three ancilla qubits. We remark that the function $f$ is assumed to be smooth; however, as discussed in~\cite{mcardle2022quantum}, the method can be extended to non-smooth functions as well, with asymptotically the same complexity. 

To apply the above result to our case, we first introduce the new variable $k = j+(i-1)N$, which satisfies $\ket{k }= \ket{j+(i-1)N} = \ket{j}\ket{i}$. Then, there is a correspondence between the states 
\begin{equation}
    \sum_{i,j=1}^{N,M} A_{ij} \ket{j} \ket{i} = \sum_{k=1}^{N^2} a_k \ket{k},    
\end{equation}
where the new amplitudes $a_k := a_{j +(i-1)N} = A_{ji}$. For any (preferably smooth) function $f: [-1,1] \rightarrow [-1,1]$, if each entry $a_k$ satisfies 
\begin{equation}
    a_k = f\left(\frac{k}{MN}\right) = f \big(j + (i-1)N\big) = A_{ji},    
\end{equation}
then, as mentioned earlier, there exists a deterministic procedure that prepares the state
\begin{align}
\begin{split}
    \frac{1}{\sqrt{MN}} \ket{00}\sum_{k=1}^{MN} f\!\left( \frac{k}{MN}\right) \ket{k}+ \ket{\rm Garbage} &= \frac{1}{\sqrt{MN}} \ket{00}\sum_{k=1}^{MN} a_k\ket{k}+ \ket{\rm Garbage} \\
    &= \frac{1}{\sqrt{MN}}\ket{00}\sum_{i,j=1}^{N,M} A_{ji} \ket{j} \ket{i}+ \ket{\rm Garbage}  \\
    &= \frac{1}{\sqrt{MN}}\ket{00} \sum_{i=1}^{N} A^i\ket{i} + \ket{\rm Garbage}.
\end{split}
    \label{c3}
\end{align}
The above state can be achieved using a quantum circuit with gate complexity $\mathcal{O}\!\left( \deg(f)\log (MN) \right)$ and three ancilla qubits. We note that, in principle, there can be many possible choices of $f$, as long as its outputs match the entries of the matrix $A$ at the corresponding inputs.

To obtain the block-encoding of $\sim A^\dagger A$ from the above state, we append another ancilla initialized in $\ket{00}$ to the state, yielding
\begin{align}
    \frac{1}{\sqrt{MN}} \sum_{i=1}^{N} \underbrace{\ket{00}}_{\rm{register}~\mathsf{1}}\underbrace{\ket{00}}_{\rm{register}~\mathsf{2}} A^i\ket{i} + \ket{00}\ket{\rm Garbage},
\end{align}
followed by two CNOT gates between registers~1 and~2. Then we obtain the state
\begin{align}
    \frac{1}{\sqrt{MN}} \ket{00} \sum_{i=1}^{N} \underbrace{\ket{00} A^i}_{\rm{register}~\mathsf{X}}\ket{i} +  \ket{\rm Garbage'}\ket{\rm Garbage},
\end{align}
where $\ket{\rm Garbage'}$ is completely orthogonal to $\ket{00}$. 
As pointed out earlier, if we trace out the first register of the vector $\left(\sum_{i=1}^{N}  A^i \ket{i}\right)/\sqrt{MN}$, we obtain the operator $A^\dagger A/(MN)$. Hence, if we trace out the register~\textsf{X} in the state above, we obtain the density operator
\begin{align}
    \ket{00}\bra{00} \otimes \frac{A^\dagger A}{MN} + (\cdots),
\end{align}
where $(\cdots)$ denotes the redundant part. The above density operator can be block-encoded via~\cref{lemma: improveddme}. According to~\cref{def: blockencode}, this density operator also serves as a block-encoding of $A^\dagger A / (MN)$, thereby achieving the desired construction.

Recall that the quantum circuit complexity of preparing the state in~\cref{c3} is $\mathcal{O}\!\left( \log (MN) \right)$ (assuming $\deg(f) = \mathcal{O}(1)$). Adding two more ancilla qubits $\ket{00}$ and using two CNOT gates increases the circuit complexity by only $\mathcal{O}(1)$. Finally, the application of~\cref{lemma: improveddme} requires one additional use of the state-preparation unitary in~\cref{c3}, leading to a total circuit complexity of $\mathcal{O}\!\left( \log (MN)\right)$.

\paragraph{Approach 2 (based on \cite{marin2023quantum}).}
In the above procedure, we used the state preparation method in~\cite{mcardle2022quantum} to prepare the state that contains $\sim \sum_{i,j=1}^{N,M} A_{ij} \ket{j}\ket{i}$ as a subvector. However, we remark that other state preparation methods can also be potentially applicable. For example, the methods in~\cite{grover2002creating, marin2023quantum} consider states of the form $\sum_{i=1}^N f(i) \ket{i}$, where $f:[0,1] \rightarrow \mathbb{R}_+$ is a continuous, positive, integrable function satisfying $\sum_{i=1}^N f(i)^2 = 1$. They show that if $f(\cdot)$ is efficiently integrable, then $\sum_{i=1}^N f(i)\ket{i}$ can be prepared with fidelity $\varepsilon$ using a $\log (N)$-qubit quantum circuit of complexity $\mathcal{O}\!\left( 2^{K(\varepsilon)}\right)$, where
\begin{equation}
    K(\varepsilon) = \max  \left\{  -\frac{1}{2} \log_2 \!\big( 4^{-\log (N)} - 96  \log (1-\varepsilon) \big) ,\, 2 \right\},
\end{equation}
which is asymptotically independent of $N$, as noted in~\cite{marin2023quantum}.

To apply this method, we first define $||A||_F = \sqrt{\sum_{i,j}A_{ij}^2}$, and, similarly to the previous approach, introduce a new variable $k = j+(i-1)N$, which satisfies $\ket{k }= \ket{j+(i-1)N} = \ket{j}\ket{i}$. If the entries $\{A_{ij}\}_{i,j=1}^N$ are positive (we will show how to handle negative entries subsequently), and there exists a positive, integrable function $f$ such that 
\begin{equation}
    f\left(\frac{k}{MN}\right) = f\left(\frac{j + (i-1) N }{MN}\right) = \frac{A_{ij}}{||A||_F} ,    
\end{equation}
then it can be used to prepare a state that contains 
\begin{equation}
    \sum_{k=1}^{MN} f\left(\frac{k}{MN}\right) \ket{k} =  \frac{1}{||A||_F}\sum_{i,j=1}^{N,M} A_{ij} \ket{j}\ket{i}    
\end{equation}
as a subvector (similarly to what was achieved in \textbf{Approach~1}, except that the function $f$ must be positive in this case), or even the exact state 
\begin{equation}
    \frac{1}{||A||_F} \sum_{i,j=1}^{N,M} A_{ij} \ket{j}\ket{i}.
\end{equation}

Given the procedure that prepares the above state, we can execute the same sequence of steps described below~\cref{c3} to obtain the block-encoding of $A^\dagger A / ||A||_F^2$. As the final and optional step, an application of~\cref{lemma: scale} with the scaling factor ${||A||_F^2}/{(MN)}$ can be used to transform the block-encoded operator
\begin{equation}
    \frac{A^\dagger A}{||A||_F^2} \rightarrow \frac{A^\dagger A}{MN}.
\end{equation}
Provided that the circuit complexity for preparing 
\begin{equation}
    \frac{1}{||A||_F} \sum_{i,j=1}^{N,M} A_{ij} \ket{j}\ket{i}    
\end{equation}
is $\mathcal{O}\!\left( \log (MN)\right)$, and~\cref{lemma: improveddme} uses this state preparation procedure $\mathcal{O}(1)$ times, the overall complexity for obtaining the block-encoding of the operator above is $\mathcal{O}\!\left( \log (MN) \right)$.

Now, we turn our attention to the case where $A$ contains both positive and negative entries. Define
\begin{equation}
    ||A^+||_F = \sqrt{\sum_{ij,\, A_{ij} \geq 0} A_{ij}^2}, \quad ||A^-||_F = \sqrt{\sum_{ij,\, A_{ij} < 0} A_{ij}^2}.    
\end{equation}
We then consider the states
\begin{align}
\begin{split}
     \ket{\phi_+ } &= \frac{1 }{ ||A^+||_F} \sum_{ij,\, A_{ij} \geq 0} A_{ij} \ket{j}\ket{i}, \\
     \ket{\phi_-} &=  \frac{1 }{ ||A^-||_F} \sum_{ij,\, A_{ij} < 0} A_{ij} \ket{j}\ket{i}.
\end{split}
\end{align}
The state $\ket{\phi_+}$ has nonnegative entries, so it can be efficiently prepared using the previously described procedure. Let $U_+$ denote the unitary that prepares this state, i.e., $U_+\ket{0}^{\log (MN)} = \ket{\phi_+}$. For the state $\ket{\phi_-}$, we consider 
\begin{equation}
    -\ket{\phi_-} = \frac{1}{||A^-||_F}\sum_{ij,\, A_{ij} < 0} (-A_{ij}) \ket{j}\ket{i}.    
\end{equation}
Since $A_{ij} < 0$ implies $-A_{ij} > 0$, the state $-\ket{\phi_-}$ can also be prepared, with the corresponding preparation unitary denoted by $U_-$. Hence, the first columns of $U_+$ and $U_-$ correspond to $\ket{\phi_+}$ and $-\ket{\phi_-}$, respectively. Then, by~\cref{lemma: sumencoding}, we can construct a block-encoding of
\begin{align}
    \left(\frac{||A^+||_F }{||A||_F }\right) U_+ - \left(\frac{||A^-||_F }{||A||_F }\right)U_-,
\end{align}
whose first column is
\begin{align}
   \frac{||A^+||_F }{||A||_F }  \ket{\phi_+} + \frac{||A^-||_F }{||A||_F }\ket{\phi_-}  
   = \frac{1}{||A||_F} \left(\sum_{ij,\, A_{ij} \geq 0} A_{ij} \ket{j}\ket{i} +  \sum_{ij,\, A_{ij} < 0} A_{ij} \ket{j}\ket{i}\right),
\end{align}
which is exactly 
\begin{equation}
    \frac{1}{||A||_F} \sum_{i,j=1}^{N,M} A_{ij} \ket{j}\ket{i}.
\end{equation}
By applying this unitary to $\ket{0}^{\log (MN)}$, we obtain the desired state. We can then proceed as outlined earlier to obtain the block-encoding of $A^\dagger A / (MN)$, with the same overall complexity of order $\mathcal{O}\!\left( \log (MN)\right)$.

Finally, we remark that in the same work~\cite{marin2023quantum}, the authors also proposed a variational approach for state preparation. As empirically demonstrated, this variational approach performs well in practice, requiring only a small amount of training time.

\paragraph{Approach 3 (based on \cite{nakaji2022approximate}).}
Meanwhile, the method proposed in~\cite{nakaji2022approximate} introduces a variational approach to prepare a state of the form $\left(\sum_{i=1}^N x_i \ket{i}\right)/||x||$, where $||x||^2 = \sum_{i=1}^N x_i^2$. As analyzed and numerically verified in~\cite{nakaji2022approximate}, a parameterized quantum circuit with complexity $\mathcal{O}\!\left( l \log N \right)$ (where $l = \mathcal{O}(1)$) is sufficient to prepare the desired state, even without explicit error tolerance. Moreover, the required number of ancilla qubits is at most $\mathcal{O}(1)$. Therefore, a direct application of this state preparation method enables us to prepare the state 
\begin{equation}
    \frac{1}{||A||_F} \sum_{i,j=1}^{N,M} A_{ij} \ket{j}\ket{i}.
\end{equation}
The procedure for obtaining the block-encoding of $A^\dagger A / ||A||_F^2$, and subsequently $A^\dagger A / (MN)$, follows straightforwardly from the steps described in the previous paragraph, achieving the same overall complexity.

\subsection{Application to our case: a possible and simple choice of $f$}  
\paragraph{Block-encoding $\sim \partial_r^\dagger \partial_r$.}
We now return to the context of~\cref{lemma: entrycomputablematrix}, where the matrix $A$ introduced earlier is replaced by the boundary operator $\partial_r$, which has size $|S_{r-1}| \times |S_r|$. As discussed above, by choosing an arbitrary (preferably smooth) function that satisfies
\begin{equation}
    f\!\left(\frac{ j + (i-1) |S_r|}{|S_{r-1}| |S_r|}\right) = (\partial_r)_{ij} \in \{-1,0,1\},    
\end{equation}
the method of~\cite{mcardle2022quantum}, along with the procedure outlined earlier, can be directly applied to obtain the block-encoding of $\partial_r^\top \partial_r / \mathrm{size}(\partial_r)$.  Since $\mathrm{size}(\partial_r)=|S_{r-1}| \times |S_r|$, this block-encoding procedure has a total circuit complexity of $\mathcal{O}\!\left(\log (|S_r||S_{r-1}|)\right)$ and, in particular, requires only $\mathcal{O}(1)$ ancilla qubits.

Alternatively, as pointed out earlier, instead of using~\cite{mcardle2022quantum}, we may employ the state preparation method in~\cite{grover2002creating, marin2023quantum} by choosing a simple, integrable, piecewise-linear function that takes discrete values as
\begin{equation}
    f\!\left(\frac{j + (i-1)|S_r|}{|S_{r-1}|\,|S_r|}\right)
    =  \frac{(\partial_r)_{ij}}{||\partial_r||_F} 
    \in \left\{  -\frac{1}{||\partial_r||_F},\,0,\,\frac{1}{||\partial_r||_F} \right\}.    
\end{equation}

The circuit complexity of this approach is $\mathcal{O}\!\left( 2^{k(\varepsilon)} \right)$ with $k(\varepsilon)$ defined earlier, and, moreover, this approach requires no additional ancilla qubits. 

We note that the second approach, via \cite{marin2023quantum}, contains the factor $||\partial_r||_F$ in the denominator, which is what we stated in~\cref{lemma: entrycomputablematrix}. Therefore, the~\cref{lemma: entrycomputablematrix} is a direct consequence of the previous  paragraph.

\paragraph{A simple choice of $f$.}
As seen above, to apply the methods in~\cite{marin2023quantum} to our case, the function $f$ must satisfy
\begin{equation}
    f\!\left(\frac{k}{|S_{r-1}|\,|S_r|}\right)
= f\!\left(\frac{j + (i-1)|S_r|}{|S_{r-1}|\,|S_r|}\right)
= \frac{(\partial_r)_{ij}}{||\partial_r||_F} ,
\end{equation}
where $k = j + (i-1)|S_r|$. In principle, there can be many valid choices for $f$.  
A particularly simple choice is as follows. We consider the $\mathbb{R}^2$ plane with data points of the form $(k, f({k}\cdot{|S_{r-1}|^{-1}|S_r|^{-1}}))$. Then, proceeding sequentially from lower to higher indices $k = 1, 2, \dots, |S_r|$, we connect consecutive data points. The resulting piecewise-linear function on $\mathbb{R}^2$ satisfies the required condition above.  

Consequently, we can apply various existing results on the approximation of piecewise-linear functions~\cite{lipman2010approximating, jimenez2016transforming, arandiga2005interpolation, amhraoui2022smoothing} to obtain the desired $f$. Although piecewise-linear, this function remains piecewise-smooth and continuous; thus, the rigorous performance guarantees of the methods in~\cite{mcardle2022quantum, marin2023quantum} continue to hold. Hence, we have completed the proof of~\cref{lemma: entrycomputablematrix}.

\section{Inverting ill-conditioned matrices}
\label{sec: invertingillconditionmatrix}
In our cohomological approach, there may be cases where computing the inverse of a matrix with a large condition number is required, such as in the derivation of expressions like~\cref{eq: big_condition_number}. In the following, we briefly introduce some known algorithms related to this task.

To overcome the practical limitations of the original HHL algorithm \cite{harrow2009quantum}, such as the dependence on efficient state preparation, post-selection-based solution extraction, and unfavorable scaling with the condition number $\kappa$, a fully unitary and observable-aware formulation of the quantum linear system algorithm (QLSA) is employed~\cite{clader2013preconditioned}. This framework enables coherent processing throughout the algorithm and supports various forms of readout that do not require full wavefunction collapse.

The input state $\ket{b}$ is not assumed to be given directly. Instead, an ancilla-assisted state $\ket{b_T}$ is prepared using controlled operations and amplitude encoding techniques. The resulting state takes the form
\begin{equation}
    \ket{b_T} = \cos\phi_b \ket{\tilde{b}}\ket{0}_a + \sin\phi_b \ket{b}\ket{1}_a,    
\end{equation}
where $\phi_b$ encodes the overlap between the constructed state and the target vector $\ket{b}$. This preparation avoids the need for an oracle that directly outputs $\ket{b}$, and instead makes use of data oracles that provide individual components $b_j$ via coherent access.

A modified QLSA is then applied in a fully unitary fashion, producing the superposition
\begin{equation}
    \ket{\Psi} = \sqrt{1 - \sin^2\phi_b \sin^2\phi_x} \ket{\Phi_0} + \sin\phi_b \sin\phi_x \ket{x}\ket{1}_a\ket{1}_a,    
\end{equation}
where $\ket{x} = A^{-1}\ket{b} / \|A^{-1}\ket{b}\|$ is the normalized solution to the linear system, and $\phi_x$ encodes the success amplitude of the inversion procedure. The desired solution is coherently embedded in the subspace where both ancilla qubits are in the state $\ket{1}$.

Information about $\ket{x}$ is extracted through observable-aware readout techniques, without full-state tomography. First, to estimate the overlap $|\braket{R}{x}|^2$ with a reference state $\ket{R}$, a swap test is performed using a similarly prepared reference superposition $\ket{R_T}$. The resulting probability amplitudes satisfy
\begin{equation}
    |{\braket{R}{x}}|^2 = \frac{P_{1110} - P_{1111}}{\sin^2\phi_b \sin^2\phi_x \sin^2\phi_r},    
\end{equation}
where $P_{111b}$ denote measurement probabilities conditioned on specific ancilla outcomes.

Second, expectation values of polynomial observables $x^n$ can be estimated by constructing Hamiltonians of the form $H_{rw} = x^n \ket{x}\bra{x}$, implemented via ancilla-assisted controlled rotations.

Third, to access individual solution components $x_j$, amplitude estimation is applied to the subspace corresponding to the computational basis state $\ket{j}$. This allows estimation of $|{x_j}|^2$ with quadratically improved sample complexity compared to classical sampling.

To mitigate the dependence on the condition number $\kappa$, a classical preconditioning step is employed. Specifically, a sparse approximate inverse (SPAI) preconditioner $M$ is constructed such that $MA \approx I$, allowing the modified linear system $MA\ket{x} = M\ket{b}$ to be solved instead of the original one. Each column $\hat{m}_k$ of $M$ is determined by solving the minimization problem of $\|\hat{A} \hat{m}_k - \hat{e}_k\|_2$, where $\hat{e}_k$ is the $k$th standard basis vector, subject to a specified sparsity pattern. This sparsity constraint ensures that the resulting preconditioner $M$ is block-encodable and compatible with efficient Hamiltonian simulation techniques. The transformed system benefits from a much smaller effective condition number, i.e., $\kappa(MA) \ll \kappa(A)$, thereby enhancing the efficiency and stability of the quantum algorithm.

\section{Proof of~\cref{lemma: traceestimation}}
\label{sec: rankestimation}

The goal is to estimate $\mathrm{rank}(A)$ given a block-encoding of a matrix $A \in \mathbb{C}^{N \times N}$. Our approach proceeds by reducing rank estimation to trace estimation after an appropriate spectral transformation.

\paragraph{Trace estimation via diagonal averaging.}
A basic identity is
\begin{equation}
    \Tr(A) = \sum_{i=1}^N \bra{i} A \ket{i}.
\end{equation}
Thus, estimating $\Tr(A)$ reduces to estimating diagonal entries of $A$ and averaging. Further, we point out a property, which also appeared in \cite{ubaru2021quantum} that for a Hermitian matrix $A$, it holds that $\rank (A) = \Tr (h (A))$ where $h(.)$ is the step function. Thus, to estimate $\mathrm{rank}(A)$, we apply n $h$ to $A$ such that $\Tr(h(A)) = \mathrm{rank}(A)$ (e.g., a spectral thresholding function). It therefore suffices to estimate $\Tr(h(A))$.

\paragraph{Quantum procedure.}
Suppose for now that we are given a block-encoding of $A$. To transform this block-encoding to the block-encoding of $h(A)$, we apply Lemma \ref{lemma: qsvt} with the polynomial $P(.)$ to be the step function $h(.)$. According to \cite{gilyen2019quantum}, for $x \in [\beta,1]$, then $h(x)$ can be approximated with an additive error $\epsilon$ by a polynomial of degree $\mathcal{O}\big( \frac{1}{\beta} \log \frac{1}{\epsilon} \big)$. Suppose for simplicity that the operator norm of $A$ is $\leq 1$, so its minimum eigenvalue (in magnitude) is $\sim \frac{1}{\kappa_A}$ where $\kappa_A$ is the condition number of $A$. So, to transform the block-encoding of $A$ to $h(A)$, we need to approximate the step function $h(.)$ by a polynomial of degree $\mathcal{O}\big( \kappa_A \log \frac{1}{\epsilon} \big)$.

Next, by using~\cref{lemma: tensorproduct}, we can construct a block-encoding of $\mathbb{I}_N \otimes h(A)$, which we denote by $U_A$.

We prepare the state
\begin{equation}
    \ket{\psi} = \ket{\mathbf{0}} \otimes \frac{1}{\sqrt{N}} \sum_{i=1}^N \ket{i}\ket{i}.
\end{equation}
Applying $U_A$ yields
\begin{equation}
    \ket{\phi}
    =
    \ket{\mathbf{0}} \otimes \frac{1}{\sqrt{N}} \sum_{i=1}^N \ket{i}\, h(A)\ket{i}
    + \ket{\mathrm{garbage}}.
\end{equation}
Taking the overlap with $\ket{\psi}$, we obtain
\begin{equation}
    \braket{\psi,\phi}
    =
    \frac{1}{N} \sum_{i=1}^N \bra{i} h(A) \ket{i}
    =
    \frac{1}{N} \Tr( h(A)).
\end{equation}
Thus, estimating this overlap yields $\Tr( h(A))/N$.

\paragraph{Overlap estimation.}
The overlap $\braket{\psi,\phi}$ can be estimated using the Hadamard test or the SWAP test. Concretely, given state-preparation unitaries $U_1, U_2$ such that
\[
    U_1 \ket{0} = \ket{\phi_1}, \qquad
    U_2 \ket{0} = \ket{\phi_2},
\]
we prepare
\begin{equation}
    \frac{1}{\sqrt{2}} \left( \ket{0}\ket{\phi_1} + \ket{1}\ket{\phi_2} \right),
\end{equation}
apply a Hadamard gate on the first qubit, and measure. The probability of obtaining $\ket{0}$ is
\begin{equation}
    p_0 = \frac{1}{2}\left(1 + \Re \braket{\phi_1,\phi_2}\right),
\end{equation}
from which $\Re \braket{\phi_1,\phi_2}$ can be inferred.

Using amplitude estimation, this overlap can be estimated to additive error $\epsilon$ with success probability at least $1-\xi$ using
\begin{equation}
    \mathcal{O}\!\left(
        \frac{1}{\sqrt{\epsilon}} \log\!\left(\frac{1}{\xi}\right)
    \right)
\end{equation}
applications of the underlying unitary.

\paragraph{Error propagation.}
We use the following elementary fact: if $0 \le x \le 1$ and $\sqrt{x}$ is estimated to additive error $\sqrt{\epsilon}$, then $x$ can be estimated to additive error $\epsilon$. Indeed, letting $\widetilde{\sqrt{x}}$ denote the estimate, we have
\begin{align}
    \big| \widetilde{\sqrt{x}}^{\,2} - x \big|
    &=
    \big| (\widetilde{\sqrt{x}} - \sqrt{x})(\widetilde{\sqrt{x}} + \sqrt{x}) \big| \\
    &\le
    |\widetilde{\sqrt{x}} - \sqrt{x}| \cdot 2 \\
    &\le
    \mathcal{O}(\sqrt{\epsilon}),
\end{align}
and tightening constants yields an $\mathcal{O}(\epsilon)$ bound.

Applying this with
\[
    x = 1 + 2 \Re \braket{\phi_1,\phi_2},
\]
we conclude that an estimate of the amplitude suffices to recover $\Re \braket{\phi_1,\phi_2}$ to additive error $\epsilon$.

\paragraph{Conclusion.} To obtain the final complexity, we mention that we first need to use~\cref{lemma: qsvt} to transform the block-encoding of $A$ to $h(A)$. As the polynomial approximation to $h(A)$ has degree $\mathcal{O}( \kappa_A \log \frac{1}{\epsilon})$, the complexity for this step is $\mathcal{O}( T_A \kappa_A \log \frac{1}{\epsilon})$ where $T_{A}$ is the cost of the block-encoding $h(A)$. Next, we use the procedure above to estimate the overlaps, which contain the desired value $\Tr(h(A))/N$. Combining the above, we obtain an estimate of $\Tr(h(A))/N$ to additive error $\epsilon$ with complexity
\[
    \mathcal{O}\!\left(
        \frac{T_{A}}{\sqrt{\epsilon}} \log (\frac{1}{\epsilon})\log\!\left(\frac{1}{\xi}\right)
    \right),
\]

Applying this procedure to $h(A)$ yields an estimate of $\mathrm{rank}(A)/N$, completing the proof.

\end{document}